\documentclass[a4paper,11pt]{article}
\usepackage{jheppub}
\usepackage{amsmath}
\usepackage{amssymb}
\usepackage{color}
\usepackage{graphicx}
\usepackage{hyperref}
\usepackage{xspace,slashed}
\usepackage[utf8]{inputenc}
\usepackage{subcaption}

\usepackage{stackengine}

\usepackage{soul}

\makeatletter
\g@addto@macro\bfseries{\boldmath}
\makeatother

\newcommand{\order}[1]{\mathcal{O}\left(#1\right)}

\newcommand{\GeV}{\;\mathrm{GeV}}
\newcommand{\TeV}{\;\mathrm{TeV}}
\newcommand{\cL}{\mathcal{L}}
\newcommand{\cN}{\mathcal{N}}
\newcommand{\as}{\alpha_s}
\newcommand{\jet}{\text{jet}}

\newcommand{\leading}{{(\ell)}}
\newcommand{\nlnb}{n\ell}
\newcommand{\nonleading}{{(\nlnb)}}
\newcommand{\nl}{\nonleading}

\usepackage{xcolor}
\definecolor{light-gray}{gray}{0.8}

\preprint{CERN-TH-2018-151}

\newcommand{\CERNaff}{CERN, Theoretical Physics Department, CH-1211
  Geneva 23, Switzerland} 
\newcommand{\MITaff}{Center for Theoretical Physics,
  Massachusetts Institute of Technology,
  Cambridge, MA 02139, USA}
\newcommand{\IPhTAff}{IPhT,
  Universit\'{e} Paris-Saclay, CNRS UMR 3681, CEA Saclay, F-91191 Gif-sur-Yvette,
  France}
\newcommand{\CNRSaff}{CNRS, UMR 7589, LPTHE, F-75005, Paris, France}
\newcommand{\OXaff}{Rudolf Peierls Centre for Theoretical Physics,
  Clarendon Laboratory, Parks Road, Oxford OX1 3PU, UK}
\newcommand{\ASCaff}{All Souls College, Oxford OX1 4AL, UK}

\title{The Lund Jet Plane}
\author[a,b]{Fr\'ed\'eric A. Dreyer,}
\affiliation[a]{\MITaff}
\affiliation[b]{\OXaff}
\author[c,b,d,*]{Gavin P.\ Salam,\note[*]{On leave from \CNRSaff}}
\affiliation[c]{\CERNaff}
\affiliation[d]{\ASCaff}
\author[e]{Gr\'egory Soyez}
\affiliation[e]{\IPhTAff}

\abstract{
  Lund diagrams, a theoretical representation of the phase space
  within jets, have long been used in discussing parton showers and
  resummations.
  We point out that they can be created for individual jets through
  repeated Cambridge/Aachen declustering, providing a powerful visual
  representation of the radiation within any given jet.
  Concentrating here on the primary Lund plane, we outline some of its
  analytical properties, highlight its scope for constraining Monte
  Carlo simulations and comment on its relation with existing
  observables such as the $z_g$ variable and the iterated soft-drop
  multiplicity.
  We then examine its use for boosted electroweak boson tagging at
  high momenta.
  It provides good performance when used as an input to machine
  learning.
  Much of this performance can be reproduced also within
  a transparent log-likelihood method, whose underlying assumption is
  that different regions of the primary Lund plane are largely
  decorrelated. 
  This suggests a potential for unique insight and experimental
  validation of the features being used by machine-learning
  approaches.
}

\begin{document}

\maketitle 
\section{Introduction}

Jets, the collimated bunches of hadrons that result from the
fragmentation of energetic quarks and gluons, are among the most
fascinating objects that are used at colliders.
The study of the internal structure of jets has become a prominent
area of research at CERN's Large Hadron Collider, both
theoretically~\cite{Larkoski:2017jix} and
experimentally~\cite{Asquith:2018igt}.
This is a reflection of its power to probe the Higgs sector of the
Standard Model (SM) and to search for physics beyond the Standard Model
(BSM), but also of the considerable scope for learning more about the
quantum chromodynamics (QCD) associated with the development of jets.

Theoretical and phenomenological work on jet substructure has taken
two main directions in the past years.
On one hand there has been extensive effort to manually construct
observables (see e.g.\
Refs.~\cite{Butterworth:2008iy,Thaler:2008ju,Kaplan:2008ie,Ellis:2009su,Ellis:2009me,Plehn:2009rk,Thaler:2010tr,Thaler:2011gf,Larkoski:2013eya,Chien:2013kca,Larkoski:2014gra,Larkoski:2014pca,Moult:2016cvt,Salam:2016yht})
that can help distinguish between different origins of jets: e.g.\
those stemming from the hadronic decay of a boosted W/Z/H boson or top
quark (signals), versus those from the fragmentation of a quark or
gluon (background).
That effort has been accompanied by extensive calculations of the
properties of those observables within perturbative QCD (e.g.\
Refs.~\cite{Seymour:1997kj,Feige:2012vc,Dasgupta:2013ihk,Dasgupta:2013via,Chien:2014nsa,Bertolini:2015pka,Dasgupta:2015lxh,Frye:2016okc,Frye:2016aiz,Dasgupta:2016ktv,Marzani:2017kqd,Marzani:2017mva,Larkoski:2017cqq,Moult:2017okx})
and experimental measurements of their distributions (e.g.\
Refs.~\cite{Khachatryan:2014vla,Aad:2015rpa,Aad:2015owa,Aaboud:2017qwh,CMS:2018yfn}).

In addition to the study of specific manually crafted observables,
several groups have highlighted the power of machine-learning (ML)
approaches to exploit jet-substructure information, using a variety of
ML architectures.
As inputs they have mainly considered discretised images of the
particles inside a
jet~\cite{Cogan:2014oua,deOliveira:2015xxd,Komiske:2016rsd,Kasieczka:2017nvn},
clustering histories from sequential recombination jet
algorithms~\cite{Louppe:2017ipp,Egan:2017ojy,Andreassen:2018apy}, or a basis of
substructure
observables~\cite{Datta:2017rhs,Datta:2017lxt,Komiske:2017aww}.
The performances that they obtain for signal versus background
discrimination are often substantially better than those based on
manually constructed observables.
This good performance comes, however, at the price of limited clarity
as to what jet substructure features are actually being exploited.
One consequence is that it is difficult to establish to what extent
widely-used modelling tools, e.g.\ parton-shower Monte Carlo
generators and detector simulations, reliably predict those features,
an aspect that is critical for the quantitative interpretation of
collider searches.\footnote{Note that some approaches attempt to
  circumvent the use of modelling
  tools~\cite{Metodiev:2017vrx,Dery:2017fap}.}

The purpose of this article is to introduce a representation of the
internal structure of jets that helps bridge the fault-line between
manually constructed observables and ML approaches.
In particular, we ask whether it is possible to organise the
information within a jet such that (a) it can be straightforwardly
measured and understood in data (b) it can be manually organised into
transparent and physically motivated discrimination observables
(without ML) and (c) it can serve as an input to ML for
signal/background discrimination, specifically one whose main
discriminating characteristics can be clearly identified and
understood.

The representation that we use is inspired by Lund
diagrams~\cite{Andersson:1988gp}, which serve as a theoretical
representation of the phase-space within jets and are often used in
discussions of Monte Carlo parton shower algorithms and resummation
of logarithmically enhanced terms in perturbation theory.
In a Lund diagram the available phase-space is mapped
to a triangle in a two dimensional (logarithmic) plane that shows the
transverse momentum and the angle of any given emission with respect
to its emitter.
Each given emission creates new phase space (a triangular leaf) for
further emissions.
One of the key observations of this paper is that Lund diagrams need
not merely be a construct for theoretical calculations.
They can be constructed for individual jets, essentially by following
the clustering tree of the
Cambridge/Aachen~\cite{Dokshitzer:1997in,Wobisch:1998wt} jet algorithm.
The pattern of emissions, notably within the first triangular
phase-space region, the primary Lund plane, carries considerable
information about the jet.

\section{Lund diagrams and the primary Lund plane}
\label{sec:lund-plane}

\begin{figure}
  \centering
  \includegraphics[width=0.8\textwidth]{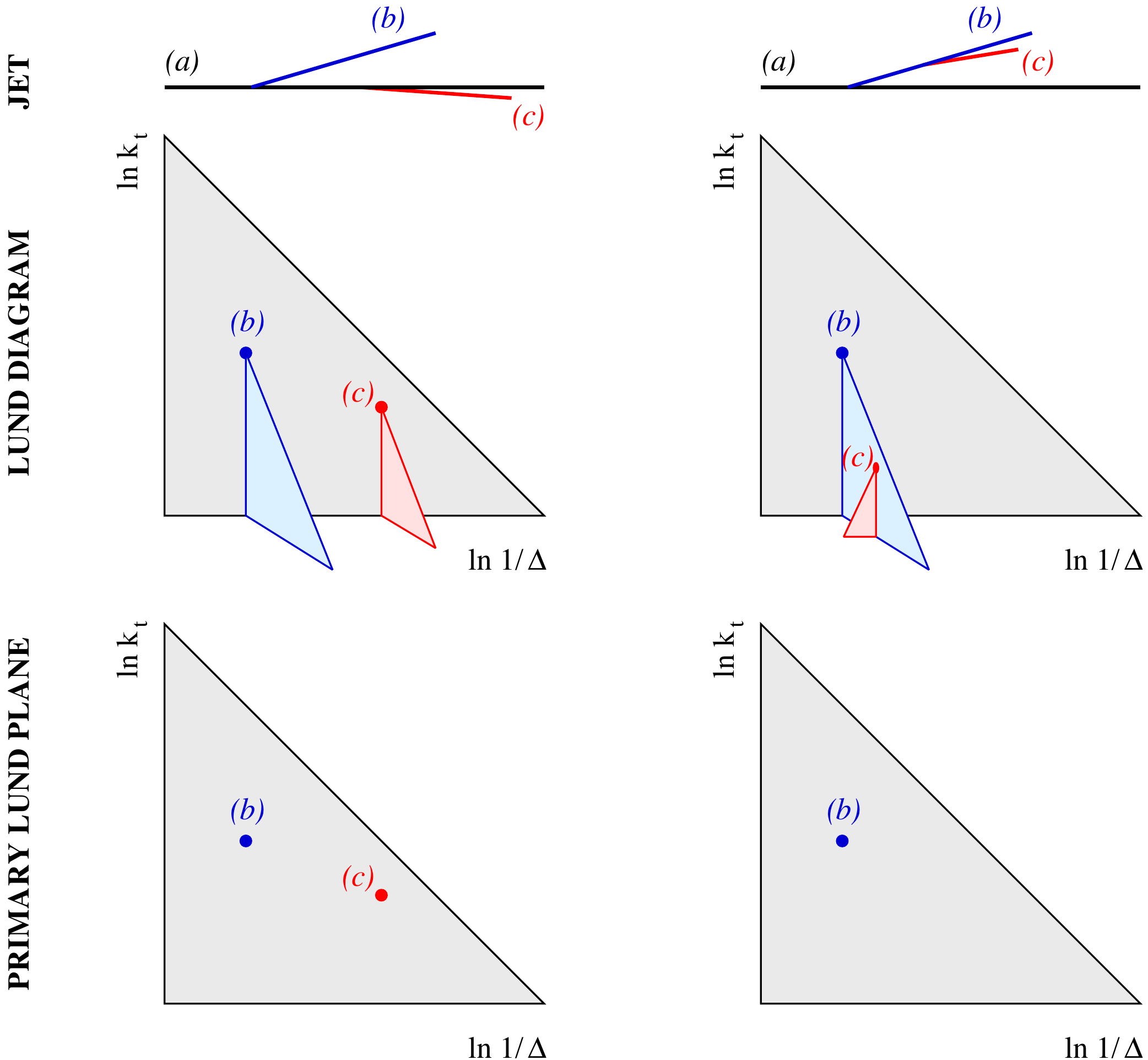}
  \caption{Different representations for two jets. Top: the particles
    inside the jet.
    Middle: the full Lund diagram.
    Bottom: the primary Lund plane.
    See text for further details.
  }
  \label{fig:lund-diagram-explanations}
\end{figure}

To help understand how the primary Lund plane is constructed,
Fig.~\ref{fig:lund-diagram-explanations} shows three representations
for each of two jets. 

The top representation shows the set of particles in the jet, with the
direction and length of each line segment schematically representing
the direction and scalar momentum of the corresponding particle.
The black particle $(a)$ is the primary particle, i.e.\ the one that
initiated the jet. Particles $(b)$ and $(c)$ are emissions inside the
jet. 

The middle representation gives the full Lund diagrams for each of the
two jets.
The phase-space for emission from each particle is represented as a
triangle in a $\ln \Delta$ and $\ln k_t$ plane, where $\Delta$ and
$k_t$ are respectively the angle and transverse momentum of an
emission with respect to its emitter.
The triangles are colour-coded to match the colours of the particles
in the upper row.
The black triangle represents the primary phase space, i.e.\ emission
from $(a)$ (our classification of which particle emits which other
ones is based on the concept of angular ordering of emissions).
Considering the left-hand jet, the blue particle $(b)$ in the jet is
represented as a blue point at the appropriate $(\Delta, k_t)$
coordinate on the (black) triangle associated with its emitter ($a$).
The blue particle has its own phase-space region, the blue triangle,
which is known as a secondary Lund triangle, or ``leaf'' where the
particle could have, but in this case didn't, emit.
Similarly for the red particle, $(c)$, which is also emitted from $(a)$.
In contrast, for the right-hand jet, $(c)$ was emitted from $(b)$ and
so its point appears on the (secondary) blue triangle associated with
particle $(b)$, while its red phase-space triangle emerges as a
tertiary triangle, or leaf, off $(b)$'s triangle.

Finally, the bottom diagram shows the primary Lund plane, which
contains just the positions of the emissions from $(a)$, but no
information about what further secondary emissions may have been
produced.
It is this simpler representation that we will use throughout most of
the article.

\subsection{Construction of the primary Lund plane}
\label{sec:lund-construction}

Our starting point for constructing the primary Lund plane is to
(re-)cluster a jet's constituents with the Cambridge-Aachen (C/A)
algorithm~\cite{Dokshitzer:1997in,Wobisch:1998wt}, which has
significant advantages over other members of the
generalised-$k_t$~\cite{Cacciari:2008gp} family (see
section~\ref{sec:sec-other-declusterings}).%
\footnote{Throughout
  this paper, we also use the C/A algorithm for the initial jet
  finding. The case where jets are clustered with the anti-$k_t$
  algorithm (and re-clustered with the C/A algorithm) is discussed in
  Appendix~\ref{sec:anti-kt}.}
The C/A algorithm identifies the pair of particles $i$ and $j$ closest
in rapidity ($y = \ln \frac{E+p_z}{E-p_z}$, with $E$ and $p_z$ the
particle's energy and longitudinal momentum with respect to the
colliding beams) and azimuth $\phi$, i.e.\ with the minimal value
of $\Delta_{ij}^2 = (y_i - y_j)^2 + (\phi_i-\phi_j)^2$.
It then recombines them into a ``pseudojet'' with momentum
$p = p_{i}+p_{j}$.
This procedure is repeated until all particles (and pseudojets) have
been recombined, or are separated by $\Delta_{ij}$ larger than some
parameter $R$.

To create a primary Lund plane representation of a jet we then work
backwards through the C/A clustering.
One starts with the full jet and then proceeds as follows:
\begin{enumerate}
\item Decluster the current object to produce two pseudojets, $p_a$
  and $p_b$, labelled such that $p_{ta}>p_{tb}$, where $p_{ti}$ is the
  transverse momentum of $i$ with respect to the colliding beams.
  We will consider $p_b$ to be the emission and $p_a+p_b$ to be the
  emitter. 
  In the limit where $p_b$ carries little momentum relative to $p_a$,
  $p_a+p_b$ and $p_a$ can be thought of being the same particle, simply
  differing through the loss of a small amount of momentum by the
  radiation of a gluon $p_b$.

\item Determine a number of variables associated with the
  declustering, e.g.\
  \begin{subequations}
    \label{eq:branching-variables}
    \begin{align}
      \label{eq:branching-variables-a}
      \Delta &\equiv \Delta_{ab},\quad
               k_t \equiv p_{tb} \Delta_{ab},\quad
               m^2 \equiv  (p_a + p_b)^2,
      \\
      \label{eq:branching-variables-b}
      z &\equiv \frac{p_{tb}}{p_{ta}\!+\!p_{tb}},
          \quad
          \kappa \equiv z\Delta\,,
          \quad
          \psi \equiv \tan^{-1} \frac{y_b\!-\!y_a}{\phi_b \!-\! \phi_a},
    \end{align}
  \end{subequations}
  In the limit $p_{tb}\ll p_{ta}$ and $\Delta \ll 1$, $k_t$ is the
  transverse momentum of particle $b$ (the emission) relative to its
  emitter, $\psi$ is an azimuthal angle around the (sub)jet axis, and $z$ is
  the momentum fraction of the branching.
  In our default definition of the Lund plane, the coordinates
  associated with this declustering will be $\ln \Delta$ and $\ln
  k_t$.
  One may also, however, make other choices of coordinates, such as
  for example $\ln \Delta$ and $\ln \kappa$, or $\ln \Delta$ and
  $\ln k_t/p_{t,\text{jet}}$ (with $p_{t,\text{jet}}$ the jet
  transverse momentum).
  We will denote the variables as a tuple
  $\mathcal{T}^{(i)} = \{k_t^{(i)}, \Delta^{(i)}, \ldots\}$ for the
  $i^\text{th}$ iteration of this step.
  
\item Repeat the procedure by going to step 1 for the harder branch,
  $p_a$.
\end{enumerate}
This procedure gives an ordered list of tuples of variables
\begin{equation}
  \label{eq:tuple-list}
  \mathcal{L}_\text{primary} = \left[\mathcal{T}^{(i)}, \ldots, \mathcal{T}^{(n)} \right]
\end{equation}
containing the kinematic variables for each of the primary branchings
off the main emitter.
The $k_t$ and $\Delta$ elements of the tuples (specifically their
logarithms) can be interpreted as set of coordinates of points in the
Lund plane, corresponding to the full set of primary branchings, as in
the lower row of Fig.~\ref{fig:lund-diagram-explanations}.
The tuple elements other than $k_t$ and $\Delta$ provide complementary
information for each point.

One could additionally follow the lower $p_t$ branch at each
declustering.
This would effectively create secondary, tertiary, etc., Lund planes
(or triangles), i.e.\ one for each emission, giving the full Lund
diagram as in the middle row of
Fig.~\ref{fig:lund-diagram-explanations}.
We postpone the study of full Lund diagrams to future work, although a
brief discussion of the use of a secondary Lund plane is given in
appendix~\ref{sec:sec-lund-plane}.

A last point in this subsection is to consider infrared and collinear
safety.
The full list of tuples produced by the primary declustering procedure
is not infrared and collinear safe.
For example, if one adds an infinitesimally soft emission, it may lead
to an additional primary declustering, and so an extra tuple in the list.
However if one considers the subset of tuples in some specified finite 
region of the Lund plane (for example all tuples with $k_t$ larger
than some cut, or all tuples in a given two-dimension bin of $k_t$ and
$\Delta$), then one recovers infrared and collinear safety.
Specifically, the extra soft emission will not modify that subset,
because either it will be declustered as a primary emission that is
outside that finite region, or it will be clustered with a harder
emission inside that region, but will not modify the kinematic
variables of the tuple associated with that harder emission's own
primary declustering.
In practice only the pattern of declusterings with
$k_t \gg \Lambda_\text{QCD}$ is amenable to perturbative calculation.

\subsection{Averaged Lund plane density and basic analytical properties}
\label{sec:agvd-lund-plane}

\begin{figure}
  \centering
  \begin{subfigure}{0.49\textwidth}
    \includegraphics[width=\textwidth,page=1]{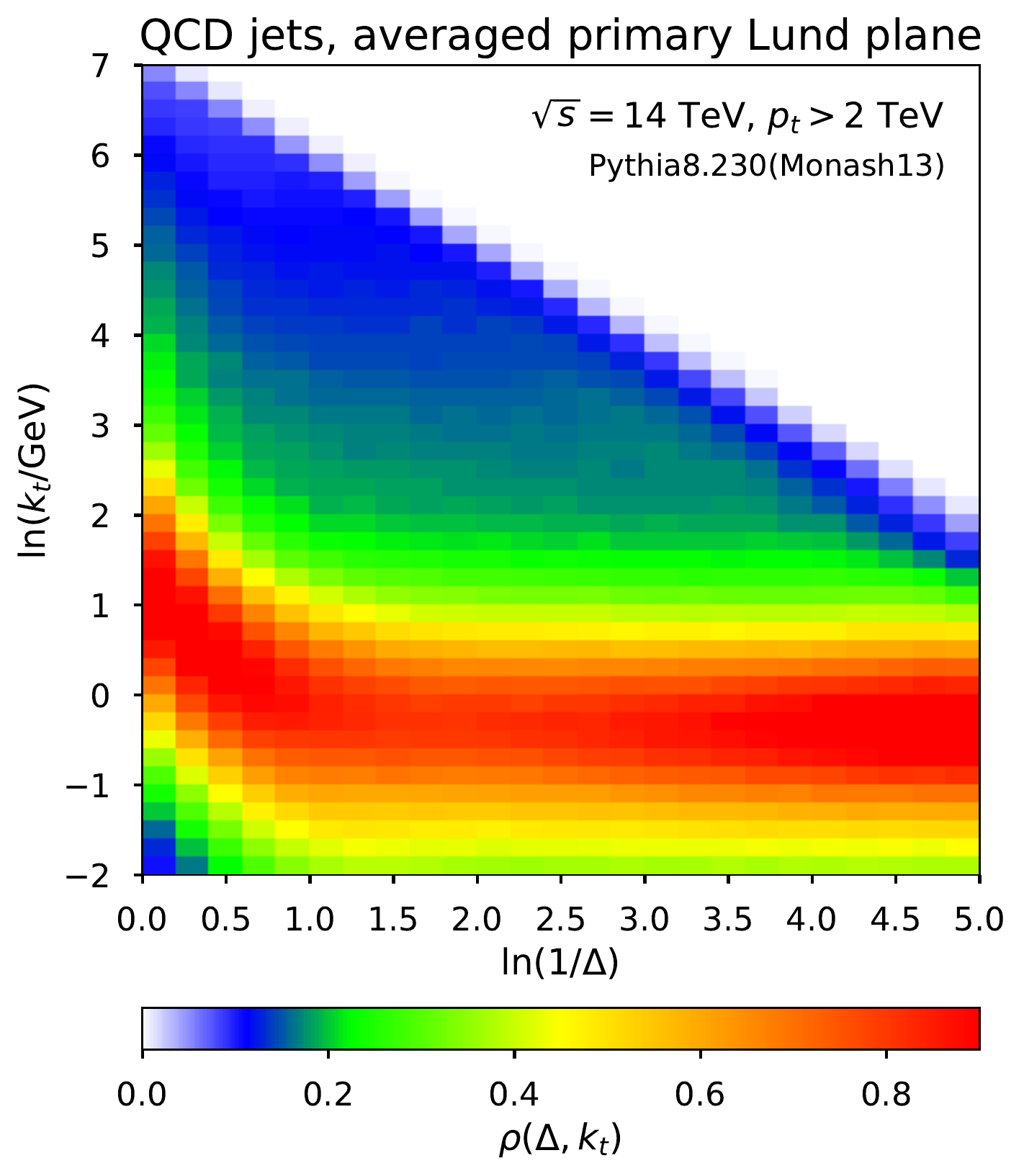}
    \caption{}
    \label{fig:avg-lund-plane}
  \end{subfigure}
  \hfill
  \begin{subfigure}{0.49\textwidth}
    \includegraphics[width=\textwidth]{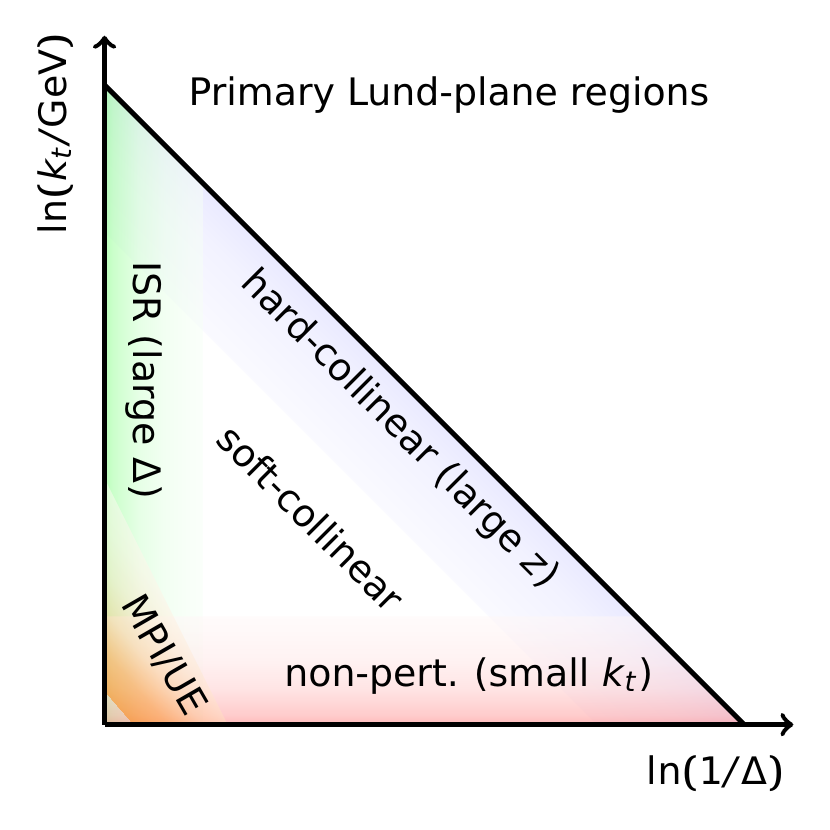}\vspace{1.5cm}
    \caption{}
    \label{fig:schematic-lund-plane}
  \end{subfigure}
    
  \caption{(a) The average primary Lund plane density, $\rho$, for jets
    clustered with the C/A algorithm and $R=1$ having $p_t > 2\TeV$
    and $|y|<2.5$, in a simulated QCD dijet sample.
    (b) Schematic representation of the different regions of the Lund
    plane. 
  }
  \label{fig:average-lund-plane}
\end{figure}

The simplest analysis of the Lund plane is to examine the average
density of points per jet and per unit area in the $\ln
k_t\,$--$\,\ln \Delta$ plane, which we denote
\begin{equation}
  \label{eq:rho-def}
  \rho(\Delta, k_t) = \frac{1}{N_\jet} \frac{dn_\text{emission}}{d\ln k_t \, d\ln
    1/\Delta}\,.
\end{equation}
One can also define a density in terms of dimensionless variables,
e.g.\
\begin{equation}
  \label{eq:rhobar-def}
  \bar \rho(\Delta, \kappa) = \frac{1}{N_\jet} \frac{dn_\text{emission}}{d\ln \kappa \, d\ln
    1/\Delta}\,.
\end{equation}
The quantity $\rho(\Delta, k_t)$ is represented in
Fig.~\ref{fig:avg-lund-plane} for a sample of (C/A, $R=1$) jets with
$p_t > 2\TeV$, simulated using the dijet process in
Pythia~8.230~\cite{Sjostrand:2014zea} with the Monash13
tune~\cite{Skands:2014pea}.
For the case of a quark-initiated jet (about $80\%$ of the jets in the
sample Fig.~\ref{fig:avg-lund-plane}), to leading order in
perturbative QCD and for $\Delta \ll 1$, one expects
\begin{equation}
  \label{eq:lund-density}
  \rho 
  \simeq
  \frac{\as(k_t) C_F }{\pi}
  {\bar z}\left(p_{gq}(\bar z) + p_{gq}(1-\bar z)\right),\;\;
  \bar z = \frac{k_t}{ p_{t,\text{jet}} \Delta},
  \qquad\qquad (\Delta \ll 1)\,,
\end{equation}
where $C_F = \frac43$, $p_{gq}(z) = \frac{1+(1-z)^2}{z}$ ($0<z<1$);
$\bar z$ is an effective momentum fraction and coincides with $z$ in
Eq.~(\ref{eq:branching-variables}) when there is a single
emission.
For $\bar z \ll 1$ the $\bar z$-dependent factor in $\rho$ is equal to
$2$ and so the density of primary Lund emissions is just proportional
to the strong coupling,
\begin{equation}
  \label{eq:lund-density-smallz}
  \rho 
  \simeq
  \frac{2\as(k_t) C_F }{\pi},
  \qquad\qquad (\Delta \ll 1,\, \bar z\ll 1)\,,
\end{equation}
The upper diagonal edge in the figure is a consequence of the
kinematic limit, $k_t < \frac12 p_{t,\jet} \Delta$.
At low scales $\as(k_t)$ gets large, which accounts for the bright red
band around $k_t = 1\GeV$.
In this region the Lund plane density is not amenable to perturbative
calculation.
Equivalently Eq.~(\ref{eq:lund-density}) receives large corrections
from non-perturbative terms proportional to powers of
$k_t/\Lambda_\text{QCD}$.
At values of $\Delta \sim 1$, initial state radiation (ISR) and
multi-parton interactions (MPI/UE) contribute to increasing the
density, which is reflected in 
the contours of constant colour bending upwards to the left.
The different regions are outlined schematically in
Fig.~\ref{fig:schematic-lund-plane}.

Beyond leading perturbative order, several further physical
effects contribute to the structure of the Lund plane.
The upper boundary gets smeared out because of degradation of the
leading subjet energy as one declusters the jet.\footnote{This
  smearing does not occur if one examines $\bar \rho(\Delta,\kappa)$,
  from Eq.~(\ref{eq:rhobar-def}), since $\kappa$ is defined in terms
  of the local $z$ fraction of the emission, which does not depend on
  earlier splittings at larger angles (while $k_t$ does).
  However, instead the non-perturbative boundary gets smeared, as does
  the relation between a given location on the plane and the invariant
  mass of the pair being declustered.}
The leading subjet can also change flavour as one moves down the
clustering tree, in particular when there is an emission close to
the upper, kinematic boundary.
This can then alter the density of emissions at smaller angles, i.e.\
subsequent declusterings.
The underlying physics of these two effects is closely connected with
small-$R$ resummations, cf.\
Refs.~\cite{Dasgupta:2014yra,Dasgupta:2016bnd}.
Non-global~\cite{Dasgupta:2001sh} and
clustering~\cite{Appleby:2002ke,Delenda:2006nf} logarithms introduce
correlations between regions of the Lund plane at similar $\Delta$
values but different $k_t$'s.
For each effect that introduces a correlation, there is typically
also an impact on the average Lund density beyond leading order.
We leave the detailed study of these contributions to future work.
Note that we expect them to contribute generically at an accuracy
$\as^n L^{n-1}$, where $L$ is some combination of $\ln k_t$ and
$\ln \Delta$.
This is the same logarithmic order as the running coupling effects
that we have explicitly highlighted in Eqs.~(\ref{eq:lund-density})
and (\ref{eq:lund-density-smallz}).
Below, in section~\ref{sec:sec-other-declusterings}, we will see
evidence that in the bulk of the Lund plane running effects dominate
numerically over these other effects.
Finally, one should keep in mind that most of these subleading effects
are, at least to some extent, included in modern Monte Carlo programs.
(The extent to which there are differences between Monte Carlo
programs is discussed next.)
In that respect these effects are expected to have only a modest impact
on the tagging studies carried out in Section~3.

\subsection{Use for measurements and constraints on Monte-Carlo
  generators}

\label{sec:MC-rho}

\begin{figure}
  \centering
  \begin{subfigure}{0.49\textwidth}
    \includegraphics[width=\textwidth,page=1]{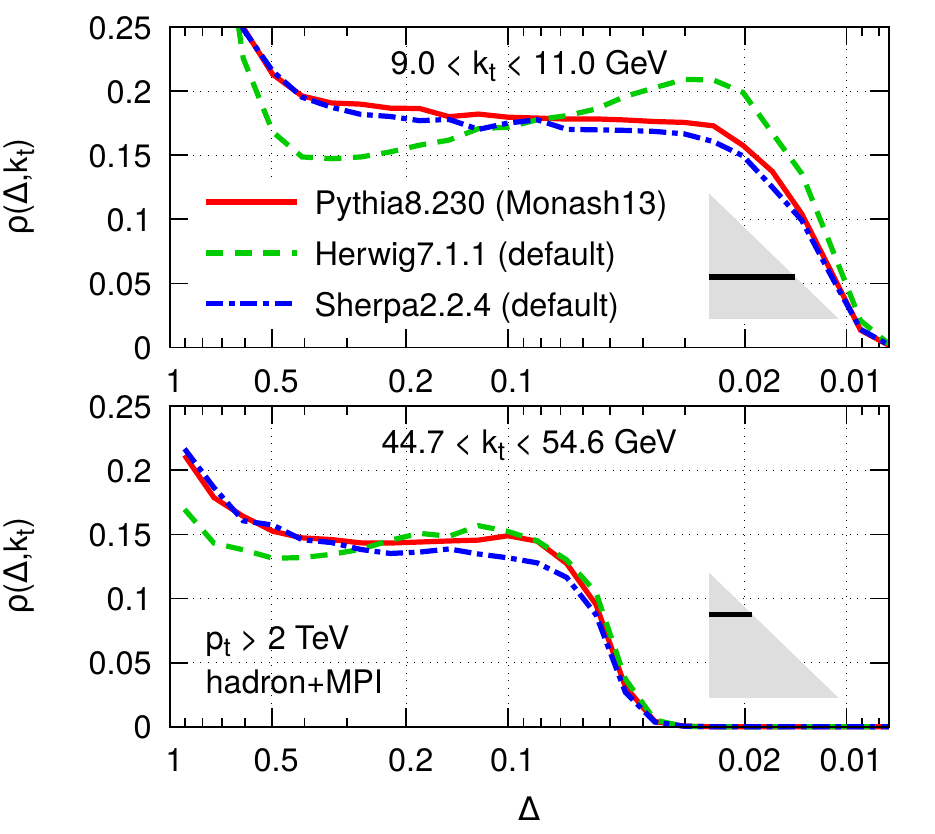}%
    \caption{}
    \label{fig:slices-fixed-kt}
  \end{subfigure}%
  \begin{subfigure}{0.49\textwidth}
    \includegraphics[width=\textwidth,page=1]{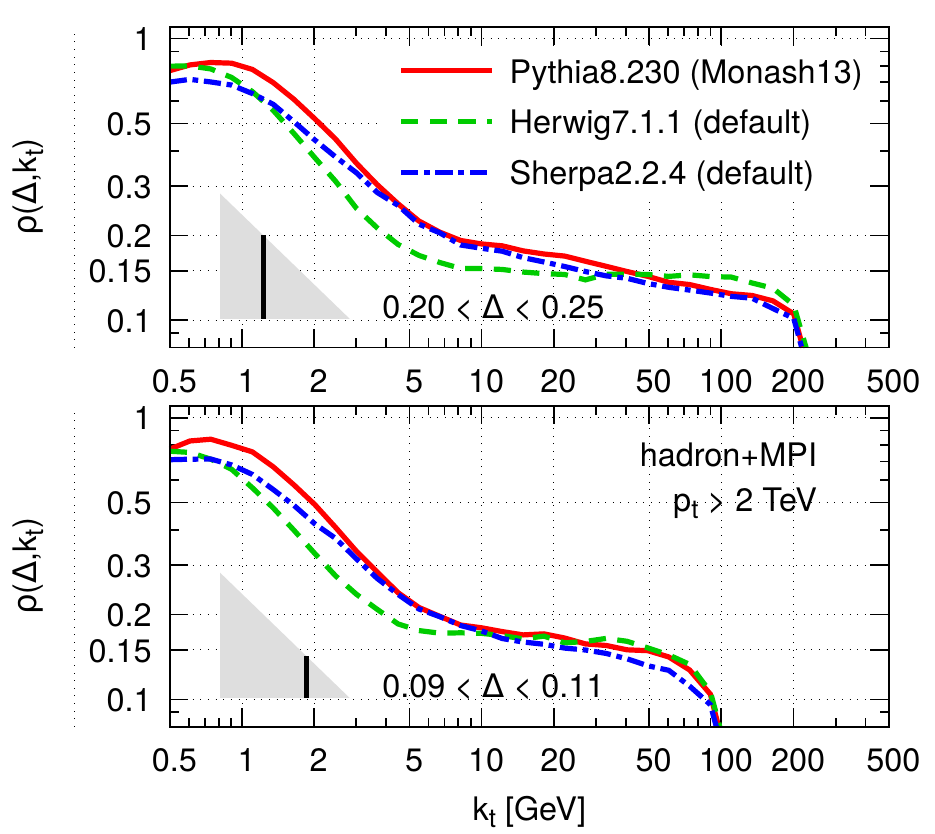}
    \caption{}
    \label{fig:slices-fixed-Delta}
  \end{subfigure}
  \caption{Emission density along slices of the Lund plane, at fixed
    $k_t$ (top) and $\Delta$ (bottom), comparing three event
    generators.
  }
  \label{fig:slices}
\end{figure}

The Lund jet plane density $\rho$ in Eq.~(\ref{eq:rho-def}) can be
directly measured experimentally and compared to analytic predictions
and parton-shower Monte-Carlo simulations.
Here we concentrate on the latter.
For such quantitative studies it is convenient to examine slices of
the Lund plane density at fixed $k_t$ and fixed $\Delta$.
Two of each are shown in Fig.~\ref{fig:slices}, illustrating the
potential of the Lund plane for providing insight into event
generators.
The figure compares the output of three different generators,
Pythia~8.230 (Monash13 tune), Sherpa~2.2.4~\cite{Gleisberg:2008ta} and
Herwig~7.1.1~\cite{Bellm:2017bvx} (angular-ordered shower).
The slices at fixed $\ln k_t$ illustrate a somewhat different trend
between the angular-ordered generator and the other two, which 
both have transverse-momentum ordered showers.
The differences span the $\pm 20{-}30\%$ range and we have found that
they are robust against non-perturbative effects (detector effects are
discussed in Appendix~\ref{sec:detector-effects}).
The slices at fixed $\Delta$ illustrate the coverage of both the
high-$k_t$, perturbative region, where the density is an infrared and
collinear safe quantity, and the low-$k_t$, non-perturbative region.
In the latter, for $k_t$ below a few GeV, one also sees differences
between generators of about $15\%$.
The ability to clearly identify separate perturbative and
non-perturbative regions provides a powerful advantage relative to
quantities such as jet-shapes that have been measured in the past, and
whose distributions tend, to some extent, to mix perturbative and
non-perturbative sensitivity.

Returning to more general considerations about the interest of
measuring the Lund plane density, an additional remark here is that at low
$k_t$, the Lund plane density could be seen as providing an effective
constraint on the strong coupling in the infrared, which one might
also be able to relate to the $\alpha_0$ parameter of
Refs~\cite{Dokshitzer:1995zt,Dokshitzer:1995qm}.

\subsection{Declustering other jet-algorithm sequences and
  higher-order effects}
\label{sec:sec-other-declusterings}

The choice of the C/A algorithm to create the clustering sequence is
related both to physical properties of the C/A algorithm and
associated higher-order perturbative structures that appear when one
calculates $\rho$ and $\bar \rho$.
To illustrate this, we write $\bar \rho$ as an expansion in powers of
$\as$.
\begin{equation}
  \label{eq:rho-expansion}
  \bar \rho(\Delta ,\kappa) = \sum_{n=1}
  \left(\frac{\as}{2\pi}\right)^n \bar \rho_n(\Delta ,\kappa)\,.
\end{equation}
The first order term, for a quark-induced jet, is given by an
expression similar to Eq.~(\ref{eq:lund-density}),
\begin{equation}
  \label{eq:rhobar1}
  \bar\rho_1(\Delta,\kappa) =
  2C_F
  {\bar z}\left(p_{gq}(\bar z) + p_{gq}(1-\bar z)\right),\;\;
  \bar z = \frac{\kappa}{\Delta},
  \qquad(\bar z<\frac12,\, \Delta \ll 1)\,,
\end{equation}
for all algorithms of the generalised-$k_t$ family. For small $\kappa$
this reduces to $4 C_F$.

At higher orders, $n > 1$, one expects that there may be logarithmic
enhancements, i.e.\ terms of the form $\as^n L^m$ where $L$ may
generically be either $\ln 1/\Delta$ or $\ln 1/\kappa$.
Ignoring potential subleading-$N_\text{\textsc{c}}$ factorisation
violation issues, or equivalently super-leading
logarithms~\cite{Forshaw:2006fk,Catani:2011st}, there are strong
reasons to believe that with the C/A algorithm the highest logarithmic
enhancement will correspond to $m = n-1$, i.e.\ at most a single
logarithmic correction factor relative to the leading order result.
Such terms will arise from: the running of the coupling, which is
naturally single logarithmic; flavour-changing hard-collinear
splittings, which have only (single) collinear logarithms; and non-global and
clustering logarithms, which have only (single) soft logarithms.
In particular the clustering logarithms are single logarithmic because
the angular-ordered nature of the algorithm matches the underlying
angular ordered pattern of soft-collinear radiation. 

\begin{figure}
  \centering
  \begin{subfigure}{0.47\textwidth}
    \includegraphics[width=\textwidth]{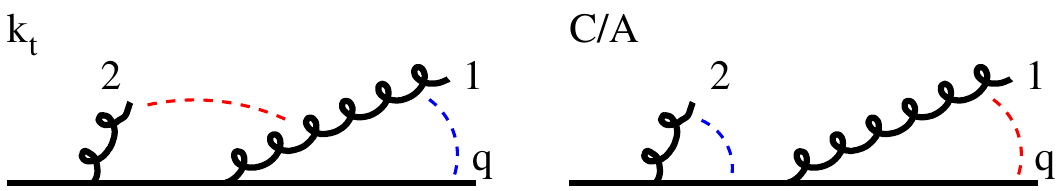}%
    \caption{}
    \label{fig:bad-config-kt}
  \end{subfigure}\hspace{0.06\textwidth}%
  \begin{subfigure}{0.47\textwidth}
    \includegraphics[width=\textwidth]{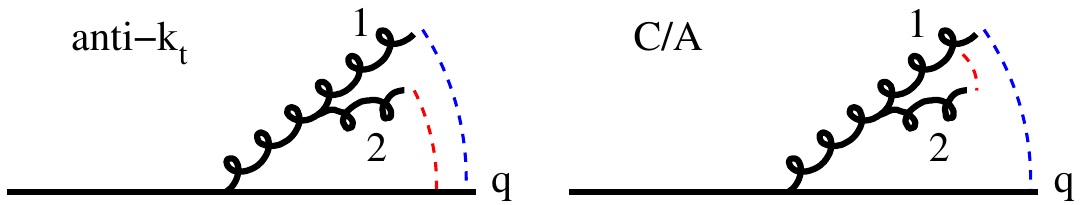}%
    \caption{}
    \label{fig:bad-config-antikt}
  \end{subfigure}%
  \caption{Configurations that lead to terms $\as^2 L^2$ for the Lund
    plane density for algorithms other than C/A, showing with a red (blue)
    dashed line the clustering that occurs first (second).
    (a) For the $k_t$-declustered Lund plane: emission $2$
    clusters with emission $1$ rather than with the quark and so fails
    to appear as part of the primary Lund plane.
    (b) For the anti-$k_t$-declustered Lund plane: emission $2$ gets
    declustered as a primary emission, even though physically it
    belongs to the secondary Lund plane associated with emission $1$,
    resulting in a spurious enhancement of the primary Lund plane
    density. }
  \label{fig:bad-configs}
\end{figure}

In contrast, if one uses the anti-$k_t$ or $k_t$ algorithms, the
clustering results in double-logarithmic enhancements, i.e.\ terms
$\as (\as L^2)^{n-1}$, or equivalently $m = 2(n-1)$.
We show this explicitly at order $\as^2$.
First consider the $k_t$ algorithm, and a configuration with two
primary emissions, with $\theta_{1q} \ll \theta_{2q} \ll 1$, $k_{t2} \ll
k_{t1}$, which implies $z_2 \ll z_1$.
Here $\theta_{ij}$ is the angle between particles $i$ and $j$, and $q$
represents the (leading) quark.
As originally pointed out in the article that proposed the Cambridge
algorithm~\cite{Dokshitzer:1997in}, when the emissions are on the same
side of the quark, then $\theta_{12} < \theta_{2q}$ and so emission
$2$ clusters with emission $1$ rather than with the quark, cf.\
Fig.~\ref{fig:bad-config-kt}. 
The resulting pseudojet retains the kinematics of emission $1$.
With the C/A algorithm, emission $2$ would have formed its own
independent primary declustering, but with the $k_t$ algorithm it does
not do so.\footnote{This was part of the motivation in
  Ref.~\cite{Dokshitzer:1997in} for inventing the Cambridge
  algorithm.}
This leads to a deficit of primary declusterings.
For a given choice of emission-$2$ kinematics, the region of
emission-$1$ kinematics where this occurs is proportional to $\ln^2
\theta_2/\kappa_2$, and results in the following double logarithmic
suppression relative to the leading-order result,
\begin{align}
  \bar\rho_2^{(k_t)}(\Delta, \kappa)
  &\simeq - (4C_F)^2
    \int \frac{d\theta_{2q}}{\theta_{2q}}
    \int \frac{d\kappa_{2}}{\kappa_{2}} 
    \int_{\kappa_2}^{\theta_2} \frac{d\kappa_{1}}{\kappa_{1}}
    \int_{\kappa_1}^{\theta_2} \frac{d\theta_{1q}}{\theta_{1q}}
    \int_{-\pi/2}^{\pi/2} \frac{d\phi_{12}}{2\pi}
    \delta\left(\ln \frac{\kappa_{2}}{\kappa}\right)
    \delta\left(\ln \frac{\theta_{2q}}{\Delta}\right)
  \nonumber
  \\
  &= - 4 C_F^2 \ln^2 \frac{\Delta}{\kappa} + \order{L}\,.
\label{eq:rho2-kt}
\end{align}
The minus sign accounts for the fact that in this region the primary
Lund plane contribution from emission $2$ is lost because $2$ clusters
with $1$.
The limits for the $\phi_{12}$ integration account for the fact that
both emissions have to be on the same side of the quark in order for
$2$ to cluster with $1$. 

For the anti-$k_t$ algorithm, the issue that arises is that secondary
splittings can end up being categorised as primary in terms of the
(de)clustering sequence.
Consider emission $1$ with an angle $\theta_{1q}$ with respect to the
quark and momentum fraction $z_1$.
It can emit a soft gluon $2$ at an angle
$\theta_{12} \ll \theta_{1q}$, carrying a momentum fraction $z_2$,
defined relative to the quark momentum, and satisfying $z_2 \ll z_1$,
cf.\ Fig.~\ref{fig:bad-config-antikt}.
With the C/A algorithm, the condition on the angles would ensure that
emission $2$ is always clustered with $1$, before $1$ clusters with
the quark, and hence emission $2$ will never on its own be considered
as a primary declustering.
For the anti-$k_t$ algorithm, emission $2$ will cluster with
emission $1$ only if $\theta_{12}^2/z_1^2 < \theta_{2q}^2 \simeq
\theta_{1q}^2$.
In the remaining region, $\theta_{12}^2 > z_1^2 \theta_{1q}^2$,
emission 2 will be clustered directly with the quark, and hence will be
considered as a primary declustering.
Fixing $\theta_{1q} \simeq \theta_{2q}$ and $z_2$, but integrating
over the $\theta_{12}$ and $z_1$, one finds a double logarithmic
enhancement, 
\begin{align}
  \bar\rho_2^{(\text{anti-}k_t)}(\Delta, \kappa)
  &\simeq 16 C_F\, C_A
    \int \frac{d\theta_{1q}}{\theta_{1q}}
    \int^{\theta_{1q}}_{\kappa} \frac{d\kappa_{1}}{\kappa_{1}} 
    \int^{\frac{\kappa_1}{\theta_{1q}}} \frac{dz_2}{z_2}
    \int^{\theta_{1q}}_{\kappa_1} \frac{d\theta_{12}}{\theta_{12}}
    \delta\left(\ln \frac{z_2 \theta_{1q}}{\kappa}\right)
    \delta\left(\ln \frac{\theta_{1q}}{\Delta}\right)
    \nonumber
  \\
  &= + 8 C_F\, C_A \ln^2 \frac{\Delta}{\kappa} + \order{L}\,.
\label{eq:rho2-antikt}
\end{align}
In setting the lower limit of the $\theta_{12}$ integral, we have made
use of the condition $\theta_{12} > z_1 \theta_{1q}  = \kappa_{1}$.
The upper bound on the $z_2$ integration comes from the constraint
$z_2 < z_1 = \kappa_1/\theta_{1q}$ and, for physical values of
$\kappa$ and $\Delta$, the solution of the $\delta$-function
constraint on $z_2$ is always below that bound.

To verify our calculations we have used the Event2
program~\cite{Catani:1996jh,Catani:1996vz} to evaluate the exact
result for $\bar\rho_2(\Delta,\kappa)$ using $e^+e^-$ versions of the
$k_t$, anti-$k_t$ and C/A 
algorithms.\footnote{Specifically, the Event2 program generates $e^+e^- \to
  \text{jets}$ events.
  We used it in $3$-jet NLO mode.
  We cluster each event with the $e^+e^-$ $k_t$ (Durham)
  algorithm~\cite{Catani:1991hj} so as to obtain exactly two jets, and
  then recluster each of those jets with the $e^+e^-$ version of the
  $k_t$, C/A or anti-$k_t$ algorithm, as defined by FastJet's
  \texttt{ee\_genkt\_algorithm}~\cite{Cacciari:2011ma} with
  $p= 1, 0, -1$ respectively.
  The resulting sequence is then used to determine the Lund-plane
  density, following the steps of section~\ref{sec:lund-construction}
  (but not first reclustering with C/A).
  For an $ij$ declustering we then define $\ln 1/\Delta = -\ln \tan
  \theta_{ij}/2$ and $\kappa = \min(E_i,E_j)/(E_i+E_j) \Delta$.
  This definition ensures that Eq.~(\ref{eq:rhobar1}) remains valid
  even for $\Delta$ of order $1$.
} 
\begin{figure}
  \centering
  \begin{subfigure}{0.33\textwidth}
    \includegraphics[width=\textwidth,page=2]{figs/double-log.pdf}%
    \caption{}
    \label{fig:event2-tests-kt}
  \end{subfigure}%
  \begin{subfigure}{0.33\textwidth}
    \includegraphics[width=\textwidth,page=3]{figs/double-log.pdf}%
    \caption{}
    \label{fig:event2-tests-antikt}
  \end{subfigure}%
  \begin{subfigure}{0.33\textwidth}
    \includegraphics[width=\textwidth,page=1]{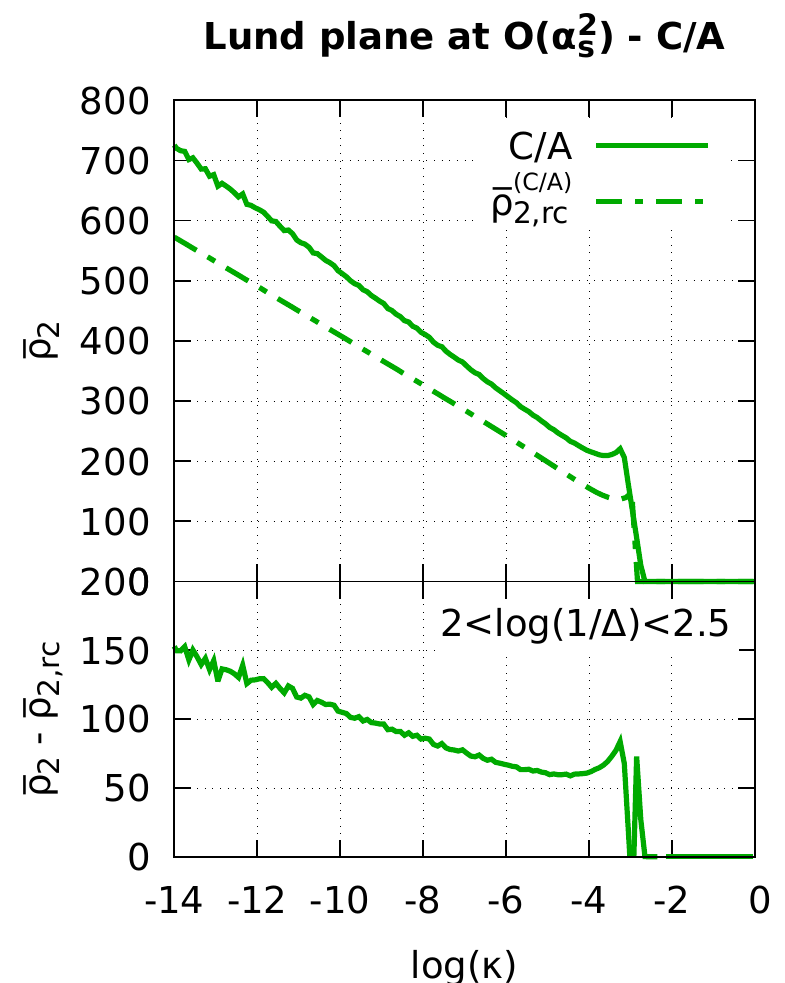}%
    \caption{}
    \label{fig:event2-tests-CA}
  \end{subfigure}%
  \caption{Evaluations with Event2 of the second-order contribution to
    the Lund plane, in a bin of $\ln 1/\Delta$, as a function of
    $\kappa$, for (de)clustering sequences obtained with the $k_t$,
    anti-$k_t$ and C/A jet algorithms.
    In (a) and (b) the dashed line corresponds to the analytic
    expectations, Eqs.~(\ref{eq:rho2-kt}) and (\ref{eq:rho2-antikt})
    for clustering-induced double-logarithms in the $k_t$ and
    anti-$k_t$ algorithms.
    In (c), for the C/A algorithm, which is seen here to be free of
    double logarithms,  the dot-dashed line corresponds to the
    (single-logarithmic) running coupling correction,
    Eq.~(\ref{eq:rc}), illustrating that it dominates the 
    second-order correction.
  }
  \label{fig:event2-tests}
\end{figure}
Fig.~\ref{fig:event2-tests}, in the upper panels, shows $\bar\rho_2$ for
the $k_t$, anti-$k_t$ and C/A algorithms, compared in the first two
cases also to the double logarithmic calculations of
Eqs.~(\ref{eq:rho2-kt}) and (\ref{eq:rho2-antikt}).
The quantity $h_{22}$ denotes the coefficient of $L^2 = \ln^2
\frac{\Delta}{\kappa}$ in those equations.
The lower panels for the $k_t$ and anti-$k_t$ algorithms show the
result after subtraction of the double-logarithmic terms.
One sees clearly that at most single logarithmic terms remain.
Fig.~\ref{fig:event2-tests} therefore validates our calculation of the
double logarithmic coefficients for the $k_t$ and anti-$k_t$
algorithms as well as our prediction of an absence of double
logarithmic corrections for the C/A algorithm.

One might ask why the presence or absence of double logarithmic
corrections should have any impact on the choice of reclustering
algorithm.
The answer lies in the origin of those double logarithms: for both
$k_t$ and anti-$k_t$, they come about because the reclustering
correlates a given part of the Lund plane with a large
(double-logarithmic) other region of the Lund plane, e.g.\ in the case
of the $k_t$ algorithm anywhere with smaller $\Delta$ and larger
$k_t$.

In contrast, with the C/A algorithm, the correlations are only within
single-logarithmic regions, for example, correlations between a given
point and the region of similar $\Delta$, or between a given point and
emissions in the band along the hard collinear boundary.
The more limited correlations for C/A reclustering should make
resummation calculations for the Lund plane more straightforward.
They are also expected to result in a cleaner input for tagging
applications, and to simplify the interpretations of comparisons
between Monte Carlo simulations and data.
Note that the single logarithmic slope observed in the case of the C/A
algorithm also contains running coupling (rc) corrections, which have
the form
\begin{equation}
  \label{eq:rc}
  \bar \rho_{2,\text{rc}}^{(\text{C/A})}(\Delta,\kappa) =
  \bar \rho_{1}(\Delta,\kappa)\,
  4\pi b_0 \ln \frac{1}{\kappa} + \order{1}\,.
\end{equation}
where $b_0 = \frac{11C_A - 2n_f}{12\pi}$.
One sees from Fig.~\ref{fig:event2-tests-CA} that this contribution
accounts for the bulk of the single-logarithmic slope seen at second
order for the C/A algorithm.
The residual single-logarithmic corrections from clustering effects
are therefore a small component of the overall single-logarithmic
contributions.
%
%

\begin{figure}
  \centering
  \includegraphics[width=0.48\textwidth,page=2]{figs/lund-reclusts-ratios.pdf}\hfill
  \includegraphics[width=0.48\textwidth,page=1]{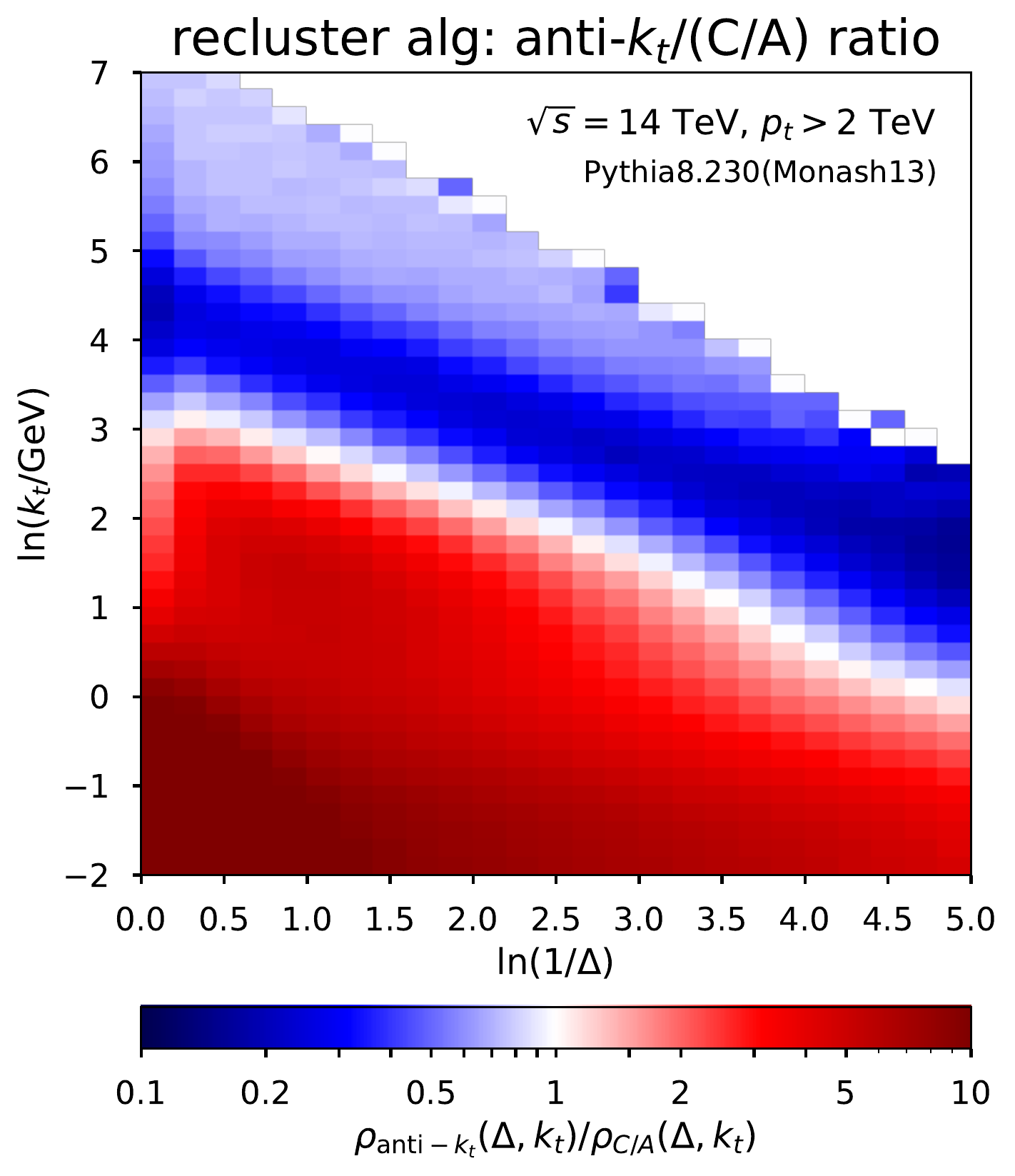}
  \caption{The $\rho(\Delta,k_t)$ results as obtained with $k_t$
    (left) and anti-$k_t$ (right) declustering, normalised to the
    result for C/A declustering.}
  \label{fig:alt-reclust-ratios}
\end{figure}

The above calculations of double logarithmic coefficients may appear to
be somewhat abstract.
Additionally therefore, it can be instructive to examine how the use
of the $k_t$ and anti-$k_t$ algorithms for (de)clustering modifies the
structure of the Lund plane in Pythia simulations.
Fig.~\ref{fig:alt-reclust-ratios} show the ratio of the averaged
primary Lund
plane density as obtained with the $k_t$ and anti-$k_t$
(de)clustering to that with our default C/A choice.
One sees large modifications, especially at lower $k_t$ values, where
the double logarithmic effects are largest.
In those regions, the modifications are in line with the expectations
from the calculations above: a suppression for the $k_t$ algorithm and
a (larger) enhancement for the anti-$k_t$ algorithm.
There are also features that go beyond our second-order
double-logarithmic calculation.
For example for the $k_t$ algorithm there is a strong enhancement
around $\ln 1/\Delta =5$ and $\ln k_t/\GeV = 2$.
For the anti-$k_t$ algorithm there is a line of deficit sloping
downwards from $\ln 1/\Delta=0$, $\ln k_t/\GeV \simeq 4.5$.
We do not have systematic analytical explanations for these features,
and their presence provides further reasons for choosing the C/A
algorithm for the declustering.

\subsection{Relations with other observables}

The declustering sequence that produces the primary Lund plane is
closely connected with a range of other jet observables.
The simplest case is that of the average multiplicity of iterated
soft-drop steps~\cite{Frye:2017yrw}, $N_\text{SD}$, which relates to
the averaged $\bar \rho$ density through
\begin{equation}
  \label{eq:N-SD}
  \langle N_\text{SD} \rangle =
   \int_{0}^\infty \frac{d\Delta'}{\Delta'}
  \int \frac{d\kappa}{\kappa} \,
  {\bar\rho}(\Delta', \kappa)
  \left[
  \Theta\left(\frac{\Delta'}{2} -\kappa\right) - 
    \Theta\left(\kappa - z_\text{cut} (\Delta')^{1+\beta}\right)\right].
\end{equation}
This relation applies to soft drop with a generic value of
$\beta$~\cite{Larkoski:2014wba}, assuming a jet radius of $1$ for
simplicity.
This is an exact relation, and it holds because the iterated soft drop
procedure simply follows the same set of declustering steps as the
primary Lund plane and counts those that satisfy the kinematic
condition that is represented in the second of the $\Theta$-functions
in Eq.~(\ref{eq:N-SD}).
The first $\Theta$-function just represents the kinematic boundary
induced by the condition $\kappa < \max(z,1-z)\Delta$.
The counting for $N_\text{SD}$ is inclusive over all primary
splittings and so the average over events that produces $\bar \rho$,
also gives the average number of iterated soft-drop steps, $\langle
N_\text{SD} \rangle$.

Further exact relations exist between various soft-drop observables
and the tuple of declustering variables defined in
section~\ref{sec:lund-construction}.
For example the soft-drop mass and $z_g$~\cite{Larkoski:2017bvj}
variables are given by $m^{(i)}$ and $z^{(i)}$
(cf.\ Eq.~(\ref{eq:branching-variables})) from the first of the entries
in the primary declustering sequence, $\mathcal{L}_\text{primary}$ of
Eq.~(\ref{eq:tuple-list}), that satisfies
$z^{(i)} \ge z_{\text{cut}}(\Delta^{(i)})^\beta $.
This is because ignoring the earlier declusterings with
$z^{(i)} < z_{\text{cut}}(\Delta^{(i)})^\beta $ is functionally
identical to the procedure of discarding (i.e.\ grooming away) the
softer branch in the soft-drop procedure.\footnote{In v1 of the arXiv
  version of this paper, we had stated connections between
  $\bar \rho(\Delta,\kappa)$ and the distributions of jet observables
  such as the soft-drop $z_g$ and the jet broadening.
  Those statements were correct only in a context with just primary
  soft branchings and so not hold in full QCD.}

\section{Application to boosted-$W$ tagging}
\label{sec:boosted-tag}

\begin{figure*}
  \centering
  \includegraphics[width=0.5\textwidth,page=2]{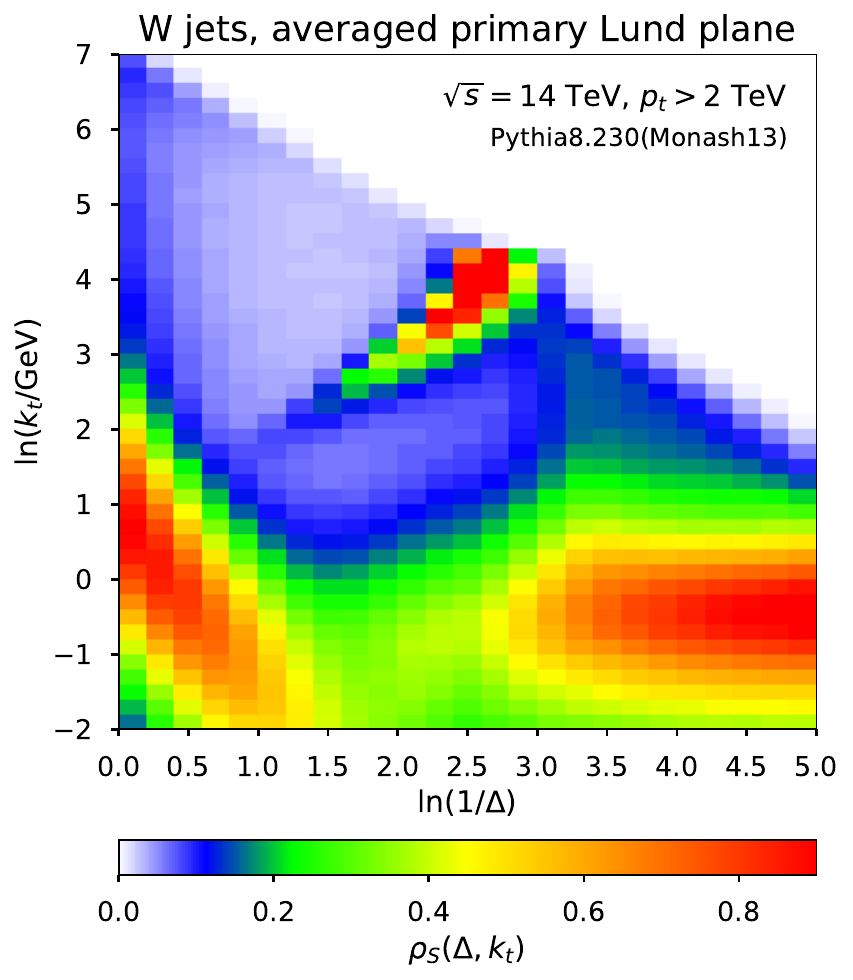}%
  \caption{Lund-plane emission density, $\rho_S(\Delta,k_t)$, for hadronically decaying
    boosted $W$ bosons, in $WW$ events, using the same jet-clustering
    and selection as in Fig.~\ref{fig:average-lund-plane}.
  }
  \label{fig:Wavg}
\end{figure*}

We now turn to the potential of the Lund plane for identifying
hadronically decaying boosted electroweak bosons, concentrating
specifically on the example of $W$ identification.
Fig.~\ref{fig:Wavg} shows the averaged (primary) Lund plane for
hadronic $W$ decays ($p_t > 2\TeV$), to be compared to
Fig.~\ref{fig:average-lund-plane} for dijets.
Two main differences are clearly visible to the human eye.
One is the diagonally oriented patch in the $W$ case, around
$\ln 1/\Delta = 2.5$ and $\ln k_t/\text{GeV}=4$, which is connected
with the fixed-mass two-pronged structure of the $W$: lines of
constant mass in the Lund plane are up-right going diagonals.
The other important feature in the $W$ case is the considerable
depletion of emissions in the upper-left region and below the $W$-mass
structure.
The depletion is principally a consequence of the colour-singlet
nature of the $W$.
It extends also into the non-perturbative region (but not
substantially affecting structures of the Lund plane associated with
the underlying event).

We investigate two broad approaches to making use of the information in
the Lund plane.
One is a log-likelihood type approach, while the other will be to use
machine learning.

\subsection{Log-likelihood use of Lund Plane}
\label{sec:LL}
The log-likelihood approach uses two main inputs: the first requires
the identification of the ``leading'' emission, $\leading$, which in
the $W$ case is likely to be associated with the two-prong decay.
We take this leading emission to be the first emission in the Lund
declustering sequence that satisfies $z > z_\text{cut}$ with
$z_\text{cut} = 0.025$, which corresponds to the emission that would
be selected by the mMDT tagger~\cite{Dasgupta:2013ihk} with the same
$z_\text{cut}$ or equivalently by the Soft-Drop (SD)
procedure~\cite{Larkoski:2014wba} with $\beta=0$ and that
$z_\text{cut}$.
We define a $\cL_\ell$ log likelihood function 
\begin{equation}
  \label{eq:L1}
  \cL_\ell(m^\leading, z^\leading) =
  \ln \left(\frac{1}{N_S}\frac{dN_S}{dm^\leading dz^\leading}
    \bigg/
    \frac{1}{N_B}\frac{dN_B}{dm^\leading dz^\leading} \right)\,,
\end{equation}
using the ratio of $dN_X/dm^\leading dz^\leading$ ($X=S,B$), the
differential distribution in the mass and $z$ variables of the leading
emission ($m^\leading$, $z^\leading$) for a simulated signal sample
$S$ ($W$ bosons) with $N_S$ jets, and the analogous quantity for a
background (QCD dijet) sample $B$.
In practice we bin logarithmically in $m^\leading$ and $z^\leading$ to construct a
discretised approximation to $\cL_\ell(m^\leading, z^\leading)$.

The second likelihood input is designed to bring sensitivity to the
pattern of non-leading $\nonleading$ emissions, i.e.\ the pattern of
additional radiation, within the primary Lund plane, that decorates
the basic two-prong structure.
It involves a function
\begin{equation}
  \label{eq:L2}
  \cL_{n\ell}(\Delta, k_t; \Delta^{(\ell)}) =
  \ln \left(\rho_S^\nonleading \big/ \rho_B^\nonleading\right)\,,
\end{equation}
where $\rho_X^{(n\ell)}$ is determined just over the non-leading
emissions,
\begin{equation}
  \label{eq:rho-nl}
  \rho_X^{(n\ell)}(\Delta, k_t; \Delta^\leading) =
  \frac{dn^\nonleading_{\text{emission},X}}{d\ln k_t \, d\ln 1\!/\Delta\,
    d\ln\Delta^\leading}
  \bigg/
  \frac{dN_X}{d\ln\Delta^\leading}\,,
\end{equation}
as a function of the angle $\Delta^{(\ell)}$ of the leading emission,
with $X=S,B$ corresponding either to the $W$ signal ($X=S$) or to the
QCD background sample ($X=B$).
Our overall log-likelihood signal-background discriminator for a given
jet is then given by
\begin{equation}
  \label{eq:Ltot}
  \cL_\text{tot} = \cL_\ell(m^\leading,z^\leading) 
  + \sum_{i\ne \ell} \cL_{n\ell} (\Delta^{(i)}, k_t^{(i)}; \Delta^\leading)
  + \cN(\Delta^\leading)\,,
\end{equation}
where the normalisation term $\cN$ is
\begin{equation}
  \label{eq:LL2-norm}
  {\cal N}(\Delta^\leading) = - \int d\ln \Delta\, d\ln k_t
  \left(\rho_S^\nl - \rho_B^\nl\right)\,,
\end{equation}
up to an overall constant.
In the sum over non-leading emissions, $i\neq \ell$ in Eq.~(\ref{eq:Ltot}),
each non-leading emission $i$ contributes information (through the
$\cL_{n\ell} (\Delta^{(i)}, k_t^{(i)}; \Delta^\leading)$ term) about
whether its corresponding region of the Lund plane tends to be more
populated by signal or background emissions.
The normalisation term ${\cal N}$ accounts for the average
difference in the number of non-leading emissions between signal and
background jets.

It is instructive to think about the conditions under which
Eq.~(\ref{eq:Ltot}) would be the optimal discriminator that can be
constructed from the sequence in primary Lund-plane declusterings: (1)
the identification of the leading emission associated with the $W$'s
two-prong structure should be correct; (2) non-leading emissions in
the Lund plane should effectively be independent of each other, which
is the basis of the sum over $i \neq \ell$ in Eq.~(\ref{eq:Ltot}); (3)
that pattern of independent emission should depend on
$\Delta^\leading$ but not on $m^\leading$.
Each of these approximations has its imperfections, but none is
expected to be particularly badly violated.

\begin{figure*}
  \centering
  \begin{subfigure}{0.48\textwidth}
    \includegraphics[width=\textwidth,page=1]{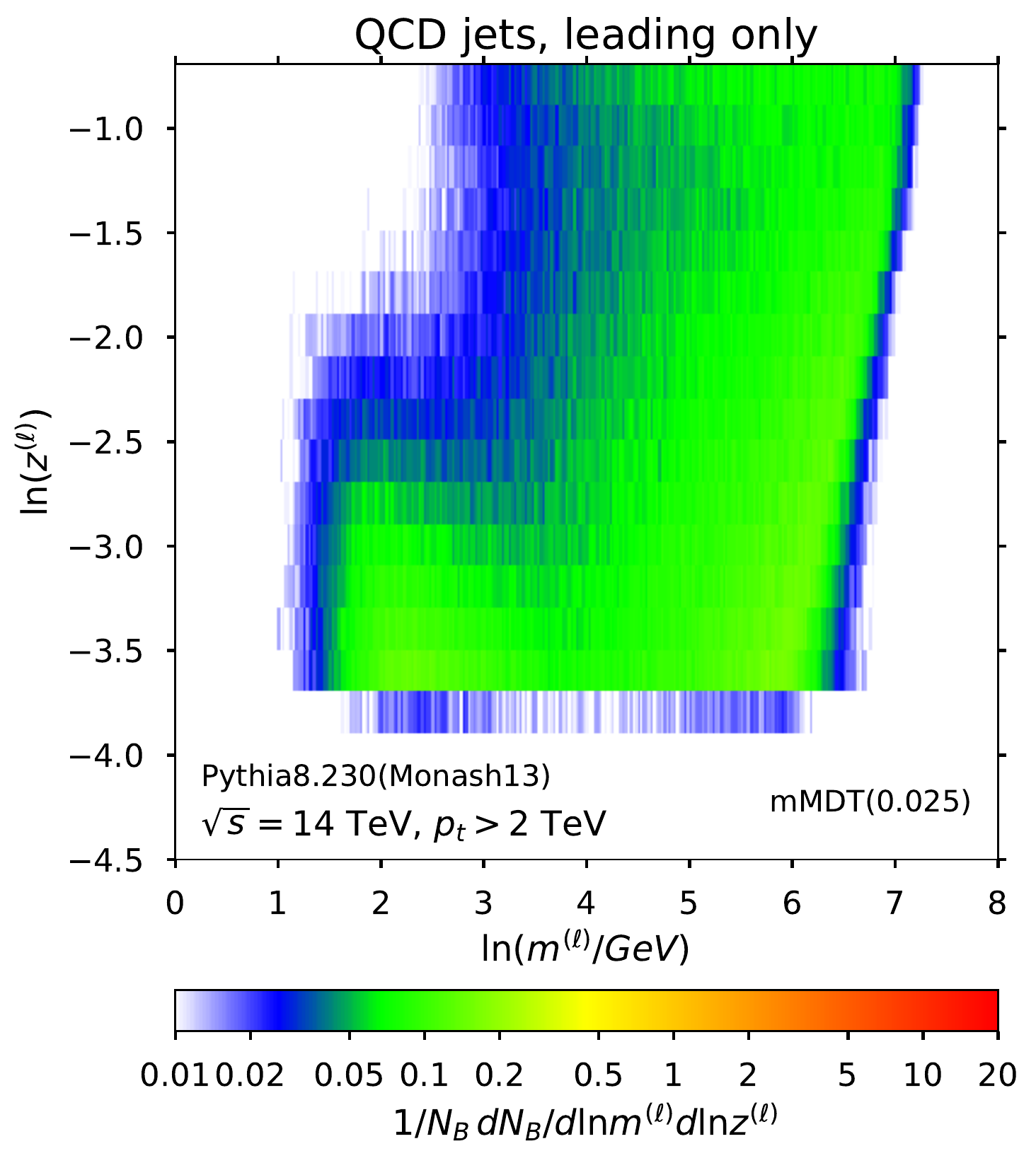}%
    \caption{}
    \label{fig:dijet-leading}
  \end{subfigure}\hfill
  \begin{subfigure}{0.48\textwidth}
    \includegraphics[width=\textwidth,page=2]{figs/plot-Wdiscrim-leading-images.pdf}%
    \caption{}
    \label{fig:W-leading}
  \end{subfigure}%
  \caption{
    Distribution of the leading emission $\frac{1}{N_X}\,\frac{dN_X}{d\ln m^{(\ell)} d\ln z^{(\ell)}}$ for (a)
    background QCD jets ($X=B$) and (b) signal $W$ jets ($X=S$).
    Note the mass peak around $\ln m^\leading = \ln M_W$ in the case of $W$
    jets and the enhancement at larger values of $\ln z^{(\ell)}$.
  }
  \label{fig:WDijet-leading}
\end{figure*}

\begin{figure*}
  \centering
  \begin{subfigure}{0.48\textwidth}
    \includegraphics[width=\textwidth,page=5]{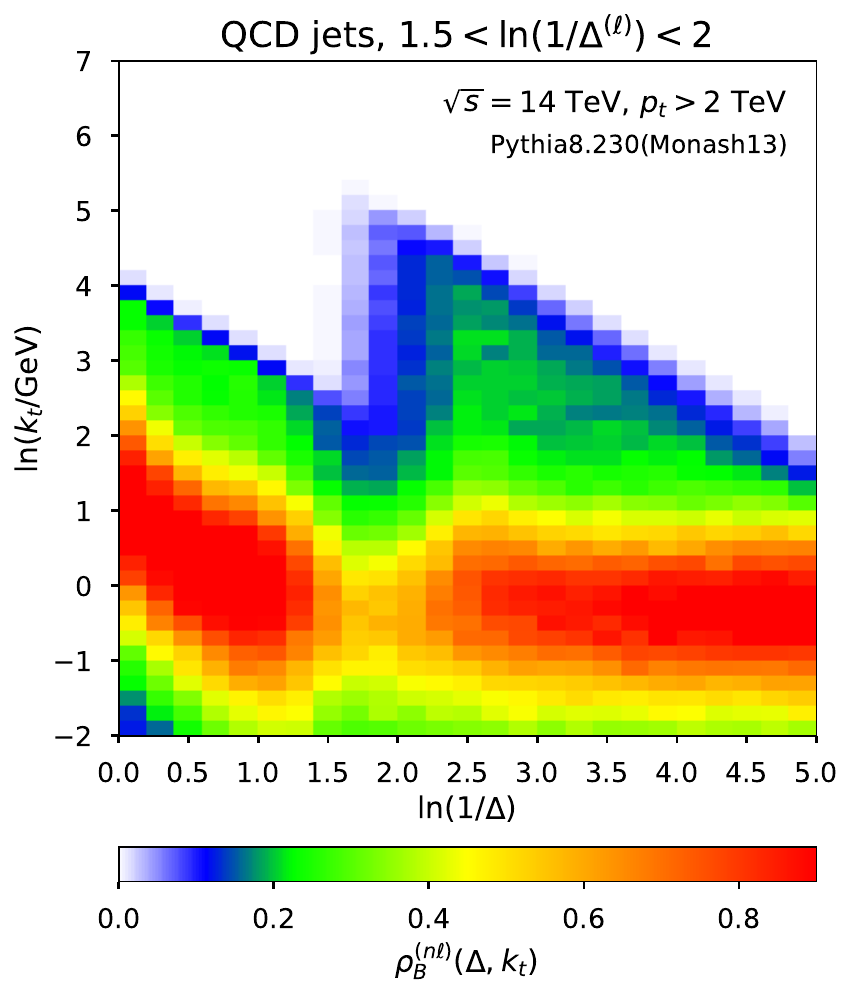}%
    \caption{}
    \label{fig:dijet-nl}
  \end{subfigure}\hfill
  \begin{subfigure}{0.48\textwidth}
    \includegraphics[width=\textwidth,page=3]{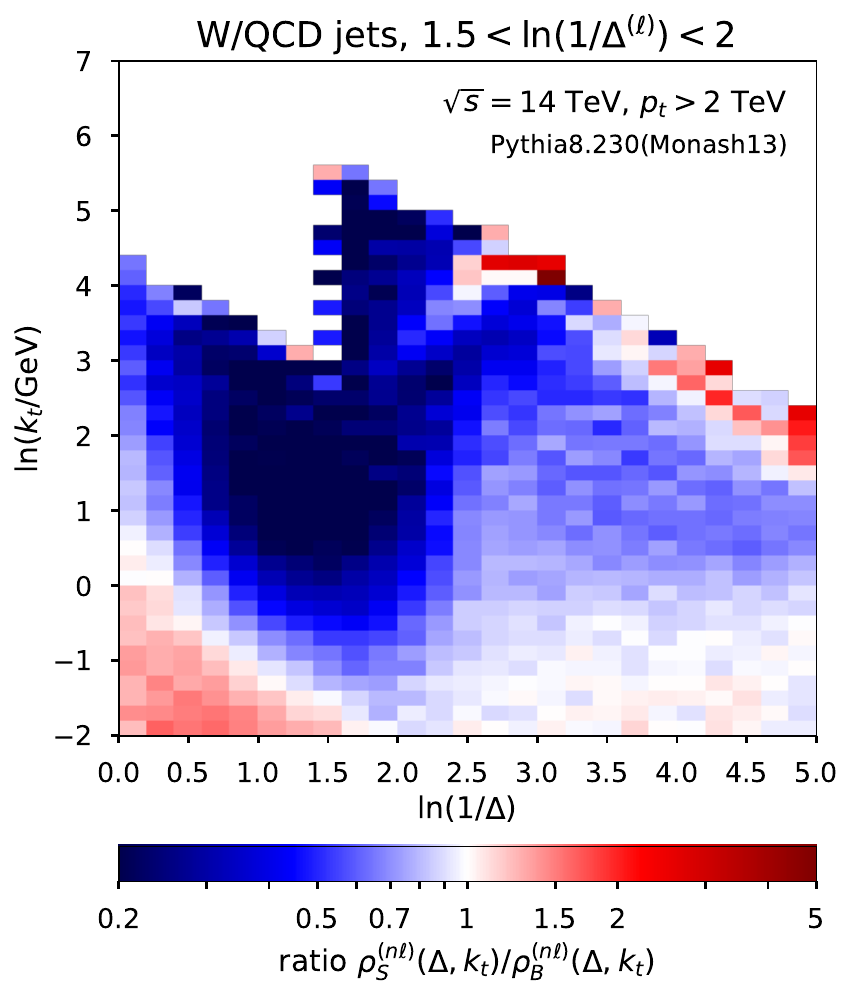}%
    \caption{}
    \label{fig:W-dijet-nl-ratio}
  \end{subfigure}%
  \caption{
    (a) Non-leading Lund-plane density for dijets, $\rho_B^{(n\ell)}(\Delta, k_t;
    \Delta^\leading)$, Eq.~(\ref{eq:rho-nl}), i.e.\ after removing the leading
    emission, specifically the first emission
    that passes an mMDT selection procedure.
    (b) Ratio of the W to QCD Lund-plane $\rho_{S,B}^{(n\ell)}(\Delta, k_t;
    \Delta^\leading)$ densities, whose logarithm is equal to the likelihood
    function $\cL_{\nlnb}$ of Eq.~(\ref{eq:L2}).
    Both plots correspond to the choice
    $1.5 < \ln 1/\Delta^\leading < 2$.
  }
  \label{fig:WDijetW}
\end{figure*}

To help illustrate how the log-likelihood approach works in practice,
we show the leading-emission distribution density,
$\frac{1}{N_X}\,\frac{dN_X}{d\ln m^{(\ell)} d\ln z^{(\ell)}}$, for
background ($X=B$) and signal ($X=S$) jets in
Fig.~\ref{fig:WDijet-leading}. 
The background is diffuse, while the signal is peaked around
$m^\leading = M_W$ and concentrated at larger $z^\leading$ values, as
one would expect.
The non-leading emission density, $\rho_B^\nl$, is shown for
background (dijets) in Fig.~\ref{fig:dijet-nl}, for jets where the
leading emission has $1.5 <\ln 1/\Delta^\leading < 2$ and
$z^\leading > z_\text{cut}$ (roughly $\ln k_t/\text{GeV} > 2.5$).
For similar rapidities and lower $k_t$'s there is a modest depletion
in the number of emissions.
This is a partial shadow cast by the leading emission: non-leading
emissions with similar $\Delta$ and $\psi$ to the leading emission
will be clustered with it.
The other main feature of note in Fig.~\ref{fig:dijet-nl} is
the empty area in the upper-left region of the plot: given that the
emission classified as leading had $1.5 <\ln 1/\Delta^\leading < 2$,
there cannot have been any emissions with a small $\Delta$ and $z >
z_\text{cut}$.

Fig.~\ref{fig:W-dijet-nl-ratio} shows the $\cL_{n\ell}$
likelihood function.
Most of its discriminating power will come from the extensive dark
blue region.
For each emission that is present in this region, one gets a negative
$\cL_{n\ell}$ contribution to $\cL_\text{tot}$, which drives
$\cL_\text{tot}$ to be more background like (i.e.\ negative).
Instead if there are few emissions, the positive contribution of the
$\cN$ term results in a more signal-like (positive) final
$\cL_\text{tot}$ value.
Note that the dark blue region of Fig.~\ref{fig:W-dijet-nl-ratio}
stretches down to $k_t$'s below a few GeV, and since that region tends
to contain a significant number of emissions in the background case
(cf.\ Fig.~\ref{fig:dijet-nl}), one expects that some of the
sensitivity in $W$ v.\ QCD discrimination will come from low-$k_t$
non-perturbative effects.
This highlights the importance of direct experimental measurements of
such regions.\footnote{
  It also highlights an important difference between our
  log-likelihood approach and the shower-deconstruction
  approach~\cite{Soper:2011cr}.
  Both are partially analytical multi-variate log-likelihood
  approaches.
  Shower-deconstruction exploits far more detailed information on
  correlations than our log-likelihood method can.
  However since the underlying shower-deconstruction likelihoods must
  be calculated perturbatively, one loses access to the substantial
  information that is contained in the non-perturbative region.
}
One can explicitly check the influence of the low-$k_t$ region on the
tagging performance by imposing a minimum $k_t$ cut in the
construction of the Lund plane. This is discussed further in
Section~\ref{sec:results} below.

In our practical implementation of the log-likelihood approach, we
will use $\ln m$ bins of size $0.025$, $\ln z$ bins of size $0.2$ for
$\cL_\ell$; for $\cL_{n\ell}$ we will take bins in $\ln k_t$ and
$\ln \Delta$ of $0.2$ and bins in $\ln \Delta^\leading$ of $0.5$.
The likelihood functions will be calculated using $500,\!000$ simulated
signal and background jets, while performance will be evaluated on an
independent sample of $200,\!000$ signal and background jets.

\subsection{Machine-learning use of Lund Plane}
\label{sec:machine-learning}

Our second approach to using the Lund-plane information for $W$
tagging is to provide it as an input to a variety of machine learning
(ML) methods.

The input can be provided in the form of a sequence of
$\{\ln 1/\Delta, \ln k_t\}$ pairs; we use this kind of sequence with
dense (DNN) and Long Short-Term Memory (LSTM) neural
networks~\cite{lstm1997}.\footnote{LSTMs for jet substructure were
  investigated recently also in Ref.~\cite{Egan:2017ojy}.}
In practice the sequence is zero-padded to form a $60\times 2$
dimensional matrix.
The DNN consists of four layers of size 200 with ReLU activation, and
a final two-dimensional layer with softmax activation.
For the LSTM network, we use a cell with 128-dimensional output
connected to a dropout layer with rate $20\%$, with a final dense
layer of dimension two and softmax activation.%
\footnote{Similar performance can also be achieved using a Gated
  Recurrent Unit~\cite{DBLP:journals/corr/ChoMGBSB14}.}
In addition to the $\ln1/\Delta$ and $\ln k_t$ variables, one could add
variables such as $\psi$, the logarithm of the subjet mass ($\ln m$),
or the particle multiplicity in the subjet.
With realistic experimental resolutions, we have found that they bring
a small additional gain in background rejection, in the $10-25\%$
range.\footnote{In contrast, at truth particle level, the gain is much
  more significant, of the order of a factor of two, mostly from the
  mass information.}
%
Keeping in mind that the variability in performance associated with
different training choices is also in the $10\%$ range, we chose not
to include these extra variables for our final results.
However they could be further investigated in future work.

Alternatively one can create a 2-dimensional Lund image for each event
(in which only a few pixels are turned on) and provide it as an input
to a convolutional neural network (CNN), where additional information
such as the azimuthal angle can be encoded through the pixel intensity
by adding new channels.
Each jet is represented as a $50\times50$ pixel image.
These images are used to train a neural network consisting of three
two-dimensional convolutional layers with ReLU activation and 128
output filters each, which are each connected to a max pooling layer
and a spatial dropout layer with rate $5\%$.
The first convolution window is of size $10\times 10$ pixels, with the
following two layers having windows of size $4\times 4$.
The last convolutional layer is connected to a dense network with 256
neurons and another dropout layer leading to a final two-dimensional
output layer with softmax activation.
This network can also be trained on jet images, where each pixel
corresponds to a bin in $(y,\phi)$-space around the jet
axis.
In this case the pixel intensity is given by the normalised scalar
$p_t$ sum of particles within that phase space region.

Our machine learning is implemented in \texttt{Keras}
2.0.8~\cite{chollet2015keras}, using \texttt{TensorFlow}
1.2.1~\cite{tensorflow2015-whitepaper} as the backend.
All model weights are initialised with a He uniform variance scaling
initialiser~\cite{DBLP:journals/corr/HeZR015}, and each training is
performed using a batch size of 128, with Adam as optimisation
algorithm~\cite{DBLP:journals/corr/KingmaB14} and a categorical
cross-entropy loss function.
The parameters for the machine learning are similar, where relevant,
to those in Refs.~\cite{deOliveira:2015xxd,Komiske:2016rsd}, though we
used a greater number of pixels for the images and larger networks.
Training is carried out on a sample of $500\,000$ jets for each of the
$W$ and background samples.
During each epoch $80\%$ of the sample is used for adapting the
network weights.
At the end of each epoch performance is tested on a validation sample
consisting of $10\%$ of the events (that were not used during the
epoch's training).
If that performance has not improved over the past 4 epochs training
halts.
The maximum number of epochs is $15$ and training typically halts at
epoch $10{-}15$.
The final performance of the network is then evaluated using a further
independent $10\%$ of the sample.

Other recent work has also made use of declustering sequences with
machine learning tools.
Ref.~\cite{Louppe:2017ipp} has used recursive neural networks on the
complete declustering tree (with various clustering algorithms), using
the momenta of the subjets at each stage as inputs.
Ref.~\cite{Egan:2017ojy}, has used the anti-$k_t$ clustering sequence
as a way of ordering all constituent particle momenta, which are then
provided in that order to an LSTM.
The Lund plane will not necessarily lead to more powerful
discriminants than these approaches, however it offers a degree of
physical insight and a scope for direct measurements of average
densities and correlations.
Both of these aspects are potentially valuable for understanding and
correctly calibrating machine learning approaches to jet tagging.
The LSTM that we have used with the Lund plane is also arguably more
straightforward to use with standard machine-learning tools such as
\texttt{Keras} than the RNN used in Ref.~\cite{Louppe:2017ipp}.

\subsection{Jet-shape discriminant}
\label{sec:jet-shape}

In addition to the log-likelihood and machine-learning based
approaches, we will also include a comparison with an optimised choice
of jet-shape discriminator.
For this purpose, we apply the SoftDrop algorithm ($\beta=2$,
$z_\text{cut}=0.05$) and use the resulting groomed jet to calculate
both the jet mass and the $D_2$ observable~\cite{Larkoski:2014gra}
with $\beta=2$ (itself very similar to $C_2$~\cite{Larkoski:2013eya}
for any given jet mass).
The jet mass and $D_2$ are then given as an input to a
boosted-decision tree (BDT) in the TMVA~\cite{Hocker:2007ht} package.
This pair of observables, used with just a mass cut, not a
BDT,\footnote{The Lund-likelihood approach can be sensibly used with a fixed
  mass window, simply by discretising the $\cL_\ell$ likelihood ratio to have
  a single mass bin.
  However it is not so straightforward to force
  machine-learning methods to use a fixed mass window.
  Therefore to obtain meaningful comparisons across all types of
  methods we must use the $D_2$ variable in combination with the full
  mass information.} was found to be close to optimal in terms of
background rejection among a comparison of 88 shape-mass combinations
in the recent Les Houches (LH) study~\cite{Bendavid:2018nar} (Fig.~III.29),
for a fixed signal efficiency of $0.4$.
Its performance is substantially above that of the default ATLAS and
CMS jet-shape discriminant choices.
We refer to it as $D_2^\text{[loose]}$.\footnote{In
  the LH report, this observable is denoted as
  $D^{(2)}_2\big[l\otimes\frac{l}{l}\big]$.}

We could also have chosen the dichroic $D_2$ variable, used as a
benchmark in~\cite{Bendavid:2018nar}, whose performance is only
slightly worse but is more resilient against non-perturbative
effects.
We will return to the question of resilience below in
section~\ref{sec:results}.

\subsection{Simulation, detector effects and reconstruction}
\label{sec:sim+reco}

In evaluating the performances of different methods, an important
consideration concerns the inclusion of realistic experimental
resolutions.
This is relevant both for reconstructing the $W$ mass and as regards
the radiation part in the rest of the Lund plane.
The baseline that we adopt for our comparisons is to use the
Delphes~\cite{deFavereau:2013fsa} fast detector simulation, version
3.4.1, with the \verb|delphes_card_CMS_NoFastJet.tcl| card to simulate
both detector effects and particle flow reconstruction.
The particle flow outputs have artefacts on angular scales associated
with the hadronic and electromagnetic calorimeter, and these have an
adverse effect on performance, both in terms of mass resolution and
availability of Lund plane information.
Accordingly we use a ``subjet-particle rescaling algorithm'' (SPRA1)
at the level of small-radius ($R_h=0.12$) subjets to retain angular
information from electromagnetic calorimeter deposits and charged
tracks at small angular scales while retaining full hadronic
calorimetry energy information at larger scales.
The SPRA algorithm is closely related to earlier methods that proposed
jet-wide, subjet and hadronic calorimeter rescaling of charged tracks
or electromagnetic calorimeter
deposits~\cite{Katz:2010mr,Son:2012mb,Schaetzel:2013vka,Larkoski:2015yqa,Bressler:2015uma,Han:2017hyv,CMS:2014joa,ATLAS:2016vmy}.
The details of the Delphes particle flow effect on the Lund plane, of
the SPRA algorithm and of the improvements it brings are given in
appendix~\ref{sec:detector-effects}.

Our results will use simulated dijet events as the background and $WW$
events as the signal, selecting jets, clustered with the C/A algorithm
with radius $R=1$ using \texttt{FastJet3.3.0}~\cite{Cacciari:2011ma},
with $p_t > 2\TeV$ and $|y|<2.5$ at a centre-of-mass energy of
$\sqrt{s}=14\TeV$.
We use Pythia~8.230 with its Monash~13 tune to generate the events.
The samples coincide with those used in the recent Les Houches
study~\cite{Bendavid:2018nar}.
We choose to concentrate on a high-$p_t$ sample for two reasons: (1)
the LHC is increasingly focusing on this region and (2) one expects
that high-$p_t$ jets contain the most information, because it is at
high-$p_t$ that the colour-singlet nature of the boson has the most
impact on the radiation pattern.
Results at lower $p_t$ are given in Appendix~\ref{sec:sec-lund-plane}.

Note that we have not included pileup in our simulations.
We therefore work within the assumption that methods for pileup
mitigation such as SoftKiller~\cite{Cacciari:2014gra},
PUPPI~\cite{Bertolini:2014bba}, constituent
subtractor~\cite{Berta:2014eza} or machine learning
approaches~\cite{Komiske:2017ubm} can successfully remove the
contamination in the regions that are critical for discrimination.
It is also possible to use area
subtraction~\cite{Cacciari:2007fd,Cacciari:2008gn}, given that at each
stage of the declustering one has (sub)jets with a well-defined area.
However, area subtraction is more likely to be susceptible to
fluctuations than other methods.
A further possibility is to supplement pileup mitigation with methods
such as filtering~\cite{Butterworth:2008iy},
trimming~\cite{Krohn:2009th} or recursive
soft-drop~\cite{Dreyer:2018tjj}, applied only to larger subjets, in
order to reduce their contamination from pileup.

\subsection{Results}
\label{sec:results}

\begin{figure}
  \centering
  \includegraphics[width=0.65\columnwidth,page=1]{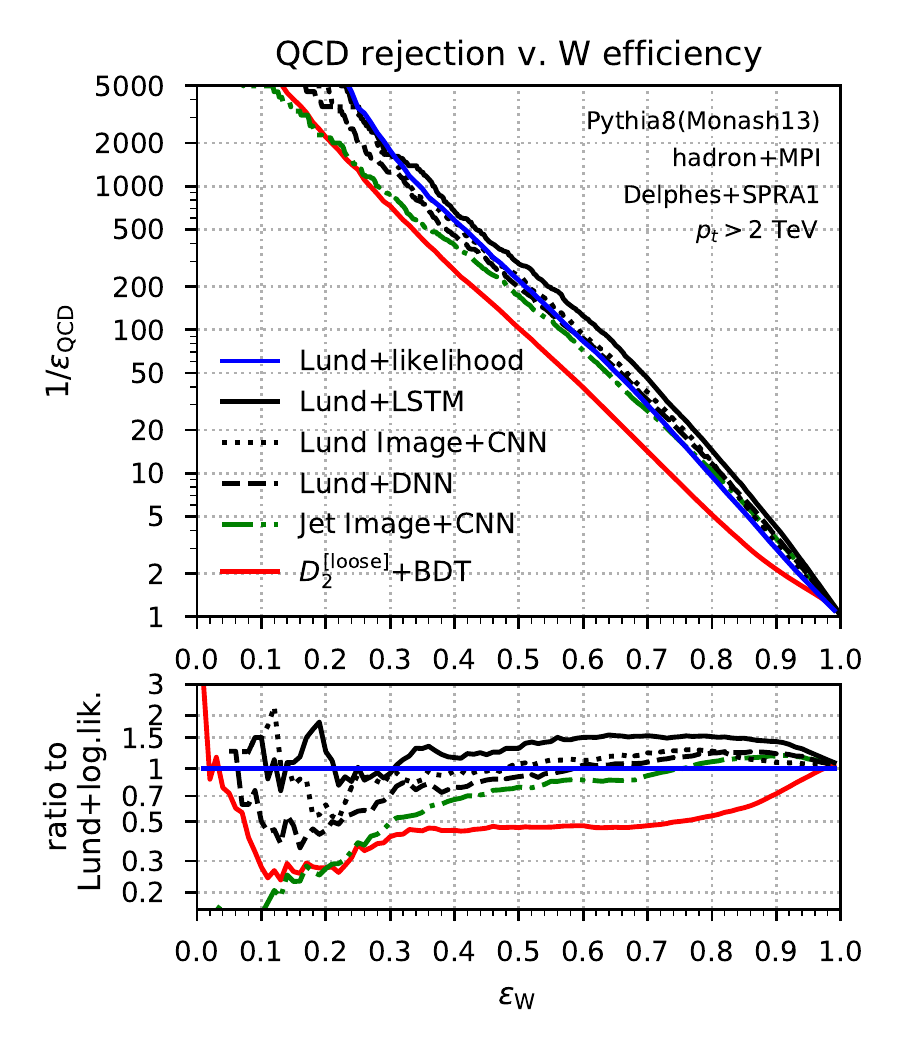}
  \caption{Background rejection ($1/\epsilon_\text{QCD}$) versus signal
    efficiency ($\epsilon_W$), per jet, for different $W$-tagging
    methods. The lower panel shows the ratio to the Lund+likelihood
    method.
  }
  \label{fig:ROCs}
\end{figure}

Results for the performance of the different tagging methods of
sections~\ref{sec:LL}--\ref{sec:jet-shape} are shown in
Fig.~\ref{fig:ROCs}.
The upper panel shows the background rejection factor for each method
as a function of signal tagging efficiency.
The lower panel shows the ratio of that rejection factor to the one
obtained for the Lund-likelihood method (the blue line).
Four of the methods are based on machine learning, as discussed in
section~\ref{sec:machine-learning}: three of them use Lund plane
inputs (Lund-image (with CNN), Lund+DNN, Lund+LSTM), while the other
uses a normal jet image (with CNN).
Finally the plot also includes the jet-shape plus mass approach,
labelled $D_2^\text{[loose]}$.%

Overall, the LSTM approach with Lund inputs performs best across nearly
the full range of signal efficiencies.
Taking $\epsilon_W=0.4$ as a reference, the Lund-likelihood is within a factor
of $0.7{-}0.8$ of the LSTM performance, while the other machine
learning methods are slightly worse than the Lund-likelihood method.
The $D_2$-based shape discrimination is a factor of two worse than the
Lund-likelihood method.\footnote{Note that $D_2$ also has access to
  information that is being discarded in the Lund-plane variables,
  specifically that associated with the substructure of secondary
  leaves off the primary plane. }
At higher (lower) signal efficiencies, the machine learning approaches
appear to perform relatively slightly better (worse).

The pattern of performances is fairly insensitive to the details both
of the Lund-likelihood procedure and the machine learning.
For example, in the Lund-likelihood approach, using the subjet that gives a
mass closest to the $W$ mass, rather than the first one to pass the
mMDT $z_\text{cut}$ condition, affects signal rejection performance
only at the $\sim 10\%$ level (making it better at high efficiencies,
worse at lower efficiencies).
Similarly, as mentioned in section~\ref{sec:machine-learning}, adding
mass and azimuth ($\psi$) information to the LSTM has only a modest
$10{-}25\%$ effect after accounting for detector effects (after
detector simulation, this gain is present only when one uses SPRA),
which does not appear to be particularly significant relative to other
training uncertainties.

We also note that the ROC curves becomes noisy at small signal
efficiencies.
This can at least in part be attributed to statistical uncertainties
associated with the finite size of our training/testing samples, in
particular when estimating the background rejection factor, where only
a small fraction of the events pass the tagger.
For example, the Lund-likelihood method uses a sample of 200\,000
events for testing. This corresponds to a statistical uncertainty of
about 30\% at $\epsilon_W=0.2$.
Correspondingly, the LSTM uses 50\,000 events for the testing phase,
meaning an expected statistical uncertainty that is twice as large
(given similar background rejection).

\subsection{Resilience to non-perturbative effects}
\label{sec:resilience}

One can argue that performance is not the only feature one may request
from a boosted object tagger.
In particular, one may require that the tagger remain relatively
insensitive to model-dependent non-perturbative effects.
Such insensitivity could translate into a reduced uncertainty on the
determination of the tagger's signal efficiency and background rejection rates.
It could also allow for the possibility of understanding the tagger's
behaviour with first-principles perturbative QCD calculations.

To carry out studies of sensitivity to non-perturbative effects, we
will compare performance between parton and hadron level.
Parton-level results cannot be sensibly passed through a detector
simulation, so the study must be carried out with actual particles
(i.e.\ partons or hadrons).
However, as we discussed in section~\ref{sec:sim+reco} and
appendix~\ref{sec:detector-effects}, real detector effects can have a
significant impact on the mass resolution in particular, which can
affect the conclusions of any multivariate study that uses the mass.
Accordingly we carry out two sets of studies in parallel.
In the first set of studies, we classify jets in terms of whether they
satisfy a loose requirement on the (possibly groomed) jet mass,
$65 < m < 105 \GeV$, and then do not further use any mass information.
Such a study is fairly realistic in terms of how much mass information
is accessible in a detector, but cannot be performed with machine
learning, because the latter is likely to ``cheat'' and learn the mass
information from other variables in the jet.
To be able to also examine machine learning, we therefore carry out a
second set of studies, in which full particle-level information is
available, allowing reconstruction of the $W$ mass peak.
All methods then exploit the unrealistically good particle-level mass
resolution on that $W$ mass peak.

\begin{figure}
  \centering
  \begin{subfigure}{0.49\textwidth}
    \includegraphics[width=\columnwidth,page=1]{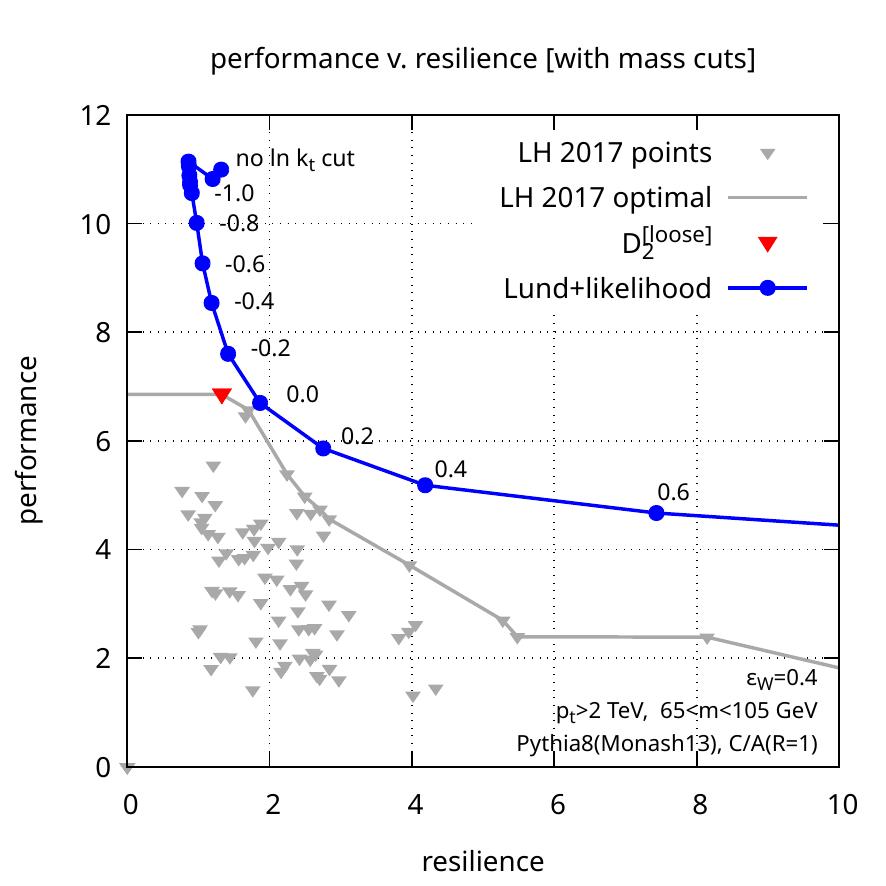}%
    \caption{}
    \label{fig:perf-resilience-mass-cut}
  \end{subfigure}
  \begin{subfigure}{0.49\textwidth}
    \includegraphics[width=\columnwidth,page=2]{figs/perf-resilience.pdf}
    \caption{}
    \label{fig:perf-resilience-full-mass}
  \end{subfigure}
  \caption{ Plots of performance
    ($\epsilon_W / \sqrt{\epsilon_\text{QCD}}$), for fixed
    (hadron+MPI) signal efficiency $\epsilon_W = 0.4$, versus
    resilience to non-perturbative effects, ($\zeta$ of
    Eq.~(\ref{eq:resi})).  %
    Grey triangles (and the red one) correspond to the full range of
    shape observables studied in the LH 2017
    study~\cite{Bendavid:2018nar}.  %
    The blue circles and black triangles correspond to the
    Lund-likelihood and Lund-LSTM methods respectively, with each
    point along a line corresponding to different lower cuts on the
    Lund-plane $\ln k_t/\text{GeV}$ value, below which declusterings
    are ignored (i.e.\ not passed to the LSTM or likelihood method;
    training is repeated for each different $k_t$ cut).
    In (a) the shape observables are used together with a cut on a
    mass variable, $65 < m < 105 \GeV$ (the mass may be groomed, or
    ungroomed, depending on the point);
    for the Lund likelihood, Eq.~(\ref{eq:L1}) is 
    evaluated with a single (groomed) mass bin, covering the same mass range as
    for the shapes, plus an
    outflow mass bin.
    In (b) shape variables are combined with the full particle-level resolution mass
    information through a boosted decision tree (BDT) and the cut
    that defines $\epsilon_W = 0.4$ is placed on the BDT output; for
    the Lund-likelihood and LSTM methods, full resolution
    Lund-plane information is used (including the mass for the
    likelihood method).
    For the Lund+likelihood method, the values
    of the $\ln k_t$ cuts are the same as for (a), i.e.\ spaced every
    0.2 units of $\ln k_t$. For
    the Lund+LSTM curves points are separated by 0.5 units of $\ln
    k_t$.
  }  \label{fig:perf-resilience}
\end{figure}

In Fig.~\ref{fig:perf-resilience}, we show the performance achieved by
the different tagging approaches versus their resilience to underlying
event and hadronisation corrections.
This is calculated following the procedure introduced in section~III.2
of the 2017 Les Houches proceedings~\cite{Bendavid:2018nar}.
The performance, $\epsilon_W/\sqrt{\epsilon_\text{QCD}}$, is plotted
versus the resilience $\zeta$, which is calculated using both
hadron+MPI-level efficiencies $\epsilon$ and parton-level efficiencies
$\epsilon'$ (all computed for a set of cuts on a shape variable, or
multi-variate tagger output, that gives a hadron+MPI-level signal
efficiency $\epsilon_W=0.4$),
\begin{equation}
  \label{eq:resi}
  \zeta = \left(\frac{\Delta \epsilon_W^2}{\langle\epsilon\rangle^2_W}
    + \frac{\Delta \epsilon_\text{QCD}^2}{\langle\epsilon\rangle^2_\text{QCD}}
  \right)^{-\tfrac{1}{2}}\,,
\end{equation}
where $\Delta \epsilon = \epsilon - \epsilon'$ and
$\langle\epsilon\rangle=\tfrac12 \left(\epsilon + \epsilon' \right)$.
The left-hand plot shows the results obtained in a specific mass-bin,
comparing our likelihood method with the results from the LH
report~\cite{Bendavid:2018nar}.\footnote{Albeit with jets obtained with
  an initial C/A clustering rather than an anti-$k_t$ clustering as in
  the original LH study.}
The right-hand plot shows the results with full mass information, and
includes results with machine learning.
Both parton-level and hadron+MPI-level efficiencies are calculated
using a discriminator determined/trained using hadron+MPI-level events
(this statement holds for all likelihood, LSTM and BDT-based
results).

Figure~\ref{fig:perf-resilience} shows grey triangles for each of the
88 combinations of a single shape variable and mass used in the LH~2017
report~\cite{Bendavid:2018nar} (the shape and mass being combined via
a BDT in the right-hand plot).
The grey line is the upper envelope of those points.
The specific $D_2^{\text{[loose]}}$ variant discussed in
section~\ref{sec:jet-shape} is highlighted in red and one can see that
it has the best performance among all shape+mass taggers.

For methods that use the Lund plane information one can impose a lower
limit, $k_{t,\text{cut}}$, on the value of $k_t$ for which Lund-plane
declusterings are considered.
Declusterings with lower $k_t$ values are simply ignored, both at the
training stage and subsequently when evaluating performance and
resilience.
For the Lund-LSTM method, the tagger is trained separately for each
$k_{t,\text{cut}}$ value.
Larger values of $k_{t,\text{cut}}$ are expected to yield taggers that
are more resilient to non-perturbative effects.
The results for the Lund-likelihood and Lund-LSTM methods are shown as
blue and black points respectively (linked by lines) in
Fig.~\ref{fig:perf-resilience}, each point corresponding to a specific
value of the $k_t$ cut.

Without a $k_t$ cut for the Lund-based taggers, performances
qualitatively mirror those in Fig.~\ref{fig:ROCs} at the corresponding
value of $\epsilon_W=0.4$: the Lund-LSTM method performs
best, then comes the Lund-likelihood method, followed by
$D_2^{\text{[loose]}}$.
Quantitative differences relative to Fig.~\ref{fig:ROCs} are a
consequence of the lack of detector simulation and the use of a broad
mass bin (Fig.~\ref{fig:perf-resilience-mass-cut}) or full mass
resolution (Fig.~\ref{fig:perf-resilience-full-mass}).
The quantitative differences are especially large in the latter case,
as one would expect (e.g.\ for the LSTM,
$\epsilon_W / \sqrt{\epsilon_\text{QCD}}\simeq 20$ at $\epsilon_W = 0.4$
translates to $1/\epsilon_\text{QCD} \sim 2500$, compared to
$1/\epsilon_\text{QCD} \simeq 700$ in Fig.~\ref{fig:ROCs}).

For the Lund-likelihood method, imposing a low $k_t$ cut,
$\ln k_{t,\text{cut}} < -1$, has little impact on the performance or
resilience relative to the situation without any cut.
Further raising the value of $k_{t,\text{cut}}$ initially leads to a
rapid loss in performance and modest improvement in resilience.
This suggests that there is information in the non-perturbative
region when discriminating boosted $W$ jets from QCD jets.
For $k_{t,\text{cut}} = 1\GeV$ ($\ln k_{t,\text{cut}}=0$), performance
is slightly better than the best shape variable at comparable
resilience.
Only for yet higher values of $k_{t,\text{cut}}$ does resilience
improve substantially, and then the Lund-likelihood performance
remains above that of the shape variables (well above for
Fig.~\ref{fig:perf-resilience-mass-cut}).
Thus it appears that the Lund-likelihood method performs well not just
in terms of raw performance, but also, with a $k_t$ cut, in terms of
performance for a given degree of sensitivity to non-perturbative
effects.

For the Lund-LSTM method, even a small cut on $k_t$ rapidly leads to a
loss of performance.
For $\ln k_{t,\text{cut}} \gtrsim -1$, its performance falls below
that of the Lund-likelihood method and that remains the case as
$k_{t,\text{cut}}$ is further increased.
In fact, for $\ln k_{t,\text{cut}}\gtrsim -0.5$, the performance of
the Lund-LSTM method even starts to fall slightly below the most
optimal shape variables.
This is somewhat puzzling and hints at potential fragility of
machine-learning approaches.

Overall we see that while an ML based approach can achieve substantially
better performance, the models obtained are not particularly
resilient to non-perturbative corrections.
We note however that other training methods, e.g.\ based on
adversarial
networks~\cite{NIPS2014_5423,Louppe:2016ylz,Shimmin:2017mfk}, could
improve the robustness of the derived taggers to specific effects such
as hadronisation, MPI and pileup.

While we have focused here on resilience to corrections from
hadronisation and MPI, one could similarly study the resilience of the
methods against pileup or detector effects.

\section{Conclusions}

The Lund plane offers a powerful new way to study and exploit the
internal structure of jets.
In contrast to traditional shape observables it connects much more
directly to individual regions of phase space.
This makes it useful across a range of applications in jet physics.
It also brings many declustering based jet observables, such as the
iterated soft-drop multiplicity and $z_g$ into a single unified
framework.

One way of studying the Lund plane is in terms of its average density,
as a function of angle and transverse momentum.
This density is amenable to calculation within both resummed and
fixed-order perturbative methods.
We limited our discussion of such a calculation to first order,
section~\ref{sec:agvd-lund-plane}, and identified a number of the
contributions that would become relevant at higher orders.
Experimentally, we believe that much of the Lund plane phase-space can
be reliably determined.
This conclusion is based on Delphes fast-detector simulations in
conjunction with subjet-particle rescaling type algorithms (SPRA,
Appendix~\ref{sec:detector-effects}), to recover information at small
angles that might otherwise be obscured by finite calorimeter
resolution.
This offers a clear potential for carrying out experimental
measurements of the pattern of radiation in both the perturbative and
non-perturbative regimes.
One application of such measurements would be to constrain Monte Carlo
simulation programs, which as we saw in section~\ref{sec:MC-rho} show
up to $30\%$ differences in their predictions of the Lund plane
density from one program to another.
Another application would be to directly identify which kinematic
regions of a jet's radiation pattern are modified in heavy-ion
collisions, thus shedding light on the mechanisms of partonic energy
loss in a hot, dense medium.\footnote{This provided the original
  motivation for introducing the Lund plane as a measurable quantity,
  in the context of the ``Novel tools and observables for jet physics
  in heavy-ion collisions / 5th Heavy Ion Jet Workshop''~\cite{Andrews:2018jcm}, and has in
  the meantime also been studied in
  Refs.~\cite{Chien:2018dfn,AndrewsALICEQM18}.
  See also \url{https://gitlab.cern.ch/gsalam/2017-lund-from-MC} for
  a corresponding implementation.
  In heavy-ion collisions, background (i.e.\ UE)
  contamination appears to be a 
  non-negligible issue, as it may be also in high-pileup $pp$
  collisions, depending on the precise pileup-mitigation scheme being
  used.
  Various potential approaches to address this were highlighted in
  section~\ref{sec:sim+reco}.
}

A use case for the Lund plane that we have explicitly examined is for
tagging boosted electroweak bosons.
Compared to the jet-image type inputs that have been the mainstay of
``visual'' machine-learning approaches to jet substructure tagging so
far (e.g.\ boosted-$W$ tagging), many of the features that can be
exploited are immediately visible to the human eye.
With certain machine-learning methods (notably LSTM's) the Lund-plane
inputs appear to yield superior $W$-tagging performance as compared to
jet images.
This is despite the fact that by discarding information about
secondary leaves of the Lund diagram, we are actually providing less
information to the machine learning methods than comes with jet images.
We note that for reliable comparisons of the relative quantitative
performance of different methods it was important to take into account
detector effects.

The fundamental information that is contained in the Lund plane, i.e.\
the kinematics of declustering sequences, has been used in other
recent work on machine learning~\cite{Louppe:2017ipp,Egan:2017ojy}.
However the Lund plane as a visualisation provides powerful insight
into the physical structure of that information and into how that
information differs according to the origin of the jet, cf.\
Fig.~\ref{fig:Wavg} for $W$ jets versus
Fig.~\ref{fig:average-lund-plane} for QCD jets.
In particular, the Lund plane's simplicity, and the relatively
moderate degree of correlation between different parts of the plane,
have the consequence that much of the performance obtained by
machine-learning algorithms can be reproduced using conceptually
simple log-likelihood approaches.
This opens the prospect for a substantial degree of experimental and
theoretical understanding of the robustness of the Lund plane
information for tagging.
That understanding may be useful also in terms of the construction of
high-performance decorrelated taggers~\cite{Dolen:2016kst}.

There is a potential for a range of other applications, including
top-tagging, quark--gluon discrimination, further improvements of
boosted electroweak boson tagging, or extensions to the recently
proposed soft-drop photon isolation approach~\cite{Hall:2018jub}.
Furthermore, Lund-plane type studies need not necessarily be
restricted to the study of final-state jets, but may also be
informative for the initial state, for example to help discriminate
different mechanisms of Higgs-boson production.

Finally, while we have restricted most of our discussion here to the
primary Lund plane (other than a brief discussion in
appendix~\ref{sec:sec-lund-plane}), one cannot help but wonder about
the potential benefits to be had from exploiting the structure of the
full Lund diagram for jets, cf.\ the middle row of
Fig.~\ref{fig:lund-diagram-explanations}.
This may be relevant both for developing our generic understanding of
the structure of jets, and for certain tagging applications, for
example with recursive neural networks (as in
Ref.~\cite{Louppe:2017ipp}) or tree-LSTM
architectures~\cite{DBLP:journals/corr/TaiSM15} to capture the full
clustering tree in the machine-learning training.

We therefore look forward to a wide range of studies with Lund-diagram
related observables in future work.
To facilitate such studies the \texttt{LundPlane} library is available
as part of the \texttt{FastJet}
\href{https://fastjet.hepforge.org/contrib/}{\texttt{contrib}}
package.

\section*{Acknowledgements}

We are grateful to the referee for helpful comments on the manuscript.
We thank Michele Selvaggi for making it straightforward to interface
an external FastJet instance to Delphes, and Jesse Thaler for
discussions on log-likelihood approaches.
We are also grateful to CERN's Techlab for computing time on a Tesla
P100, and to the NVIDIA Corporation for the donation of a Titan Xp GPU
used for this research.
Our introduction of the Lund plane as a measurable observable was
originally motivated by discussions in the context of the ``Novel
tools and observables for jet physics in heavy-ion collisions / 5th
Heavy Ion Jet Workshop''.
A study of the Lund plane in terms of its sensitivity to jet quenching
has appeared as part of the workshop proceedings~\cite{Andrews:2018jcm} and has also been
examined in Refs.~\cite{Chien:2018dfn,AndrewsALICEQM18}.
F.D.\ is supported by the SNF grant P2SKP2\_165039 and by the Office of
High Energy Physics of the U.S. Department of Energy (DOE) under grant
DE-SC-0012567.
F.D.\ thanks the University of Zurich and the Pauli Center for
Theoretical Studies, G.P.S.\ thanks the IPhT Saclay, and G.S.\ thanks CERN
for hospitality while this work was being completed.
G.S.\ is supported in part by the French Agence Nationale de la
Recherche, under grant ANR-15-CE31-0016.

\appendix
\section{The Lund plane for (C/A-reclustered)  anti-$k_t$ jets}
\label{sec:anti-kt}

\begin{figure}
  \centering
  \begin{subfigure}{0.45\textwidth}
    \includegraphics[width=\textwidth,page=1]{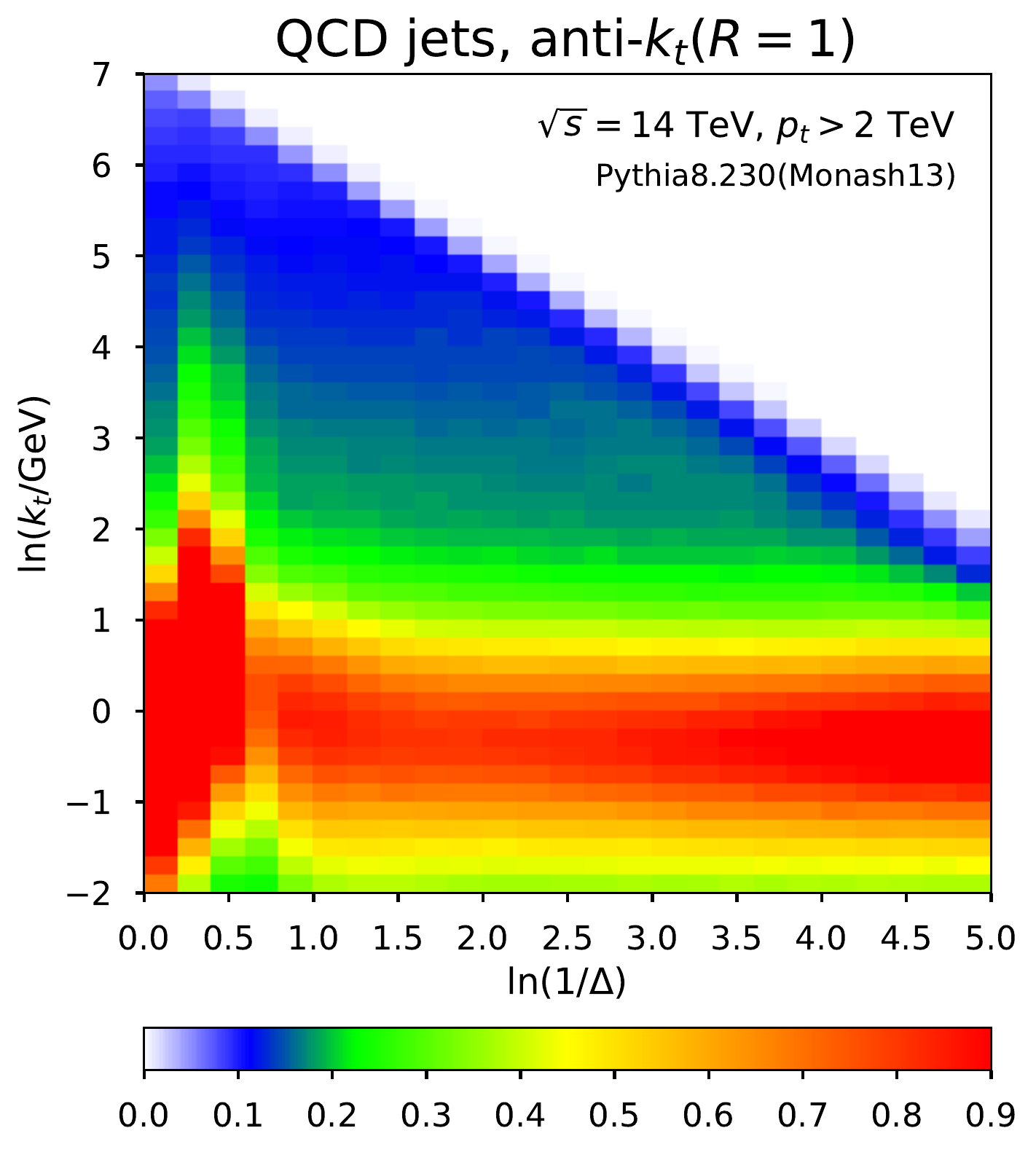}
    \caption{}
    \label{fig:antikt-lund-abs}
  \end{subfigure}
  \begin{subfigure}{0.45\textwidth}
    \includegraphics[width=\textwidth,page=3]{figs/cam-v-antikt.pdf}
    \caption{}
    \label{fig:antikt-lund-rel}
  \end{subfigure}
  \caption{The averaged primary Lund plane density (a) for jets initially obtained with anti-$k_t$
    clustering (whose constituents are then reclustered with the
    Cambridge/Aachen algorithm) and its ratio (b) to the averaged Lund
    plane density for jets originally obtained with Cambridge/Aachen
    clustering (Fig.~\ref{fig:average-lund-plane}).
    Note the structure around $\Delta=0$ and $\ln k_t = 0$ that is
    present here and not in Fig.~\ref{fig:average-lund-plane}.  }
  \label{fig:antikt-lund}
\end{figure}

Throughout the main text we have determined the Lund plane for jets
obtained with an initial Cambridge/Aachen clustering.
This choice has the advantage of procedural simplicity and also
makes it relatively straightforward to interpret the structures that
appear in different regions of the Lund plane.
One could, however, instead use the anti-$k_t$~\cite{Cacciari:2008gp}
algorithm to find the initial jets, and then recluster their
constituents with the C/A algorithm in order to obtain the Lund plane.
We expect that this might be the preferred experimental approach given
that much effort into experimental jet calibration gets directed to
the anti-$k_t$ algorithm.
The averaged primary Lund plane obtained for C/A-reclustered
anti-$k_t$ jets is shown in Fig.~\ref{fig:antikt-lund}.
It is almost identical to Fig.~\ref{fig:average-lund-plane}, except
near $\Delta=0$ and $\ln k_t = 0$, where the anti-$k_t$ jets appear to
have an additional structure: an up-right going diagonal structure for
$\ln k_t/\text{GeV} \sim 0$ and $\ln R/\Delta \lesssim 0.75$.

\def\stackalignment{r}
\begin{figure}
  \centering
  \makebox[0pt][l]{\small\stackanchor[8pt]{C/A $\rightarrow$}{C/A-reclust. anti-$k_t$ $\downarrow$}}\phantom{%
  \raisebox{-0.35\height}{\includegraphics[width=0.090\textwidth,page=1]{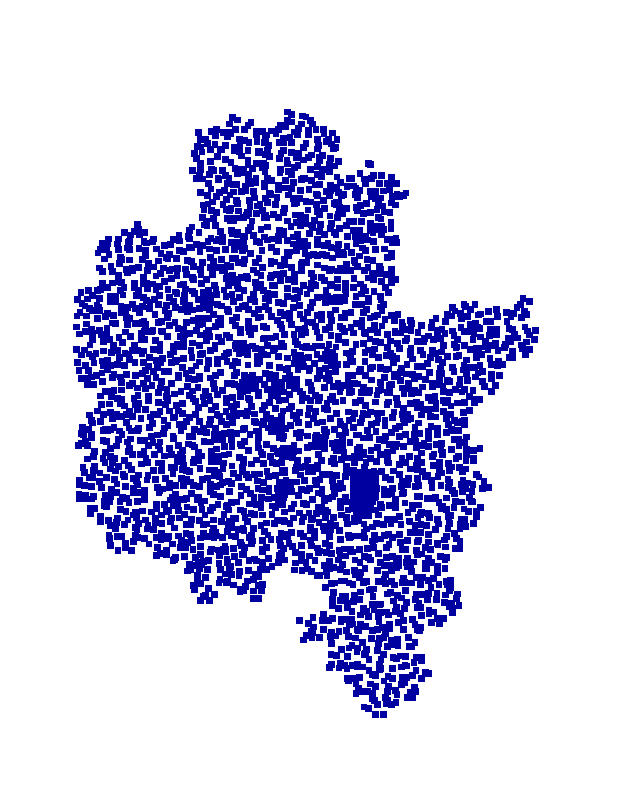}}$\to$%
  \raisebox{-0.35\height}{\includegraphics[width=0.090\textwidth,page=2]{figs/subjet-plots-cam.pdf}}$\to$}%
  \raisebox{-0.35\height}{\includegraphics[width=0.090\textwidth,page=1]{figs/subjet-plots-cam.pdf}}$\to$%
  \raisebox{-0.35\height}{\includegraphics[width=0.090\textwidth,page=2]{figs/subjet-plots-cam.pdf}}$\to$%
  \raisebox{-0.35\height}{\includegraphics[width=0.090\textwidth,page=3]{figs/subjet-plots-cam.pdf}}$\to$%
  \raisebox{-0.35\height}{\includegraphics[width=0.090\textwidth,page=4]{figs/subjet-plots-cam.pdf}}$\to$%
  \raisebox{-0.35\height}{\includegraphics[width=0.090\textwidth,page=5]{figs/subjet-plots-cam.pdf}}$\to$%
  \raisebox{-0.35\height}{\includegraphics[width=0.090\textwidth,page=6]{figs/subjet-plots-cam.pdf}}$\to\cdots$\\
  \raisebox{-0.35\height}{\includegraphics[width=0.090\textwidth,page=1]{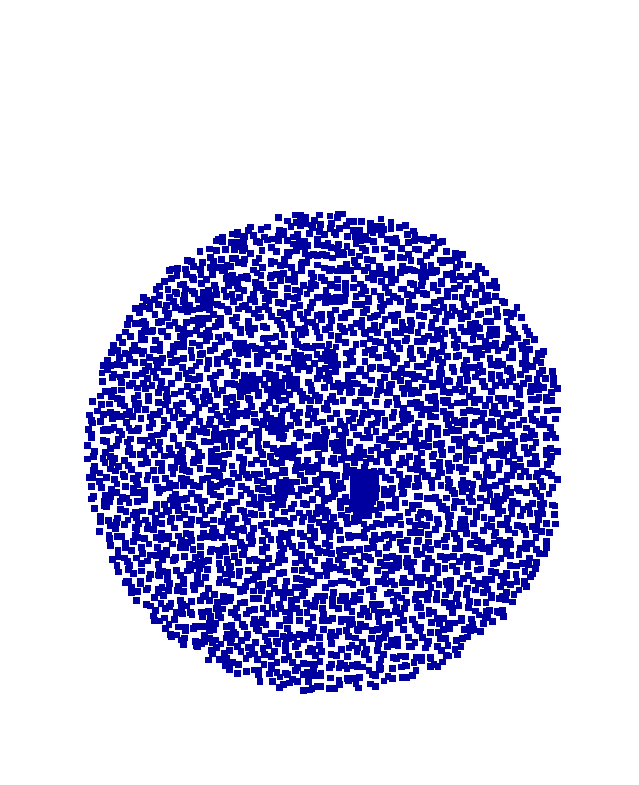}}$\to$%
  \raisebox{-0.35\height}{\includegraphics[width=0.090\textwidth,page=2]{figs/subjet-plots-antikt.pdf}}$\to$%
  \raisebox{-0.35\height}{\includegraphics[width=0.090\textwidth,page=3]{figs/subjet-plots-antikt.pdf}}$\to$%
  \raisebox{-0.35\height}{\includegraphics[width=0.090\textwidth,page=4]{figs/subjet-plots-antikt.pdf}}$\to$%
  \raisebox{-0.35\height}{\includegraphics[width=0.090\textwidth,page=5]{figs/subjet-plots-antikt.pdf}}$\to$%
  \raisebox{-0.35\height}{\includegraphics[width=0.090\textwidth,page=6]{figs/subjet-plots-antikt.pdf}}$\to$%
  \raisebox{-0.35\height}{\includegraphics[width=0.090\textwidth,page=7]{figs/subjet-plots-antikt.pdf}}$\to$%
  \raisebox{-0.35\height}{\includegraphics[width=0.090\textwidth,page=8]{figs/subjet-plots-antikt.pdf}}$\to\cdots$\\
  \caption{
    Illustration of declustering sequences for a
    Cambridge/Aachen jet (upper row) and the same neighbourhood of an event
    clustered with the anti-$k_t$ algorithm and subsequently
    reclustered with Cambridge/Aachen (lower row).
    The jets include ghost
    particles~\cite{Cacciari:2007fd,Cacciari:2008gn}
    so as to illustrate the area of
    the jets involved at each declustering stage.
    The plot shows each stage of the declustering, with the softer
    subjet shown in red ($b$ in the notation of section
    \ref{sec:lund-construction}) and the harder subjet shown in blue
    ($a$).
  }
  \label{fig:declustering-ghosted}
\end{figure}

A reasonable hypothesis is that this structure is associated with the
clustering of soft (mostly underlying-event) particles near the edge
of the jet.
To help understand this in more detail,
Fig.~\ref{fig:declustering-ghosted} shows the rapidity-azimuth
distribution of particles in a single C/A jet (upper row) versus a
C/A-reclustered anti-$k_t$ jet (lower row), at various stages of the
declustering.
At each stage, the softer subjet is shown in red.
The declustering steps are aligned such that the last step shown
corresponds to a similar pair of subjets in the two sequences.
At the earlier stages, for C/A, one sees that large values of $\Delta$
are associated with large-area softer (red) subjets.
In contrast, with reclustered anti-$k_t$ jets, the softer subjets for
large values of $\Delta$ tend to have smaller areas, constrained by
the circular boundary of the original anti-$k_t$ jet.
Smaller areas imply a smaller amount of $p_t$ coming from the
underlying event.
Therefore in the Lund plane the peak of subjets at large $\Delta$
(small $\ln R/\Delta$) should come at lower $k_t$ for the reclustered
anti-$k_t$ jets than for the C/A jets.

\begin{figure}
  \centering
  \includegraphics[width=0.58\textwidth,page=1]{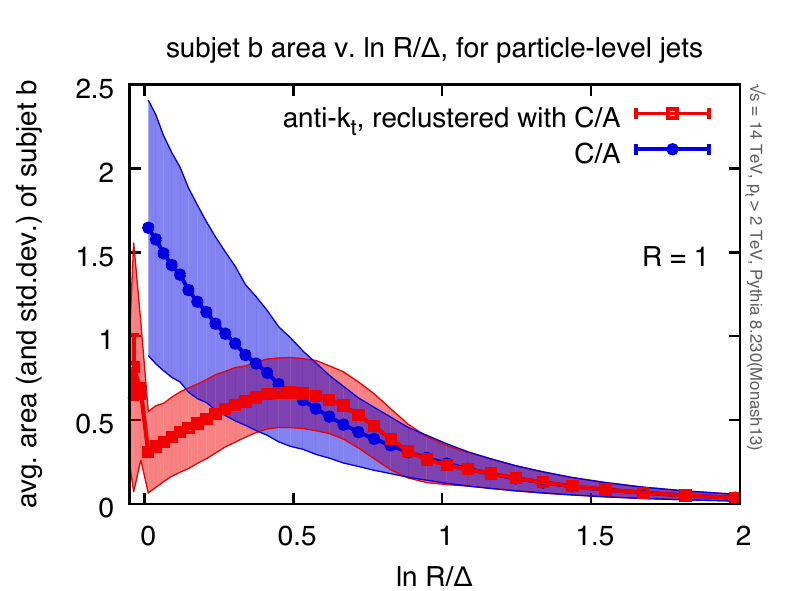}
  \caption{Average area (points), and standard deviation (band), of
    the softer subjet (subjet $b$) in the Lund plane declusterings,
    shown as a function of $\Delta$ for C/A jets (blue) and for
    C/A-reclustered anti-$k_t$ jets (red).}
  \label{fig:jet-areas-v-Delta}
\end{figure}

It is natural to ask whether the pattern of softer-subjet area versus
$\Delta$ seen in Fig.~\ref{fig:declustering-ghosted} holds beyond the
case of a single jet.
To answer this question, Fig.~\ref{fig:jet-areas-v-Delta} shows the
average subjet area as a function of $\ln R/\Delta$ for C/A jets (in
blue) and C/A-reclustered anti-$k_t$ jets (in red).
The bands represent the standard-deviation of the jet areas.
The event sample and jet selection are identical to those used in
Fig.~\ref{fig:antikt-lund}.
In the C/A case, the softer subjet area increases for smaller
$\ln R/\Delta$, i.e.\ for increasing $\Delta$, and is consistent with an
area $A_b$ that scales as $A_b \sim \Delta^2$.
If the density of $p_t$ per unit area from the underlying event is
$\rho$, then one expects the $k_t$ of typical softer subjets to go as
$k_t \sim \rho A_b \Delta \sim \Delta^3$.
In contrast, for $\ln R/\Delta < 0.5$, the typical area for
reclustered anti-$k_t$ softer subjets tends to decrease as
$\ln R/\Delta$ decreases.
The scaling near $\Delta = 0$ is found to be roughly $A_b \sim
\Delta^{-2.6}$, which would lead to $k_t \sim \Delta^{-1.6}$.
Note, however that the area scaling behaviours given here include a
component where the subjet is moderately hard and so the scaling
behaviours for pure underlying-event jets may differ in their details.

A point to note about reclustered anti-$k_t$ jets is that it is
possible to have $\Delta > R$, cf.\ the points at negative $\ln
R/\Delta$ in Fig.~\ref{fig:jet-areas-v-Delta}.
This occurs only rarely and tends to be driven by specific
configurations of hard particles in the jet.

A final comment is that since the difference in structure between C/A
and C/A-reclustered anti-$k_t$ jets' Lund planes is in a region that
is anyway dominated by soft particles from the underlying event, we
expect it to have little impact on discrimination power if one uses
reclustered anti-$k_t$ rather than C/A jets.
Explicit studies with the Lund-likelihood method bear out this
expectation.

\section{Moderate energy jets and secondary Lund planes}
\label{sec:sec-lund-plane}

For most of this article, jets have been considered with a $p_t>2\TeV$
cut on the transverse momentum.
However, when considering lower energy jets, the peak in the primary
Lund plane associated with the $W$ splitting, cf.\ Fig.~\ref{fig:Wavg}, moves to the left and the
shadow region to \emph{its} left, associated with the colour-singlet nature
of the $W$, become less visible.
This is expected to reduce the performance achieved by taggers based just
on the primary Lund plane variables (though the larger fraction of
gluon-induced background jets at lower $p_t$ may partially counteract
this).

\begin{figure}
  \centering
  \includegraphics[width=0.65\columnwidth,page=2]{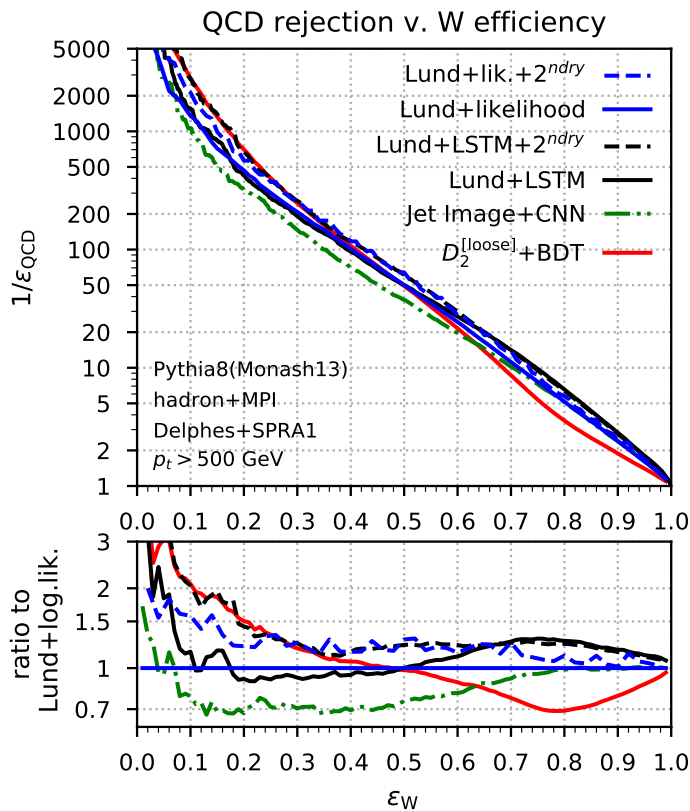}
  \caption{Background rejection ($1/\epsilon_\text{QCD}$) versus
    signal efficiency ($\epsilon_W$), with a transverse momentum cut
    on the jets of $500\GeV$. The lower panel shows the ratio to the
    Lund+likelihood method.
    The solid curves are to be compared to the
    corresponding ones in Fig.~\ref{fig:ROCs} for jets with $p_t >
    2\TeV$ (note the different scale in the lower panel).
  }
  \label{fig:ROC-500gev}
\end{figure}

The reduced performance at lower $p_t$ from primary-Lund plane methods
is visible in Fig.~\ref{fig:ROC-500gev}, which shows the signal
efficiency against the background rejection for $p_t > 500\GeV$ jets,
using the primary Lund-plane log-likelihood method, the primary
Lund-plane LSTM method and also the $D_2^{\text{[loose]}}$ observable
(sensitive also to emissions beyond primary Lund-plane ones).
Those curves are to be compared to the corresponding ones in
Fig.~\ref{fig:ROCs} for $p_t > 2 \TeV$ jets.
Using $\epsilon_W = 0.4$ as a reference point, one sees that all
methods are worse in background rejection at lower $p_t$.
The loss is a factor of $5-7$ for the primary-Lund-plane based
methods, while it's only a little more than a factor of $2$ for the
$D_2^{\text{[loose]}}$ observable, with the result that for
$\epsilon_W = 0.4$ $D_2^{\text{[loose]}}$'s performance at
moderate-$p_t$ is comparable to that of the primary Lund-plane
methods.
At higher (lower) signal efficiencies, $D_2^{\text{[loose]}}$ does
worse (better) than the primary-Lund-plane methods.

\begin{figure}
  \centering
  \includegraphics[width=1.0\textwidth]{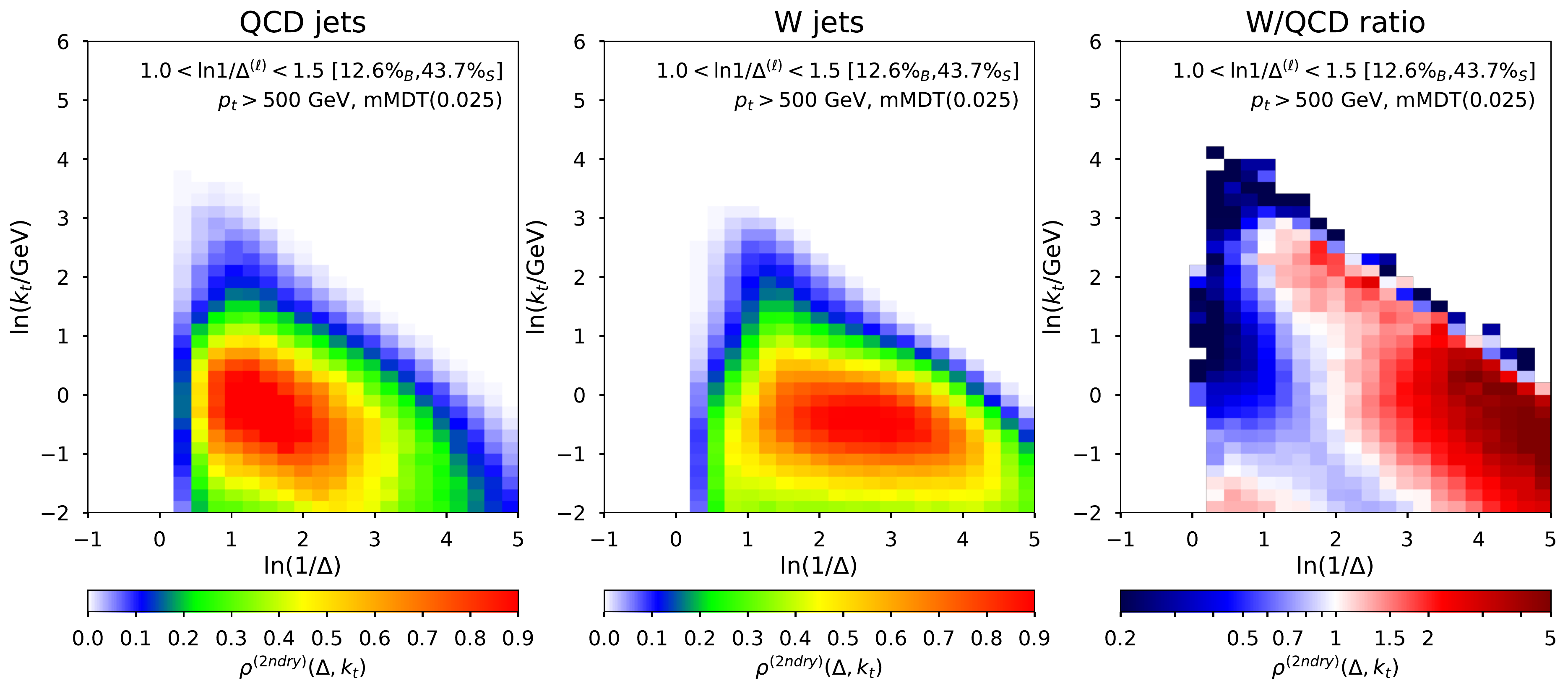}
  \caption{Averaged density,
    $\rho^{(2\text{ndry})}(\Delta,k_t)$, of the secondary Lund plane
    associated with the leading (mMDT(0.025)) emission, 
    for jets in which the leading emission's angle satisfied
    $1<\ln 1/\Delta^{(\ell)} < 1.5$.
    From left to right: for jets in dijet events, for jets in $WW$
    events, and the ratio of the two.
    The percentages in square brackets indicate the fractions of jets
    for which the leading emission is in the given bin, for background
    and signal respectively.}
  \label{fig:Lund-secondary}
\end{figure}

The $D_2$ observable effectively takes into account information not
just from the primary but also secondary Lund planes, information that
is discarded by the primary Lund-plane log likelihood and LSTM
methods.
Fig.~\ref{fig:Lund-secondary} shows the secondary Lund planes for the
leading primary emission, defined as in section~\ref{sec:LL} as the
first emission in the C/A declustering sequence that satisfies
$z > 0.025$ (cf.\ the ``mMDT(0.025)'' label in the figures).
The left-hand plot corresponds to QCD jets, the middle one to $W$
jets.
The right-hand plot, which shows the $W/$QCD secondary Lund-plane
ratio, helps illustrate the nature of the discriminating information
contained in the secondary Lund plane.
In particular, the leading emission in QCD jets will tend to have more
large-angle radiation, while for W jets the emissions are more likely
to be relatively hard and collinear.

To help test the hypothesis that $D_2$ is doing well effectively
because of the secondary information, we have attempted to explicitly
add secondary Lund-plane information also to the log-likelihood and
LSTM methods.

For the log-likelihood method we found that the best performance came
with two steps: replacing the mMDT(0.025) identification of the
leading emission with an identification based on finding the primary
emission that had the smallest value of\linebreak
$|\ln (p_{ta} p_{tb} \Delta^2 / m_W^2) \ln z |$; the additional
likelihood for the secondary plane, similar to the non-leading primary
plane likelihood, is combined with the primary $\cL_\text{tot}$ from
Eq.~(\ref{eq:Ltot}) not by direct addition, but via a 2-dimensional
map of the two likelihoods.\footnote{A two-dimensional map of the two
  likelihoods helps take into 
  account the correlations between the primary and secondary
  likelihoods. These correlations can, for example, come from radiation
  from the leading parton which gets clustered as a secondary emission
  or vice-versa.}
The fact that this last step was needed suggests that there may be
scope for better understanding correlations between the primary and
secondary Lund planes.

For the LSTM method, we start by identifying the secondary Lund plane
using the same method as in the previous log-likelihood approach.
The primary and secondary Lund plane declusterings are then given
as input to two separate LSTMs with a dropout layer, with 128 units for the
primary plane and 64 for the secondary one.
The output of the LSTMs are then concatenated and passed to a dense
layer with 100 dimensions, and then given to a final two-dimensional
layer with softmax activation.

The performances of the methods including the leading emission's
secondary Lund plane are shown as dashed lines in
figure~\ref{fig:ROC-500gev}, and are labelled ``+$2^\text{ndry}$''.
They improve the background rejection by $20-30\%$ in the
$\epsilon_W\sim 0.3-0.6$ range, such that our Lund-based tagging
including the secondary plane now outperforms $D_2$ down to about
$\epsilon_W\sim0.3$.
With the LSTM approach the performance is matched also at lower values
of $\epsilon_W$.
Note that at high $p_t$ we did not find a significant benefit in
including secondary Lund plane information.

We leave a more extensive study of the secondary Lund plane and its
impact on jet tagging for future work, keeping in mind also that
today's parton showers may not always correctly reproduce the patterns of
correlations between emissions~\cite{Dasgupta:2018nvj}.

\section{Detector effects and the subjet-particle rescaling algorithm (SPRA)}
\label{sec:detector-effects}

\begin{figure}
  \centering
  \includegraphics[width=0.6\columnwidth]{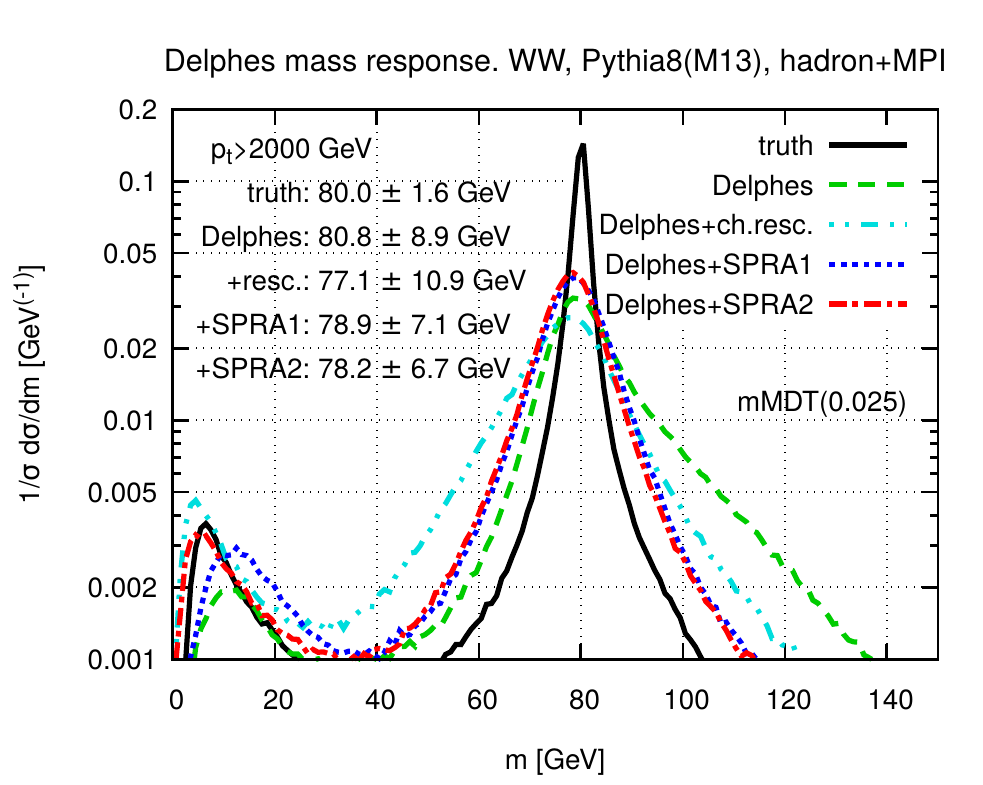}
  \caption{Reconstructed mMDT ($z_{\text{cut}} = 0.025$) $W$ mass peak
    at particle level (``truth''), with Delphes particle
    flow~\cite{deFavereau:2013fsa}, and additionally with the
    rescaling algorithms described in the text to improve
    resolution.
    The average and standard deviation for the reconstructed $W$ peak
    at the different levels are shown based on a fit of a Gaussian
    distribution between $50$ and $110\GeV$.
    The generation and selection of jets is as described in
    section~\ref{sec:sim+reco}, using the $WW$ process, in particular
    selecting jets with the requirement $p_t > 2\TeV$, $|y|<2.5$.  }
  \label{fig:mass-resolution}
\end{figure}

At particle (``truth'') level, in a $W$ sample, with
$p_{t,\text{jet}} > 2\TeV$, the mass of the (sub)jet selected by the
mMDT procedure is very sharply peaked around the true $W$ mass, with
an effective resolution of about $1.5\GeV$.
In contrast if one uses the Delphes fast detector
simulation~\cite{deFavereau:2013fsa}, with particle flow (PF) and the
\verb|delphes_card_CMS_NoFastJet.tcl| card, one finds an mMDT mass
resolution of about $9\GeV$.
This is illustrated in Fig.~\ref{fig:mass-resolution} (we return to
the SPRA curves below).
Such a large difference in resolution between particle and
detector-level can have a big impact on conclusions about performance,
especially given that both ML and likelihood-based approaches tend to derive
considerable discrimination power from the mass variable.
This is true even if the mass is not directly passed as an input
to a ML approach, because the mass can quite effectively be deduced from the
Lund-plane $k_t$ and $\Delta$ variables.

\begin{figure*}
  \centering
  \begin{subfigure}{0.45\textwidth}
    \includegraphics[width=\textwidth,page=1]{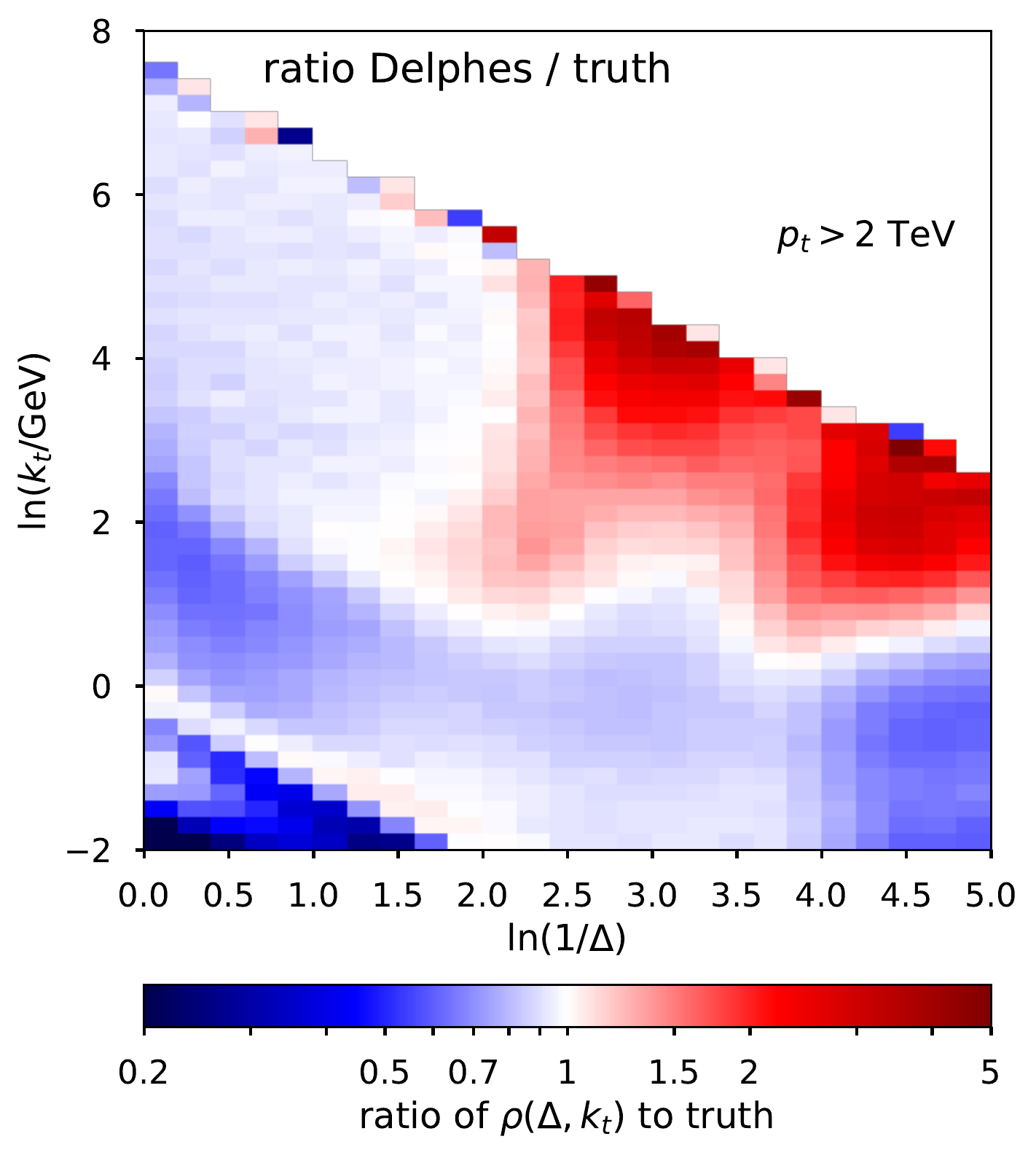}
    \caption{}
    \label{fig:detector-lund-Delphes}
  \end{subfigure}
  \begin{subfigure}{0.45\textwidth}
    \includegraphics[width=\textwidth,page=2]{figs/delphes-lund-images.pdf}
    \caption{}
    \label{fig:detector-lund-Delphes-chgrescaled}
  \end{subfigure}
  \\ 
  \begin{subfigure}{0.45\textwidth}
    \includegraphics[width=\textwidth,page=3]{figs/delphes-lund-images.pdf}
    \caption{}
    \label{fig:detector-lund-Delphes-SPRA1}
  \end{subfigure}
  \begin{subfigure}{0.45\textwidth}
    \includegraphics[width=\textwidth,page=4]{figs/delphes-lund-images.pdf}
    \caption{}
    \label{fig:detector-lund-Delphes-SPRA2}
  \end{subfigure}
  \caption{Ratio of detector-level Lund-plane densities to the truth
    density.
    (a) With Delphes' particle flow algorithm.
    (b) Delphes particle flow supplemented with charged rescaling
    (after rescaling, neutral particles are discarded).
    (c,d) Delphes particle flow supplemented with SPRA1, SPRA2
    respectively.
    The generation and selection of jets is as described in
    section~\ref{sec:sim+reco}, using the dijet process, selecting jets
    with the requirement $p_t > 2\TeV$, $|y|<2.5$. 
  }
  \label{fig:detector-lund}
\end{figure*}

Detector effects can also have a significant impact on the Lund plane
at angular scales commensurate with the hadronic calorimeter spacing
and lower.
This is illustrated in Fig.~\ref{fig:detector-lund-Delphes}, which
shows the ratio of the Lund plane for dijets as obtained with Delphes
particle flow relative to the particle-level truth.
In the lower-left corner there is a prominent dark blue region
indicating missing Lund plane subjets at detector level: this can be
interpreted as a consequence of finite $p_t$ thresholds (a given $p_t$
maps onto a downwards-right going diagonal).
In the right-hand part of the Lund plane, just below the kinematic
limit, there are two prominent enhanced regions (in red), with
corresponding deficits at lower $k_t$: their positions in
$\ln 1/\Delta$ coincide with the intrinsic angular resolution scales
of the hadronic (HCal) and electromagnetic calorimeters (ECal), which
in the central part of the detector have $\eta,\phi$ segmentations of
$0.087\times0.087$ and $0.0174\times 0.0174$ respectively.
Those segmentations translate to $\ln 1/\Delta$ values of about $2.4$
and $4.1$.

The origin of the enhancements is relatively straightforward to
understand.
Consider the structure associated with the hadronic calorimeter scale,
$\Delta_\text{HCal}$.
On average, about $10\%$ of particles in jets are undecayed neutral
hadrons (for example $K_L$).
Schematically, the particle flow algorithm can identify the energy
deposit from such particles as the difference between the energy in a
given hadronic calorimeter tower and that observed in the charged
tracks that enter the tower.
The Delphes PF implementation assigns that energy difference to a point in
$\eta,\phi$ that is randomly distributed over the calorimeter tower
area.
If the neutral hadron has a true separation
$\Delta_\text{true} \ll \Delta_\text{HCal}$ and transverse momentum
$k_{t,\text{true}}$ relative to the jet core, the reconstructed
separation and transverse momentum will be
\begin{equation}
  \label{eq:reco-v-true}
  \Delta_\text{reco} \sim \Delta_\text{HCal}\gg \Delta_\text{true} \,,\quad
  k_{t,\text{reco}} \sim
  \frac{\Delta_\text{HCal}}{\Delta_\text{reco}} k_{t,\text{true}} \gg k_{t,\text{true}}\,,
\end{equation}
where the scaling of the transverse momentum, $k_t$, relative to the
jet core, follows from the assumption that the neutral hadron's
transverse momentum $p_t$ relative to the beam direction is correctly
determined ($k_t = p_t \Delta $, cf.\
Eq.~(\ref{eq:branching-variables-a})).
For a jet core containing $\order{10{-}20}$ particles, there will
typically be at least one neutral hadron, which
will be reconstructed with an angular scale $\Delta_\text{HCal}$ and a
transverse momentum that is larger than its true transverse momentum.
It is this that creates the red bump around
$\ln 1/\Delta \simeq \ln 1/\Delta_\text{HCal}\simeq 2.4$: with the
particle flow algorithm there is nearly always something in that
region, whereas at truth level there isn't.
This is arguably also the origin of the long tail to high masses for
the Delphes curve in Fig.~\ref{fig:mass-resolution}.
An analogous phenomenon occurs on the electromagnetic calorimeter
granularity scale.
The depletions at lower $k_t$ values may be shadows induced by the
enhanced regions. (Given an emission at some $\Delta$, a fraction of
emissions at lower $k_t$ and similar $\Delta$ and $\psi$ get clustered
with it and so are assigned to its secondary Lund plane).

The limitations induced by detector granularity have been addressed in
a range of past work.
Katz, Son and Tweedie~\cite{Katz:2010mr} were the first to document
the issue in the context of substructure and they proposed a solution,
whereby $5\times5$ groups of ECal cells associated with a single HCal cell
were rescaled to match the total $\text{HCal} + \text{ECal}$ energy.
Ref.~\cite{Son:2012mb} extended the procedure, applying the scaling
within minijets.
Nowadays, CMS has an approach referred to as ``split PF
photons+neutrals''~\cite{CMS:2014joa}, which is conceptually similar
insofar as it distributes neutral-hadron energy across the ECal cells
(it also includes tracking improvements for high-energy jets).
Schaetzel and Spannowsky investigated rescaling the charged tracks in
jet to match the jet's total calorimeter energy (``track-flow'' in the
nomenclature of Ref.~\cite{Han:2017hyv}; here we will call it
charge-rescaling).
A procedure that is functionally equivalent,
track-assisted mass, has been studied by ATLAS~\cite{ATLAS:2016vmy} to
improve its mass resolution.
Other studies of the question include
Refs.~\cite{Larkoski:2015yqa,Bressler:2015uma,Han:2017hyv,Elder:2018mcr}.

To mitigate the impact of detector granularity, we adopt a subjet
particle rescaling algorithm (SPRA) that is similar in spirit to that
of Ref.~\cite{Son:2012mb}.
We have two variants, SPRA1 and SPRA2.
For SPRA1, we take a jet and recluster its Delphes particle-flow
objects into subjets using the C/A algorithm with radius $R_h= 0.12$,
commensurate with the hadronic calorimeter granularity.\footnote{As
  pointed out in Ref.~\cite{Son:2012mb}, strictly one should choose
  $\sqrt{2} < R_h/\Delta_\text{HCal} < 2$, but we found little
  difference between a value in that range and our choice that is
  slightly below the lower edge of this range. }
Taking each subjet in turn we scale each PF charged-particle ($h^\pm$)
and photon ($\gamma$) candidate that it contains by a factor $f_1$,
\begin{equation}
  \label{eq:SPRA1}
  f_{1} = \frac{\sum_{i\in\text{subjet}} p_{t,i}}%
               {\sum_{i\in\text{subjet}(h^{\pm},\gamma)} p_{t,i}}.
\end{equation}
and discard the other particles (i.e.\ neutral hadron candidates).
If the subjet does not contain any photon or charged-particle
candidates, we instead retain all of the subjet's particles with their
original momenta.
After having applied this procedure to each subjet, we then recluster
the full set of resulting particles, i.e.\ from all subjets, into a
single large jet and evaluate the mass and Lund plane on that set of
particles.

The SPRA2 variant is a similar but carries out two levels of
rescaling.
The motivation behind the double rescaling is to reduce the effects of
the granularity of both the hadronic and electromagnetic calorimeters.
After applying the SPRA1 algorithm, we recluster the resulting
particles with a radius $R_e = 0.03$, commensurate with the
electromagnetic calorimeter granularity.
Taking each ($R_e$) subjet in turn, we scale each PF charged-particle
($h^\pm$) candidate that it contains by a factor
$f_2$,
\begin{equation}
  \label{eq:SPRA2}
  f_{2} = \frac{\sum_{i\in\text{subjet}} p_{t,i}}%
               {\sum_{i\in\text{subjet}(h^{\pm})} p_{t,i}}.
\end{equation}
and discard the other particles (i.e.\ photons and possibly some
neutral hadron candidates).
If the subjet does not contain any charged-particle candidates, we
instead retain all of the subjet's particles with their original momenta.
Again, after having applied the procedure to each subjet, we take all
the particles and produce a single large-radius jet from them.

Fig.~\ref{fig:mass-resolution} shows that there is some gain in mass
resolution from the SPRA1 algorithm, from about $9\GeV$ to $7 \GeV$.
There is, however, only limited gain in going to the double rescaling,
SPRA2, algorithm.

\begin{figure}
  \centering
    \begin{subfigure}{0.49\textwidth}
      \includegraphics[width=\textwidth,page=1]{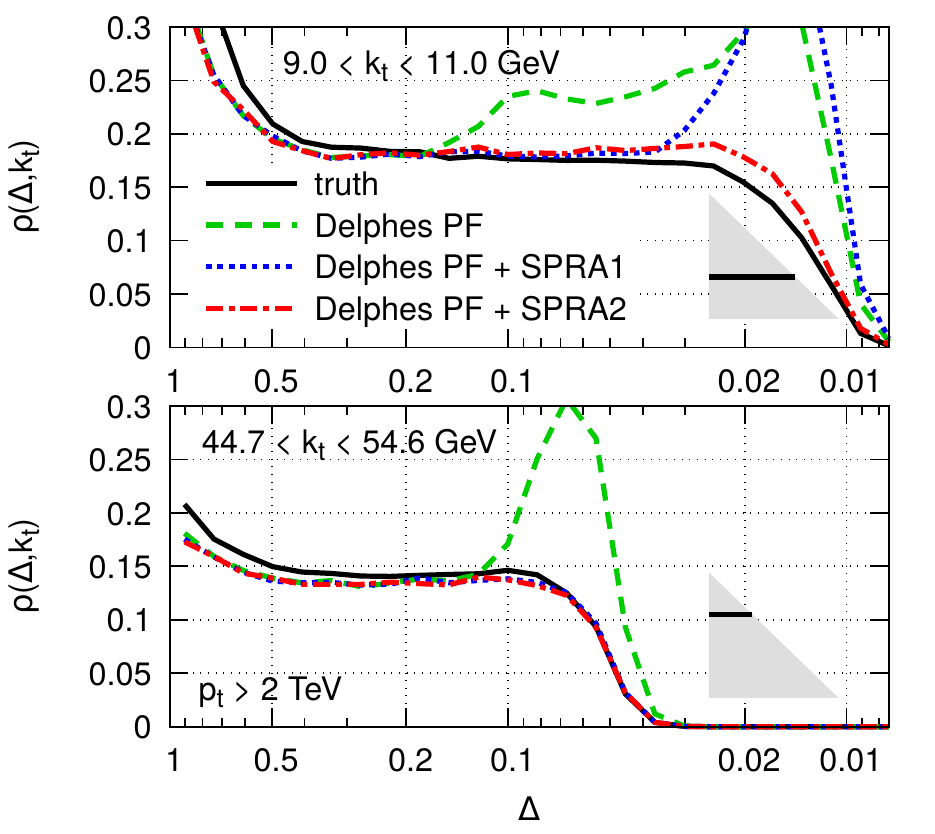}%
    \caption{}
    \label{fig:slices-fixed-kt-delphes}
  \end{subfigure}%
  \begin{subfigure}{0.49\textwidth}
    \includegraphics[width=\textwidth,page=1]{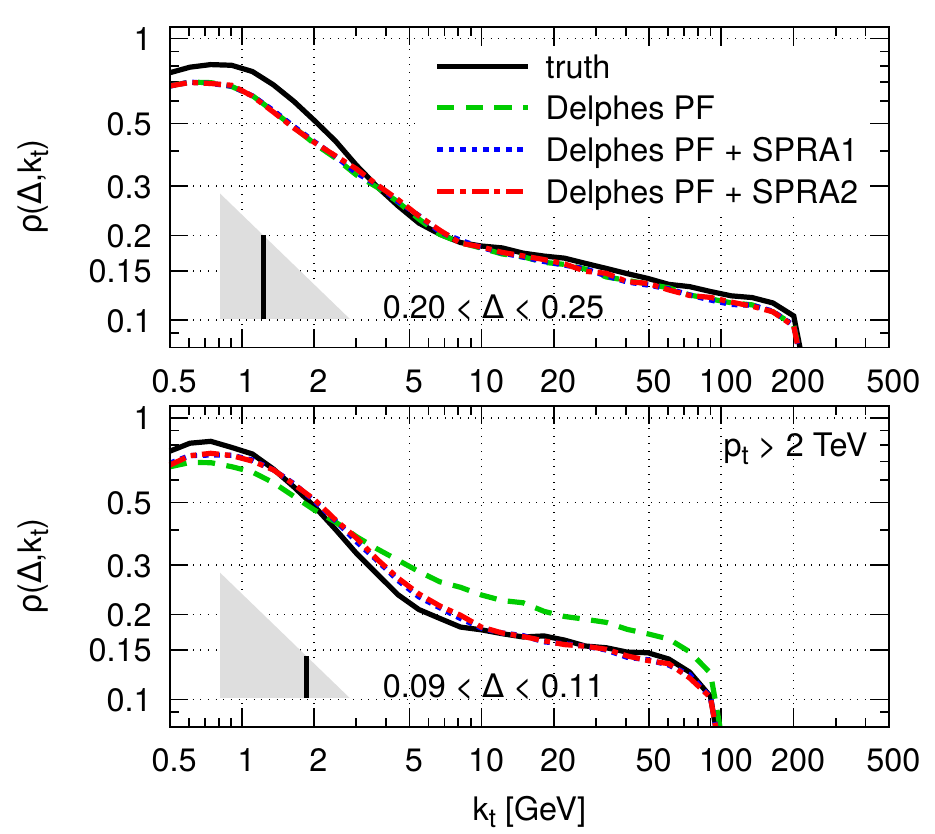}
    \caption{}
    \label{fig:slices-fixed-Delta-delphes}
  \end{subfigure}
  \caption{Lund-plane slices comparing the truth result to Delphes PF
    with and without the SPRA rescalings.
    The slices are shown at fixed $k_t$ as a function of $\Delta$
    (left) and at fixed $\Delta$ versus $k_t$ (right).
    The artefacts visible in the top-left plot at scales of the
    hadronic ($\Delta \sim 0.087$) and electromagnetic
    ($\Delta \sim 0.0174$) calorimeters are well brought under control
    by the SPRA1 and SPRA2 rescalings respectively.
    The generation and selection of jets is as described in
    section~\ref{sec:sim+reco}, using the dijet process, selecting jets
    with the requirement $p_t > 2\TeV$, $|y|<2.5$. 
  }
  \label{fig:lund-slices-Delphes}
\end{figure}

If we now consider the Lund plane reconstruction,
Fig.~\ref{fig:detector-lund}, we see that SPRA1
(Fig.~\ref{fig:detector-lund-Delphes-SPRA1}) eliminates most of the
structure associated with the hadronic calorimeter scale, while SPRA2
(Fig.~\ref{fig:detector-lund-Delphes-SPRA2}) also alleviates some of
the structure associated with the electromagnetic calorimeter scale.
Overall, the conclusion is that with the help of the SPRA algorithms,
a large part of the Lund plane can be measured fairly reliably,
with detector effects that remain limited to within $20{-}30\%$.
This is visible also in the plots of Lund-plane slices in
Fig.~\ref{fig:lund-slices-Delphes}.

An even simpler approach is to adopt a jet-wide track rescaling, where
every charged track is multiplied by a factor
\begin{equation}
  \label{eq:chg-resc}
  f_\text{chg} = \frac{p_{t,\text{jet}}}%
               {\sum_{i\in\text{jet}(h^{\pm})} p_{t,i}}.
\end{equation}
and only those rescaled tracks are used as an input.
This performs fairly well, cf.\ the Lund-plane density ratio in
Fig.~\ref{fig:detector-lund-Delphes-chgrescaled}. 

\begin{figure}
  \centering
  \includegraphics[width=0.6\columnwidth]{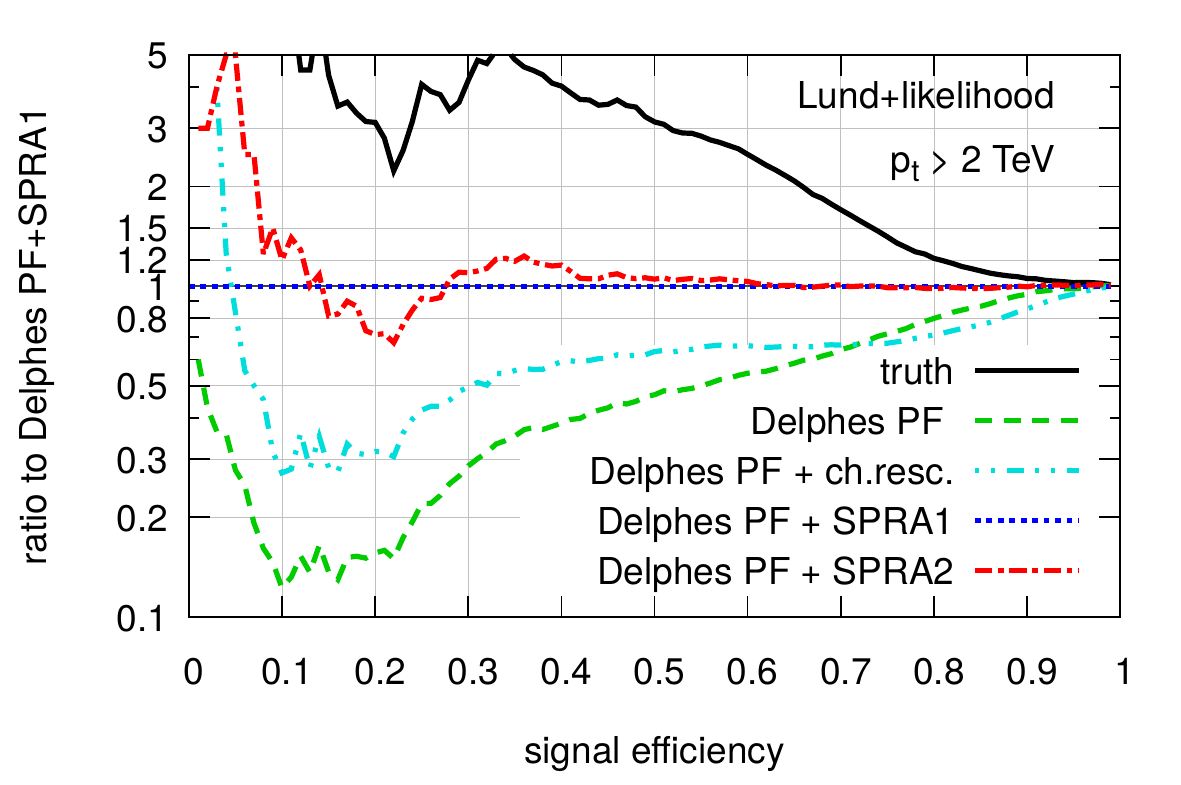}
  \caption{Performance of the Lund-likelihood discriminator, at truth
    level and with Delphes PF, with and without the SPRA algorithms,
    and also with charge rescaling.
    As a function of signal efficiency, the plot shows the ratio of
    background rejection relative to that obtained for Delphes
    PF+SPRA1.
    The log-likelihood maps have been determined separately for each
    setup (truth, Delphes PF, etc.).
    The generation and selection of jets is as described in
    section~\ref{sec:sim+reco}, using the dijet process, selecting jets
    with the requirement $p_t > 2\TeV$, $|y|<2.5$. 
  }
  \label{fig:perf-delphes-v-truth}
\end{figure}

A final test of the SPRA algorithms is shown in
Fig.~\ref{fig:perf-delphes-v-truth}, which compares the background
rejection power of the log-likelihood $W$-tagger, as a function of
signal efficiency, for truth particles and for Delphes PF with and without
SPRA.
A first observation is that for a signal efficiency of $0.5$, the
background rejection is about six times worse with Delphes PF as
compared to truth particles (the factor is even larger with machine
learning taggers).
The SPRA1 algorithm brings about a factor of two improvement relative
to plain PF.
The further gain from the SPRA2 algorithm is limited (and perhaps not
statistically significant).
Accordingly for our main $W$-tagger performance results in
section~\ref{sec:results} we use SPRA1, which is arguably also the
most similar to the procedure used by CMS in Ref.~\cite{CMS:2014joa}.
However at higher $p_t$'s or for measurements of the Lund plane, it is
probably advantageous to use SPRA2.

Fig.~\ref{fig:perf-delphes-v-truth} also shows charge-rescaling, which
performs less well than the SPRA approaches, though still better than
Delphes particle flow alone.
Our understanding of the worse performance relative to the SPRA
approaches is as follows: the performance observed in
Fig.~\ref{fig:perf-delphes-v-truth} combines the performance for the
Lund plane as seen in Fig.~\ref{fig:detector-lund} with the
performance for the jet mass that is observed in
Fig.~\ref{fig:mass-resolution}.
While charge rescaling does well in Fig.~\ref{fig:detector-lund}, it
performs somewhat worse than either of the SPRAs in
Fig.~\ref{fig:mass-resolution}.
Note also that the region of Fig.~\ref{fig:detector-lund} where charge
rescaling performs better than SPRA1, $\ln 1/\Delta \gtrsim 3.5$, is a
region that does not contribute dominantly to the discrimination
power, cf.\ Fig.~\ref{fig:W-dijet-nl-ratio}.

\bibliographystyle{JHEP}
\bibliography{lund}

\providecommand{\href}[2]{#2}\begingroup\raggedright\begin{thebibliography}{100}

\bibitem{Larkoski:2017jix}
A.~J. Larkoski, I.~Moult, and B.~Nachman, {\it {Jet Substructure at the Large
  Hadron Collider: A Review of Recent Advances in Theory and Machine
  Learning}},  \href{http://arxiv.org/abs/1709.04464}{{\tt arXiv:1709.04464}}.

\bibitem{Asquith:2018igt}
L.~Asquith et~al., {\it {Jet Substructure at the Large Hadron Collider :
  Experimental Review}},  \href{http://arxiv.org/abs/1803.06991}{{\tt
  arXiv:1803.06991}}.

\bibitem{Butterworth:2008iy}
J.~M. Butterworth, A.~R. Davison, M.~Rubin, and G.~P. Salam, {\it {Jet
  substructure as a new Higgs search channel at the LHC}},  {\em Phys. Rev.
  Lett.} {\bf 100} (2008) 242001, [\href{http://arxiv.org/abs/0802.2470}{{\tt
  arXiv:0802.2470}}].

\bibitem{Thaler:2008ju}
J.~Thaler and L.-T. Wang, {\it {Strategies to Identify Boosted Tops}},  {\em
  JHEP} {\bf 07} (2008) 092, [\href{http://arxiv.org/abs/0806.0023}{{\tt
  arXiv:0806.0023}}].

\bibitem{Kaplan:2008ie}
D.~E. Kaplan, K.~Rehermann, M.~D. Schwartz, and B.~Tweedie, {\it {Top Tagging:
  A Method for Identifying Boosted Hadronically Decaying Top Quarks}},  {\em
  Phys. Rev. Lett.} {\bf 101} (2008) 142001,
  [\href{http://arxiv.org/abs/0806.0848}{{\tt arXiv:0806.0848}}].

\bibitem{Ellis:2009su}
S.~D. Ellis, C.~K. Vermilion, and J.~R. Walsh, {\it {Techniques for improved
  heavy particle searches with jet substructure}},  {\em Phys. Rev.} {\bf D80}
  (2009) 051501, [\href{http://arxiv.org/abs/0903.5081}{{\tt
  arXiv:0903.5081}}].

\bibitem{Ellis:2009me}
S.~D. Ellis, C.~K. Vermilion, and J.~R. Walsh, {\it {Recombination Algorithms
  and Jet Substructure: Pruning as a Tool for Heavy Particle Searches}},  {\em
  Phys. Rev.} {\bf D81} (2010) 094023,
  [\href{http://arxiv.org/abs/0912.0033}{{\tt arXiv:0912.0033}}].

\bibitem{Plehn:2009rk}
T.~Plehn, G.~P. Salam, and M.~Spannowsky, {\it {Fat Jets for a Light Higgs}},
  {\em Phys. Rev. Lett.} {\bf 104} (2010) 111801,
  [\href{http://arxiv.org/abs/0910.5472}{{\tt arXiv:0910.5472}}].

\bibitem{Thaler:2010tr}
J.~Thaler and K.~Van~Tilburg, {\it {Identifying Boosted Objects with
  N-subjettiness}},  {\em JHEP} {\bf 03} (2011) 015,
  [\href{http://arxiv.org/abs/1011.2268}{{\tt arXiv:1011.2268}}].

\bibitem{Thaler:2011gf}
J.~Thaler and K.~Van~Tilburg, {\it {Maximizing Boosted Top Identification by
  Minimizing N-subjettiness}},  {\em JHEP} {\bf 02} (2012) 093,
  [\href{http://arxiv.org/abs/1108.2701}{{\tt arXiv:1108.2701}}].

\bibitem{Larkoski:2013eya}
A.~J. Larkoski, G.~P. Salam, and J.~Thaler, {\it {Energy Correlation Functions
  for Jet Substructure}},  {\em JHEP} {\bf 06} (2013) 108,
  [\href{http://arxiv.org/abs/1305.0007}{{\tt arXiv:1305.0007}}].

\bibitem{Chien:2013kca}
Y.-T. Chien, {\it {Telescoping jets: Probing hadronic event structure with
  multiple R ’s}},  {\em Phys. Rev.} {\bf D90} (2014), no.~5 054008,
  [\href{http://arxiv.org/abs/1304.5240}{{\tt arXiv:1304.5240}}].

\bibitem{Larkoski:2014gra}
A.~J. Larkoski, I.~Moult, and D.~Neill, {\it {Power Counting to Better Jet
  Observables}},  {\em JHEP} {\bf 12} (2014) 009,
  [\href{http://arxiv.org/abs/1409.6298}{{\tt arXiv:1409.6298}}].

\bibitem{Larkoski:2014pca}
A.~J. Larkoski, J.~Thaler, and W.~J. Waalewijn, {\it {Gaining (Mutual)
  Information about Quark/Gluon Discrimination}},  {\em JHEP} {\bf 11} (2014)
  129, [\href{http://arxiv.org/abs/1408.3122}{{\tt arXiv:1408.3122}}].

\bibitem{Moult:2016cvt}
I.~Moult, L.~Necib, and J.~Thaler, {\it {New Angles on Energy Correlation
  Functions}},  {\em JHEP} {\bf 12} (2016) 153,
  [\href{http://arxiv.org/abs/1609.07483}{{\tt arXiv:1609.07483}}].

\bibitem{Salam:2016yht}
G.~P. Salam, L.~Schunk, and G.~Soyez, {\it {Dichroic subjettiness ratios to
  distinguish colour flows in boosted boson tagging}},  {\em JHEP} {\bf 03}
  (2017) 022, [\href{http://arxiv.org/abs/1612.03917}{{\tt arXiv:1612.03917}}].

\bibitem{Seymour:1997kj}
M.~H. Seymour, {\it {Jet shapes in hadron collisions: Higher orders,
  resummation and hadronization}},  {\em Nucl. Phys.} {\bf B513} (1998)
  269--300, [\href{http://arxiv.org/abs/hep-ph/9707338}{{\tt hep-ph/9707338}}].

\bibitem{Feige:2012vc}
I.~Feige, M.~D. Schwartz, I.~W. Stewart, and J.~Thaler, {\it {Precision Jet
  Substructure from Boosted Event Shapes}},  {\em Phys. Rev. Lett.} {\bf 109}
  (2012) 092001, [\href{http://arxiv.org/abs/1204.3898}{{\tt
  arXiv:1204.3898}}].

\bibitem{Dasgupta:2013ihk}
M.~Dasgupta, A.~Fregoso, S.~Marzani, and G.~P. Salam, {\it {Towards an
  understanding of jet substructure}},  {\em JHEP} {\bf 09} (2013) 029,
  [\href{http://arxiv.org/abs/1307.0007}{{\tt arXiv:1307.0007}}].

\bibitem{Dasgupta:2013via}
M.~Dasgupta, A.~Fregoso, S.~Marzani, and A.~Powling, {\it {Jet substructure
  with analytical methods}},  {\em Eur. Phys. J.} {\bf C73} (2013), no.~11
  2623, [\href{http://arxiv.org/abs/1307.0013}{{\tt arXiv:1307.0013}}].

\bibitem{Chien:2014nsa}
Y.-T. Chien and I.~Vitev, {\it {Jet Shape Resummation Using Soft-Collinear
  Effective Theory}},  {\em JHEP} {\bf 12} (2014) 061,
  [\href{http://arxiv.org/abs/1405.4293}{{\tt arXiv:1405.4293}}].

\bibitem{Bertolini:2015pka}
D.~Bertolini, J.~Thaler, and J.~R. Walsh, {\it {The First Calculation of
  Fractional Jets}},  {\em JHEP} {\bf 05} (2015) 008,
  [\href{http://arxiv.org/abs/1501.01965}{{\tt arXiv:1501.01965}}].

\bibitem{Dasgupta:2015lxh}
M.~Dasgupta, L.~Schunk, and G.~Soyez, {\it {Jet shapes for boosted jet
  two-prong decays from first-principles}},  {\em JHEP} {\bf 04} (2016) 166,
  [\href{http://arxiv.org/abs/1512.00516}{{\tt arXiv:1512.00516}}].

\bibitem{Frye:2016okc}
C.~Frye, A.~J. Larkoski, M.~D. Schwartz, and K.~Yan, {\it {Precision physics
  with pile-up insensitive observables}},
  \href{http://arxiv.org/abs/1603.06375}{{\tt arXiv:1603.06375}}.

\bibitem{Frye:2016aiz}
C.~Frye, A.~J. Larkoski, M.~D. Schwartz, and K.~Yan, {\it {Factorization for
  groomed jet substructure beyond the next-to-leading logarithm}},  {\em JHEP}
  {\bf 07} (2016) 064, [\href{http://arxiv.org/abs/1603.09338}{{\tt
  arXiv:1603.09338}}].

\bibitem{Dasgupta:2016ktv}
M.~Dasgupta, A.~Powling, L.~Schunk, and G.~Soyez, {\it {Improved jet
  substructure methods: Y-splitter and variants with grooming}},  {\em JHEP}
  {\bf 12} (2016) 079, [\href{http://arxiv.org/abs/1609.07149}{{\tt
  arXiv:1609.07149}}].

\bibitem{Marzani:2017kqd}
S.~Marzani, L.~Schunk, and G.~Soyez, {\it {The jet mass distribution after Soft
  Drop}},  {\em Eur. Phys. J.} {\bf C78} (2018), no.~2 96,
  [\href{http://arxiv.org/abs/1712.05105}{{\tt arXiv:1712.05105}}].

\bibitem{Marzani:2017mva}
S.~Marzani, L.~Schunk, and G.~Soyez, {\it {A study of jet mass distributions
  with grooming}},  {\em JHEP} {\bf 07} (2017) 132,
  [\href{http://arxiv.org/abs/1704.02210}{{\tt arXiv:1704.02210}}].

\bibitem{Larkoski:2017cqq}
A.~J. Larkoski, I.~Moult, and D.~Neill, {\it {Factorization and Resummation for
  Groomed Multi-Prong Jet Shapes}},  {\em JHEP} {\bf 02} (2018) 144,
  [\href{http://arxiv.org/abs/1710.00014}{{\tt arXiv:1710.00014}}].

\bibitem{Moult:2017okx}
I.~Moult, B.~Nachman, and D.~Neill, {\it {Convolved Substructure: Analytically
  Decorrelating Jet Substructure Observables}},  {\em JHEP} {\bf 05} (2018)
  002, [\href{http://arxiv.org/abs/1710.06859}{{\tt arXiv:1710.06859}}].

\bibitem{Khachatryan:2014vla}
{\bf CMS} Collaboration, V.~Khachatryan et~al., {\it {Identification techniques
  for highly boosted W bosons that decay into hadrons}},  {\em JHEP} {\bf 12}
  (2014) 017, [\href{http://arxiv.org/abs/1410.4227}{{\tt arXiv:1410.4227}}].

\bibitem{Aad:2015rpa}
{\bf ATLAS} Collaboration, G.~Aad et~al., {\it {Identification of boosted,
  hadronically decaying W bosons and comparisons with ATLAS data taken at
  $\sqrt{s} = 8$ TeV}},  {\em Eur. Phys. J.} {\bf C76} (2016), no.~3 154,
  [\href{http://arxiv.org/abs/1510.05821}{{\tt arXiv:1510.05821}}].

\bibitem{Aad:2015owa}
{\bf ATLAS} Collaboration, G.~Aad et~al., {\it {Search for high-mass diboson
  resonances with boson-tagged jets in proton-proton collisions at $ \sqrt{s}=8
  $ TeV with the ATLAS detector}},  {\em JHEP} {\bf 12} (2015) 055,
  [\href{http://arxiv.org/abs/1506.00962}{{\tt arXiv:1506.00962}}].

\bibitem{Aaboud:2017qwh}
{\bf ATLAS} Collaboration, M.~Aaboud et~al., {\it {A measurement of the
  soft-drop jet mass in pp collisions at $\sqrt{s} = 13$ TeV with the ATLAS
  detector}},  \href{http://arxiv.org/abs/1711.08341}{{\tt arXiv:1711.08341}}.

\bibitem{CMS:2018yfn}
{\bf CMS} Collaboration, C.~Collaboration, ``{Measurement of jet substructure
  observables in $\mathrm{t \bar t}$ events from pp collisions at
  $\sqrt{s}=13~\mathrm{TeV}$}.'' CMS-PAS-TOP-17-013, 2018.

\bibitem{Cogan:2014oua}
J.~Cogan, M.~Kagan, E.~Strauss, and A.~Schwarztman, {\it {Jet-Images: Computer
  Vision Inspired Techniques for Jet Tagging}},  {\em JHEP} {\bf 02} (2015)
  118, [\href{http://arxiv.org/abs/1407.5675}{{\tt arXiv:1407.5675}}].

\bibitem{deOliveira:2015xxd}
L.~de~Oliveira, M.~Kagan, L.~Mackey, B.~Nachman, and A.~Schwartzman, {\it
  {Jet-images — deep learning edition}},  {\em JHEP} {\bf 07} (2016) 069,
  [\href{http://arxiv.org/abs/1511.05190}{{\tt arXiv:1511.05190}}].

\bibitem{Komiske:2016rsd}
P.~T. Komiske, E.~M. Metodiev, and M.~D. Schwartz, {\it {Deep learning in
  color: towards automated quark/gluon jet discrimination}},  {\em JHEP} {\bf
  01} (2017) 110, [\href{http://arxiv.org/abs/1612.01551}{{\tt
  arXiv:1612.01551}}].

\bibitem{Kasieczka:2017nvn}
G.~Kasieczka, T.~Plehn, M.~Russell, and T.~Schell, {\it {Deep-learning Top
  Taggers or The End of QCD?}},  {\em JHEP} {\bf 05} (2017) 006,
  [\href{http://arxiv.org/abs/1701.08784}{{\tt arXiv:1701.08784}}].

\bibitem{Louppe:2017ipp}
G.~Louppe, K.~Cho, C.~Becot, and K.~Cranmer, {\it {QCD-Aware Recursive Neural
  Networks for Jet Physics}},  \href{http://arxiv.org/abs/1702.00748}{{\tt
  arXiv:1702.00748}}.

\bibitem{Egan:2017ojy}
S.~Egan, W.~Fedorko, A.~Lister, J.~Pearkes, and C.~Gay, {\it {Long Short-Term
  Memory (LSTM) networks with jet constituents for boosted top tagging at the
  LHC}},  \href{http://arxiv.org/abs/1711.09059}{{\tt arXiv:1711.09059}}.

\bibitem{Andreassen:2018apy}
A.~Andreassen, I.~Feige, C.~Frye, and M.~D. Schwartz, {\it {JUNIPR: a Framework
  for Unsupervised Machine Learning in Particle Physics}},
  \href{http://arxiv.org/abs/1804.09720}{{\tt arXiv:1804.09720}}.

\bibitem{Datta:2017rhs}
K.~Datta and A.~Larkoski, {\it {How Much Information is in a Jet?}},  {\em
  JHEP} {\bf 06} (2017) 073, [\href{http://arxiv.org/abs/1704.08249}{{\tt
  arXiv:1704.08249}}].

\bibitem{Datta:2017lxt}
K.~Datta and A.~J. Larkoski, {\it {Novel Jet Observables from Machine
  Learning}},  {\em JHEP} {\bf 03} (2018) 086,
  [\href{http://arxiv.org/abs/1710.01305}{{\tt arXiv:1710.01305}}].

\bibitem{Komiske:2017aww}
P.~T. Komiske, E.~M. Metodiev, and J.~Thaler, {\it {Energy flow polynomials: A
  complete linear basis for jet substructure}},  {\em JHEP} {\bf 04} (2018)
  013, [\href{http://arxiv.org/abs/1712.07124}{{\tt arXiv:1712.07124}}].

\bibitem{Metodiev:2017vrx}
E.~M. Metodiev, B.~Nachman, and J.~Thaler, {\it {Classification without labels:
  Learning from mixed samples in high energy physics}},  {\em JHEP} {\bf 10}
  (2017) 174, [\href{http://arxiv.org/abs/1708.02949}{{\tt arXiv:1708.02949}}].

\bibitem{Dery:2017fap}
L.~M. Dery, B.~Nachman, F.~Rubbo, and A.~Schwartzman, {\it {Weakly Supervised
  Classification in High Energy Physics}},  {\em JHEP} {\bf 05} (2017) 145,
  [\href{http://arxiv.org/abs/1702.00414}{{\tt arXiv:1702.00414}}].

\bibitem{Andersson:1988gp}
B.~Andersson, G.~Gustafson, L.~Lonnblad, and U.~Pettersson, {\it {Coherence
  Effects in Deep Inelastic Scattering}},  {\em Z. Phys.} {\bf C43} (1989) 625.

\bibitem{Dokshitzer:1997in}
Y.~L. Dokshitzer, G.~D. Leder, S.~Moretti, and B.~R. Webber, {\it {Better jet
  clustering algorithms}},  {\em JHEP} {\bf 08} (1997) 001,
  [\href{http://arxiv.org/abs/hep-ph/9707323}{{\tt hep-ph/9707323}}].

\bibitem{Wobisch:1998wt}
M.~Wobisch and T.~Wengler, {\it {Hadronization corrections to jet
  cross-sections in deep inelastic scattering}},  in {\em {Monte Carlo
  generators for HERA physics. Proceedings, Workshop, Hamburg, Germany,
  1998-1999}}, pp.~270--279, 1998.
\newblock \href{http://arxiv.org/abs/hep-ph/9907280}{{\tt hep-ph/9907280}}.

\bibitem{Cacciari:2008gp}
M.~Cacciari, G.~P. Salam, and G.~Soyez, {\it {The Anti-k(t) jet clustering
  algorithm}},  {\em JHEP} {\bf 04} (2008) 063,
  [\href{http://arxiv.org/abs/0802.1189}{{\tt arXiv:0802.1189}}].

\bibitem{Sjostrand:2014zea}
T.~Sjöstrand, S.~Ask, J.~R. Christiansen, R.~Corke, N.~Desai, P.~Ilten,
  S.~Mrenna, S.~Prestel, C.~O. Rasmussen, and P.~Z. Skands, {\it {An
  Introduction to PYTHIA 8.2}},  {\em Comput. Phys. Commun.} {\bf 191} (2015)
  159--177, [\href{http://arxiv.org/abs/1410.3012}{{\tt arXiv:1410.3012}}].

\bibitem{Skands:2014pea}
P.~Skands, S.~Carrazza, and J.~Rojo, {\it {Tuning PYTHIA 8.1: the Monash 2013
  Tune}},  {\em Eur. Phys. J.} {\bf C74} (2014), no.~8 3024,
  [\href{http://arxiv.org/abs/1404.5630}{{\tt arXiv:1404.5630}}].

\bibitem{Dasgupta:2014yra}
M.~Dasgupta, F.~Dreyer, G.~P. Salam, and G.~Soyez, {\it {Small-radius jets to
  all orders in QCD}},  {\em JHEP} {\bf 04} (2015) 039,
  [\href{http://arxiv.org/abs/1411.5182}{{\tt arXiv:1411.5182}}].

\bibitem{Dasgupta:2016bnd}
M.~Dasgupta, F.~A. Dreyer, G.~P. Salam, and G.~Soyez, {\it {Inclusive jet
  spectrum for small-radius jets}},  {\em JHEP} {\bf 06} (2016) 057,
  [\href{http://arxiv.org/abs/1602.01110}{{\tt arXiv:1602.01110}}].

\bibitem{Dasgupta:2001sh}
M.~Dasgupta and G.~P. Salam, {\it {Resummation of nonglobal QCD observables}},
  {\em Phys. Lett.} {\bf B512} (2001) 323--330,
  [\href{http://arxiv.org/abs/hep-ph/0104277}{{\tt hep-ph/0104277}}].

\bibitem{Appleby:2002ke}
R.~B. Appleby and M.~H. Seymour, {\it {Nonglobal logarithms in interjet energy
  flow with kt clustering requirement}},  {\em JHEP} {\bf 12} (2002) 063,
  [\href{http://arxiv.org/abs/hep-ph/0211426}{{\tt hep-ph/0211426}}].

\bibitem{Delenda:2006nf}
Y.~Delenda, R.~Appleby, M.~Dasgupta, and A.~Banfi, {\it {On QCD resummation
  with k(t) clustering}},  {\em JHEP} {\bf 12} (2006) 044,
  [\href{http://arxiv.org/abs/hep-ph/0610242}{{\tt hep-ph/0610242}}].

\bibitem{Gleisberg:2008ta}
T.~Gleisberg, S.~Hoeche, F.~Krauss, M.~Schonherr, S.~Schumann, F.~Siegert, and
  J.~Winter, {\it {Event generation with SHERPA 1.1}},  {\em JHEP} {\bf 02}
  (2009) 007, [\href{http://arxiv.org/abs/0811.4622}{{\tt arXiv:0811.4622}}].

\bibitem{Bellm:2017bvx}
J.~Bellm et~al., {\it {Herwig 7.1 Release Note}},
  \href{http://arxiv.org/abs/1705.06919}{{\tt arXiv:1705.06919}}.

\bibitem{Dokshitzer:1995zt}
Y.~L. Dokshitzer and B.~R. Webber, {\it {Calculation of power corrections to
  hadronic event shapes}},  {\em Phys. Lett.} {\bf B352} (1995) 451--455,
  [\href{http://arxiv.org/abs/hep-ph/9504219}{{\tt hep-ph/9504219}}].

\bibitem{Dokshitzer:1995qm}
Y.~L. Dokshitzer, G.~Marchesini, and B.~R. Webber, {\it {Dispersive approach to
  power behaved contributions in QCD hard processes}},  {\em Nucl. Phys.} {\bf
  B469} (1996) 93--142, [\href{http://arxiv.org/abs/hep-ph/9512336}{{\tt
  hep-ph/9512336}}].

\bibitem{Forshaw:2006fk}
J.~R. Forshaw, A.~Kyrieleis, and M.~H. Seymour, {\it {Super-leading logarithms
  in non-global observables in QCD}},  {\em JHEP} {\bf 08} (2006) 059,
  [\href{http://arxiv.org/abs/hep-ph/0604094}{{\tt hep-ph/0604094}}].

\bibitem{Catani:2011st}
S.~Catani, D.~de~Florian, and G.~Rodrigo, {\it {Space-like (versus time-like)
  collinear limits in QCD: Is factorization violated?}},  {\em JHEP} {\bf 07}
  (2012) 026, [\href{http://arxiv.org/abs/1112.4405}{{\tt arXiv:1112.4405}}].

\bibitem{Catani:1996jh}
S.~Catani and M.~H. Seymour, {\it {The Dipole formalism for the calculation of
  QCD jet cross-sections at next-to-leading order}},  {\em Phys. Lett.} {\bf
  B378} (1996) 287--301, [\href{http://arxiv.org/abs/hep-ph/9602277}{{\tt
  hep-ph/9602277}}].

\bibitem{Catani:1996vz}
S.~Catani and M.~H. Seymour, {\it {A General algorithm for calculating jet
  cross-sections in NLO QCD}},  {\em Nucl. Phys.} {\bf B485} (1997) 291--419,
  [\href{http://arxiv.org/abs/hep-ph/9605323}{{\tt hep-ph/9605323}}]. [Erratum:
  Nucl. Phys.B510,503(1998)].

\bibitem{Catani:1991hj}
S.~Catani, Y.~L. Dokshitzer, M.~Olsson, G.~Turnock, and B.~R. Webber, {\it {New
  clustering algorithm for multi - jet cross-sections in e+ e- annihilation}},
  {\em Phys. Lett.} {\bf B269} (1991) 432--438.

\bibitem{Cacciari:2011ma}
M.~Cacciari, G.~P. Salam, and G.~Soyez, {\it {FastJet User Manual}},  {\em Eur.
  Phys. J.} {\bf C72} (2012) 1896, [\href{http://arxiv.org/abs/1111.6097}{{\tt
  arXiv:1111.6097}}].

\bibitem{Frye:2017yrw}
C.~Frye, A.~J. Larkoski, J.~Thaler, and K.~Zhou, {\it {Casimir Meets Poisson:
  Improved Quark/Gluon Discrimination with Counting Observables}},  {\em JHEP}
  {\bf 09} (2017) 083, [\href{http://arxiv.org/abs/1704.06266}{{\tt
  arXiv:1704.06266}}].

\bibitem{Larkoski:2014wba}
A.~J. Larkoski, S.~Marzani, G.~Soyez, and J.~Thaler, {\it {Soft Drop}},  {\em
  JHEP} {\bf 05} (2014) 146, [\href{http://arxiv.org/abs/1402.2657}{{\tt
  arXiv:1402.2657}}].

\bibitem{Larkoski:2017bvj}
A.~Larkoski, S.~Marzani, J.~Thaler, A.~Tripathee, and W.~Xue, {\it {Exposing
  the QCD Splitting Function with CMS Open Data}},  {\em Phys. Rev. Lett.} {\bf
  119} (2017), no.~13 132003, [\href{http://arxiv.org/abs/1704.05066}{{\tt
  arXiv:1704.05066}}].

\bibitem{Soper:2011cr}
D.~E. Soper and M.~Spannowsky, {\it {Finding physics signals with shower
  deconstruction}},  {\em Phys. Rev.} {\bf D84} (2011) 074002,
  [\href{http://arxiv.org/abs/1102.3480}{{\tt arXiv:1102.3480}}].

\bibitem{lstm1997}
S.~Hochreiter and J.~Schmidhuber, {\it Long short-term memory},  {\em Neural
  computation} {\bf 9} (1997) 1735--80.

\bibitem{DBLP:journals/corr/ChoMGBSB14}
K.~Cho, B.~van Merrienboer, {\c{C}}.~G{\"{u}}l{\c{c}}ehre, F.~Bougares,
  H.~Schwenk, and Y.~Bengio, {\it Learning phrase representations using {RNN}
  encoder-decoder for statistical machine translation},  {\em CoRR} {\bf
  abs/1406.1078} (2014) [\href{http://arxiv.org/abs/1406.1078}{{\tt
  arXiv:1406.1078}}].

\bibitem{chollet2015keras}
F.~Chollet, ``Keras.'' \url{https://keras.io}, 2015.

\bibitem{tensorflow2015-whitepaper}
M.~Abadi et~al., {\it {TensorFlow}: Large-scale machine learning on
  heterogeneous systems},  2015.
\newblock Software available from tensorflow.org.

\bibitem{DBLP:journals/corr/HeZR015}
K.~He, X.~Zhang, S.~Ren, and J.~Sun, {\it Delving deep into rectifiers:
  Surpassing human-level performance on imagenet classification},  {\em CoRR}
  {\bf abs/1502.01852} (2015) [\href{http://arxiv.org/abs/1502.01852}{{\tt
  arXiv:1502.01852}}].

\bibitem{DBLP:journals/corr/KingmaB14}
D.~P. Kingma and J.~Ba, {\it Adam: {A} method for stochastic optimization},
  {\em CoRR} {\bf abs/1412.6980} (2014)
  [\href{http://arxiv.org/abs/1412.6980}{{\tt arXiv:1412.6980}}].

\bibitem{Hocker:2007ht}
A.~Hocker et~al., {\it {TMVA - Toolkit for Multivariate Data Analysis}},  {\em
  PoS} {\bf ACAT} (2007) 040, [\href{http://arxiv.org/abs/physics/0703039}{{\tt
  physics/0703039}}].

\bibitem{Bendavid:2018nar}
J.~R. Andersen et~al., {\it {Les Houches 2017: Physics at TeV Colliders
  Standard Model Working Group Report}},  2018.
\newblock \href{http://arxiv.org/abs/1803.07977}{{\tt arXiv:1803.07977}}.

\bibitem{deFavereau:2013fsa}
{\bf DELPHES 3} Collaboration, J.~de~Favereau, C.~Delaere, P.~Demin,
  A.~Giammanco, V.~Lemaître, A.~Mertens, and M.~Selvaggi, {\it {DELPHES 3, A
  modular framework for fast simulation of a generic collider experiment}},
  {\em JHEP} {\bf 02} (2014) 057, [\href{http://arxiv.org/abs/1307.6346}{{\tt
  arXiv:1307.6346}}].

\bibitem{Katz:2010mr}
A.~Katz, M.~Son, and B.~Tweedie, {\it {Jet Substructure and the Search for
  Neutral Spin-One Resonances in Electroweak Boson Channels}},  {\em JHEP} {\bf
  03} (2011) 011, [\href{http://arxiv.org/abs/1010.5253}{{\tt
  arXiv:1010.5253}}].

\bibitem{Son:2012mb}
M.~Son, C.~Spethmann, and B.~Tweedie, {\it {Diboson-Jets and the Search for
  Resonant Zh Production}},  {\em JHEP} {\bf 08} (2012) 160,
  [\href{http://arxiv.org/abs/1204.0525}{{\tt arXiv:1204.0525}}].

\bibitem{Schaetzel:2013vka}
S.~Schaetzel and M.~Spannowsky, {\it {Tagging highly boosted top quarks}},
  {\em Phys. Rev.} {\bf D89} (2014), no.~1 014007,
  [\href{http://arxiv.org/abs/1308.0540}{{\tt arXiv:1308.0540}}].

\bibitem{Larkoski:2015yqa}
A.~J. Larkoski, F.~Maltoni, and M.~Selvaggi, {\it {Tracking down hyper-boosted
  top quarks}},  {\em JHEP} {\bf 06} (2015) 032,
  [\href{http://arxiv.org/abs/1503.03347}{{\tt arXiv:1503.03347}}].

\bibitem{Bressler:2015uma}
S.~Bressler, T.~Flacke, Y.~Kats, S.~J. Lee, and G.~Perez, {\it {Hadronic
  Calorimeter Shower Size: Challenges and Opportunities for Jet Substructure in
  the Superboosted Regime}},  {\em Phys. Lett.} {\bf B756} (2016) 137--141,
  [\href{http://arxiv.org/abs/1506.02656}{{\tt arXiv:1506.02656}}].

\bibitem{Han:2017hyv}
Z.~Han, M.~Son, and B.~Tweedie, {\it {Top-Tagging at the Energy Frontier}},
  {\em Phys. Rev.} {\bf D97} (2018), no.~3 036023,
  [\href{http://arxiv.org/abs/1707.06741}{{\tt arXiv:1707.06741}}].

\bibitem{CMS:2014joa}
{\bf CMS} Collaboration, C.~Collaboration, {\it {V Tagging Observables and
  Correlations}}, .

\bibitem{ATLAS:2016vmy}
{\bf ATLAS} Collaboration, T.~A. collaboration, {\it {Jet mass reconstruction
  with the ATLAS Detector in early Run 2 data}}, .

\bibitem{Cacciari:2014gra}
M.~Cacciari, G.~P. Salam, and G.~Soyez, {\it {SoftKiller, a particle-level
  pileup removal method}},  {\em Eur. Phys. J.} {\bf C75} (2015), no.~2 59,
  [\href{http://arxiv.org/abs/1407.0408}{{\tt arXiv:1407.0408}}].

\bibitem{Bertolini:2014bba}
D.~Bertolini, P.~Harris, M.~Low, and N.~Tran, {\it {Pileup Per Particle
  Identification}},  {\em JHEP} {\bf 10} (2014) 059,
  [\href{http://arxiv.org/abs/1407.6013}{{\tt arXiv:1407.6013}}].

\bibitem{Berta:2014eza}
P.~Berta, M.~Spousta, D.~W. Miller, and R.~Leitner, {\it {Particle-level pileup
  subtraction for jets and jet shapes}},  {\em JHEP} {\bf 06} (2014) 092,
  [\href{http://arxiv.org/abs/1403.3108}{{\tt arXiv:1403.3108}}].

\bibitem{Komiske:2017ubm}
P.~T. Komiske, E.~M. Metodiev, B.~Nachman, and M.~D. Schwartz, {\it {Pileup
  Mitigation with Machine Learning (PUMML)}},  {\em JHEP} {\bf 12} (2017) 051,
  [\href{http://arxiv.org/abs/1707.08600}{{\tt arXiv:1707.08600}}].

\bibitem{Cacciari:2007fd}
M.~Cacciari and G.~P. Salam, {\it {Pileup subtraction using jet areas}},  {\em
  Phys. Lett.} {\bf B659} (2008) 119--126,
  [\href{http://arxiv.org/abs/0707.1378}{{\tt arXiv:0707.1378}}].

\bibitem{Cacciari:2008gn}
M.~Cacciari, G.~P. Salam, and G.~Soyez, {\it {The Catchment Area of Jets}},
  {\em JHEP} {\bf 04} (2008) 005, [\href{http://arxiv.org/abs/0802.1188}{{\tt
  arXiv:0802.1188}}].

\bibitem{Krohn:2009th}
D.~Krohn, J.~Thaler, and L.-T. Wang, {\it {Jet Trimming}},  {\em JHEP} {\bf 02}
  (2010) 084, [\href{http://arxiv.org/abs/0912.1342}{{\tt arXiv:0912.1342}}].

\bibitem{Dreyer:2018tjj}
F.~A. Dreyer, L.~Necib, G.~Soyez, and J.~Thaler, {\it {Recursive Soft Drop}},
  {\em JHEP} {\bf 06} (2018) 093, [\href{http://arxiv.org/abs/1804.03657}{{\tt
  arXiv:1804.03657}}].

\bibitem{NIPS2014_5423}
I.~Goodfellow, J.~Pouget-Abadie, M.~Mirza, B.~Xu, D.~Warde-Farley, S.~Ozair,
  A.~Courville, and Y.~Bengio, {\it Generative adversarial nets},  in {\em
  Advances in Neural Information Processing Systems 27} (Z.~Ghahramani,
  M.~Welling, C.~Cortes, N.~D. Lawrence, and K.~Q. Weinberger, eds.),
  pp.~2672--2680.
\newblock Curran Associates, Inc., 2014.

\bibitem{Louppe:2016ylz}
G.~Louppe, M.~Kagan, and K.~Cranmer, {\it {Learning to Pivot with Adversarial
  Networks}},  \href{http://arxiv.org/abs/1611.01046}{{\tt arXiv:1611.01046}}.

\bibitem{Shimmin:2017mfk}
C.~Shimmin, P.~Sadowski, P.~Baldi, E.~Weik, D.~Whiteson, E.~Goul, and
  A.~Søgaard, {\it {Decorrelated Jet Substructure Tagging using Adversarial
  Neural Networks}},  {\em Phys. Rev.} {\bf D96} (2017), no.~7 074034,
  [\href{http://arxiv.org/abs/1703.03507}{{\tt arXiv:1703.03507}}].

\bibitem{Andrews:2018jcm}
H.~A. Andrews et~al., {\it {Novel tools and observables for jet physics in
  heavy-ion collisions}},  \href{http://arxiv.org/abs/1808.03689}{{\tt
  arXiv:1808.03689}}.

\bibitem{Chien:2018dfn}
Y.-T. Chien and R.~Kunnawalkam~Elayavalli, {\it {Probing heavy ion collisions
  using quark and gluon jet substructure}},
  \href{http://arxiv.org/abs/1803.03589}{{\tt arXiv:1803.03589}}.

\bibitem{AndrewsALICEQM18}
{\bf ALICE} Collaboration, H.~Andrews, {\it Exploring phase space of jet
  splittings at alice using grooming and recursive techniques},  2018.
\newblock Talk at Quark Matter 2018, Venice, Italy,
  \url{https://indico.cern.ch/event/656452/contributions/2869941/attachments/1649044/2636550/HarryAndrews_QuarkMatter18Final.pdf}.

\bibitem{Dolen:2016kst}
J.~Dolen, P.~Harris, S.~Marzani, S.~Rappoccio, and N.~Tran, {\it {Thinking
  outside the ROCs: Designing Decorrelated Taggers (DDT) for jet
  substructure}},  {\em JHEP} {\bf 05} (2016) 156,
  [\href{http://arxiv.org/abs/1603.00027}{{\tt arXiv:1603.00027}}].

\bibitem{Hall:2018jub}
Z.~Hall and J.~Thaler, {\it {Photon isolation and jet substructure}},
  \href{http://arxiv.org/abs/1805.11622}{{\tt arXiv:1805.11622}}.

\bibitem{DBLP:journals/corr/TaiSM15}
K.~S. Tai, R.~Socher, and C.~D. Manning, {\it Improved semantic representations
  from tree-structured long short-term memory networks},  {\em CoRR} {\bf
  abs/1503.00075} (2015) [\href{http://arxiv.org/abs/1503.00075}{{\tt
  arXiv:1503.00075}}].

\bibitem{Dasgupta:2018nvj}
M.~Dasgupta, F.~A. Dreyer, K.~Hamilton, P.~F. Monni, and G.~P. Salam, {\it
  {Logarithmic accuracy of parton showers: a fixed-order study}},
  \href{http://arxiv.org/abs/1805.09327}{{\tt arXiv:1805.09327}}.

\bibitem{Elder:2018mcr}
B.~T. Elder and J.~Thaler, {\it {Aspects of Track-Assisted Mass}},
  \href{http://arxiv.org/abs/1805.11109}{{\tt arXiv:1805.11109}}.

\end{thebibliography}\endgroup

\end{document}